\documentclass[aps,floatfix,eqsecnum,showpacs,preprint,tightenlines]{revtex4-1}
\usepackage{epsfig}

\usepackage{amsmath}
\usepackage[labelformat=simple]{subcaption}

\usepackage{enumerate}
\begin{document}


\title{Theoretical uncertainties of the elastic nucleon-deuteron scattering 
observables}
\author{R. Skibi{\'n}ski}
\affiliation{M. Smoluchowski Institute of Physics, Jagiellonian University, PL-30348 Krak\'ow, Poland}
\author{Yu. Volkotrub}
\affiliation{M. Smoluchowski Institute of Physics, Jagiellonian University, PL-30348 Krak\'ow, Poland}
\author{J. Golak}
\affiliation{M. Smoluchowski Institute of Physics, Jagiellonian University, PL-30348 Krak\'ow, Poland}
\author{K. Topolnicki}
\affiliation{M. Smoluchowski Institute of Physics, Jagiellonian University, PL-30348 Krak\'ow, Poland}
\author{H. Wita{\l}a}
\affiliation{M. Smoluchowski Institute of Physics, Jagiellonian University, PL-30348 Krak\'ow, Poland}

\date{\today}

\begin{abstract}

Theoretical uncertainties 
of various types 
are discussed for the nucleon-deuteron elastic scattering observables at the incoming nucleon laboratory 
energies up to 200 MeV.
We are especially interested in the statistical errors 
arising from uncertainties of parameters of a nucleon-nucleon interaction.
The obtained uncertainties of the differential cross section and numerous scattering
observables are in general small, grow with the reaction energy and amount up to a few percent at 200 MeV.
We compare these uncertainties with the other types of theoretical errors 
like truncation errors, numerical uncertainties and 
uncertainties arising from using the various models of nuclear interaction. 
We find the latter ones 
to be dominant source of uncertainties of modern predictions for the three-nucleon scattering observables.
To perform above mentioned studies we use the One-Pion-Exchange Gaussian potential derived by the Granada group,
for which the covariance matrix of its parameters is known, 
and solve the Faddeev equation
for the nucleon-deuteron elastic scattering. 
Thus beside studying theoretical uncertainties we also 
show a description of the nucleon-deuteron elastic scattering data by the One-Pion-Exchange Gaussian model
and compare it with results obtained with other nucleon-nucleon potentials, 
including chiral N$^4$LO forces from the Bochum-Bonn and Moscow(Idaho)-Salamanca groups.
In this way we confirm the usefulness and high quality of the One-Pion-Exchange Gaussian force.   

\end{abstract}

\pacs{21.45.-v, 13.75.Cs, 25.40.Cm}

\maketitle


\section{Introduction}

One of the main goals of nuclear physics is to establish properties of the nuclear interactions.
After many years of investigations we are now in position to study details of the
nuclear forces both from the theoretical as well as the experimental sides. It has been found that
the three-nucleon (3$N$) system, which allows to probe also the off-energy-shell properties of 
the nuclear potential, is especially important for such studies. Moreover, to obtain 
a precise description of the 3$N$ data one has to supplement the two-nucleon (2$N$) interaction 
by a 3$N$ force acting in this system. Currently the structure of 3$N$ force is
still unclear and many efforts are directed to fix 3$N$ force properties. However, in order to obtain 
trustable and precise information from a comparison of 3$N$ data with predictions based 
on theoretical models it is necessary to take into account, or at least to estimate, 
in addition to the uncertainties of data also the errors of theoretical predictions.

The precision of the experimental data has significantly increased and achieved in 
recent measurements a high 
level, 
see e.g. Refs.~\cite{Sekiguchi, Weisel, Przewoski, Kistryn, Howell} for examples of  
state-of-the-art experimental studies in the three-nucleon sector. Precision of these and other 
experiments has become so high that the question about the uncertainties of the theoretical
predictions is very timely~\cite{PRA_Editors_remark}. In the past the theoretical uncertainties for observables 
in three-nucleon reactions were estimated by comparing predictions based on various models of
nuclear interactions~\cite{Witala2001} or by performing benchmark calculations using the
same interaction but various theoretical approaches~\cite{benchmarks,COR90,HUB95,FRI90b,FRI95}.
Such a strategy was dictated by a) a common belief that a poor knowledge about the nuclear forces, 
reflected by the existence of very different models of nuclear interaction, is a dominant source of the 
theoretical uncertainty,
b) lack of knowledge about the correlations between nucleon-nucleon ($NN$) potential parameters, 
c) using inconsistent models of 2$N$ and 3$N$ forces,
and last but not least d) a magnitude of uncertainties of experimental data available at the time.
Nowadays these arguments, at least partially, are no longer valid due to
the above mentioned progress in experimental techniques, progress in the derivation of consistent 2$N$ and 3$N$ 
interactions, e.g. within the Chiral Effective Field 
Theory ($\chi$EFT) ~\cite{Epelbaum_review, Bernard1, Bernard2, Machleidt-review, Machleidt2017} and
due to availability of new models of nuclear forces, where free parameters are fixing by  
performing a careful statistical analysis~\cite{Navarro_Corase,Navarro_OPE}.
As a consequence, the estimation of theoretical uncertainties has become again an important issue in
theoretical studies.

An extensive introduction to an error estimation for theoretical models was given 
in Ref.~\cite{Dobaczewski}, followed by a special issue of J. Phys. G: Nucl. Part. Phys.~\cite{SpecialIssue}. 
In the latter reference 
many applications of the error estimation to nuclear systems and processes are discussed.
However, omitting the few-nucleon reactions, the authors focus mainly on models used in direct fitting to data or on
models used in nuclear structure studies. 
Among the other papers focused on the estimation of theoretical uncertainties of $NN$ interaction we refer the reader to
works by A.Ekstr\"om et al.~\cite{Ekstrom}, R.Navarro P\'erez et al.~\cite{Navarro_OPE,Navarro_TPE,Navarro-review} and 
to a recent work by P.Reinert et al.~\cite{Reinert}. Simultaneously, the Bayesian approach to 
estimate uncertainties in the 2$N$ system was derived in Ref.~\cite{Furnstahl} with some applications shown 
again in Ref.~\cite{SpecialIssue}.
Beyond the 2$N$ system, the uncertainty of theoretical models has been recently studied in the context 
of nuclear structure calculations 
for which such an evaluation is important also from a practical point of view. 
Namely, predictions for many-nucleon systems require not only a huge amount of advanced
computations but also
rely, e.g. in the case of the No-Core shell model~\cite{NCSM}, on extrapolations to large model spaces.
A knowledge of precision of the theoretical models is important for efficient use of available 
computer resources.

Studies of theoretical uncertainties in few-nucleon reactions are less advanced. Beside the above mentioned 
attempts to estimate their magnitudes by means of benchmark calculations most efforts
in the field  were orientated to estimate uncertainties present in the $\chi$EFT approach~\cite{Epelbaum-review}.
In this case three sources of theoretical uncertainties have been investigated:
the truncation of the chiral expansion at a finite order (what results in the so-called truncation errors), 
the introduction of regulator functions (what results in a cut-off dependence), and
the procedure of fixing values of low-energy constants.  
A simple prescription how to estimate the truncation errors was proposed by E.Epelbaum and collaborators for 
the 2$N$ system~\cite{imp1} and adopted also for 3$N$ systems, for the case where predictions were based 
on a two-body interaction~\cite{Binder} only.
It was found that both for pure nuclear systems~\cite{Binder}, as well as for electroweak processes~\cite{Skibinski_electroweak}
the magnitude of truncation errors strongly decreases with the order of chiral expansion and 
at the fifth order (N$^4$LO) it becomes relatively small.
The prescription of Ref.~\cite{imp1} is in agreement with the Bayesian approach~\cite{Furnstahl},
see also the recent work~\cite{Melendez} for a discussion of the Bayesian truncation errors for the $NN$ observables.
The dependence of the chiral predictions on used regulator functions and their parameters has been
studied since the first applications of chiral potentials to the 2$N$ and 3$N$ systems~\cite{Epelbaum_start, Epelbaum_3N, Machleidt_force}.
The regulator dependence of chiral forces was broadly discussed in the past, see e.g.~\cite{Machleidt_regul} and various 
regulator functions were proposed.
The non-local regularization in the momentum space was initially used and estimations of the theoretical 
uncertainties of the 2$N$ and many-body observables related to regulators were made by comparing predictions 
obtained with various values of regularization parameters. 
It was found that the non-local regularization leads to an unwanted dependence of observables on the 
parameters used. This dependence was especially strong for predictions for the nucleon-deuteron ($Nd$) 
elastic scattering based on 
2$N$ and 3$N$ forces at the next-to-next-to-next-to leading order (N$^3$LO) of chiral expansion~\cite{Witala_lenpic1}
and for the electromagnetic processes in the 3$N$ systems when also the leading meson-exchange currents were
taken into account~\cite{Rozpedzik, Skibinski_chiral_elmag_mec}.
These results were one of the reasons for introducing another, the so-called "semi-local" method 
of regularization of chiral forces.
Such an improved method was presented and applied to the $NN$ system in Refs.~\cite{imp1,imp2},
leading to weak cut-off dependence of predictions in two-body system 
at chiral orders above the leading order.
Similar picture of weak dependence of predictions based on the chiral forces of Refs.~\cite{imp1,imp2} 
was found for $Nd$ elastic scattering~\cite{Binder} and for various electroweak processes~\cite{Skibinski_electroweak}.
Also the nuclear structure calculations confirmed this observation~\cite{improved_in_structure,lenpic_long_paper}. 

The estimation of the theoretical uncertainties arising from an uncertainty of the potential parameters 
(which we will call in the following also a statistical error)
has not been studied yet, to the best of our knowledge, in $Nd$ scattering. Within this paper we investigate 
how such statistical uncertainties propagate from the $NN$ potential parameters to the $Nd$ scattering observables.
We also compare them with the remaining theoretical uncertainties for the same observables.  
To this end we use, for the first time in $Nd$ scattering, the One-Pion-Exchange (OPE) Gaussian $NN$ interaction derived recently 
by the Granada group~\cite{Navarro_OPE}.
The knowledge of the covariance matrix of the OPE-Gaussian potential parameters is a distinguishing 
feature of this interaction. This is also crucial for our investigations as we use a statistical 
approach to estimate theoretical uncertainties.
Namely, given the covariance matrix for the potential parameters, we sample 50 sets of the potential parameters
and, after calculating for each set the 3$N$ observables, we study statistical properties of the obtained 
predictions. The OPE-Gaussian interaction is described briefly in Sec.~\ref{formalism} 
and our method to obtain statistical errors is discussed step by step in Sec.~\ref{results}. 
The OPE-Gaussian force has been already used, within the same method, to estimate the statistical 
uncertainty of the $^3$H binding energy~\cite{Perez_CONF} which was found to be around 15 keV ($\approx 0.16\%$).

The paper is organized as follows: in Sec.~\ref{formalism} we show the essential elements of our formalism,
describe its numerical realization and give some more information on the OPE-Gaussian potential and the chiral models used.
In Sec.~\ref{results} we present predictions for the $Nd$ elastic scattering observables obtained with the OPE-Gaussian force
and compare them with predictions based on the AV18 $NN$ potential~\cite{AV18}. We also discuss
various estimators of uncertainties in hand for the 3$N$ scattering observables. In Sec.~\ref{results_various} we compare,
for a few chosen observables, the theoretical uncertainties arising from various sources, including the truncation errors and
the regulator dependence. Here, beside the OPE-Gaussian potential and other semi-phenomenological $NN$ forces, 
we also use the chiral interaction of Ref.~\cite{imp1,imp2} and, 
for the first time in the $Nd$ scattering, the chiral N$^4$LO interaction
recently derived by the Moscow(Idaho)-Salamanca group~\cite{Machleidt2017}.
Finally, we summarize in Sec.~\ref{Summary}.
  
\section{Formalism}
\label{formalism}
The formalism of the momentum space Faddeev equation is one of the standard techniques to investigate 3$N$ reactions 
and has been described in detail many times, see e.g.~\cite{Glockle-raport,Glockle-book}.
Thus we 
only briefly remind the reader of its key elements. 

For a given $NN$ interaction $V$ we solve
the Lippmann-Schwinger equation $t = V + V\tilde{G_0}t$ 
to obtain matrix elements of the 2$N$ $t$ operator,
with $\tilde{G_0}$ being the 2$N$ free propagator. 
These matrix elements enter the 3$N$ Faddeev scattering equation which, neglecting the 3$N$ force, 
takes the following form
\begin{equation}
T \vert \phi \rangle = tP \vert \phi \rangle + tPG_0 T \vert \phi \rangle.
\label{eq_Fadd_NN_3N}
\end{equation}
The initial state $\vert \phi \rangle$ is composed of a deuteron 
and a momentum eigenstate of the projectile nucleon, $G_0$ is the free 3$N$ propagator 
and $P$ is a permutation operator. 

The transition amplitude for the elastic $Nd$ scattering process
$\langle \phi' \vert U \vert \phi \rangle$ contains the final channel state $\vert \phi' \rangle$
and is obtained as 
\begin{equation}
\langle \phi' \vert U \vert \phi \rangle = \langle \phi' \vert PG_0^{-1} \vert \phi \rangle 
+ \langle \phi' \vert PT \vert \phi \rangle\;,
\label{eq.U}
\end{equation}
from which observables can be obtained in the standard way~\cite{Glockle-raport}.

Equation~(\ref{eq_Fadd_NN_3N}) is solved in the partial wave basis comprising all 
3$N$ states with the two-body subsystem total angular momentum $j \leq 5$
and the total 3$N$ angular momentum $J \leq \frac{25}{2}$.

Since we obtained the bulk of our results with the OPE-Gaussian interaction~\cite{Navarro_OPE},
we briefly remind now the reader of a structure of this potential. A basic concept at the heart of this force 
is analogous to the one stated behind the well-known AV18 interaction~\cite{AV18}.
The OPE-Gaussian potential $V(\vec{r})$ is composed of the long-range 
$V_{long}(\vec{r})$ and the short-range $V_{short}(\vec{r})$ parts
\begin{equation}
\label{label1}
V(\vec{r}) = V_{short}(\vec{r})\theta (r_{c}-r)+V_{long}(\vec{r})\theta (r-r_{c}),
\end{equation}
where $r_{c}$=3~fm and the $V_{long}(\vec{r})$ part contains the OPE force and the electromagnetic corrections.
The $V_{short}(\vec{r})$ component is built from 18 operators $\hat{O}_{n}$, among which 16 
are the same as in the AV18 model. Each of them is multiplied by a linear combination of the Gaussian 
functions $F_{k}(r)= \exp{(-r^2 / (2a_k^2))}$, with $a_k=\frac{a}{1+k}$, and the strength coefficients $V_{k,n}$:
\begin{equation}
V_{short}(\vec{r})= \sum\limits_{n=1}^{18} \hat{O}_{n}\left[  \sum\limits_{k=1}^{4} V_{k,n} F_{k}(r)\right]. 
\label{eq.OPEGshort}
\end{equation}
The free parameter $a$ present in the $F_{k}(r)$ functions together with the parameters $V_{k,n}$ 
have been fixed from the data.
It is worth noting that to this end 
the "3$\sigma$ self-consistent database"~\cite{Navarro_Corase} was used. It incorporates 6713 proton-proton and 
neutron-proton data, gathered within 
the years 1950 to 2013, in the laboratory energy range $E_{lab}$ up to 350 MeV. 
The careful statistical revision of data and the fitting  
procedure
allowed the authors of Ref.~\cite{Navarro_OPE}
to confirm good statistical properties of their $\chi^2$ fit, e.g. by checking the normality of residuals.
The $\chi^2/data$ for the OPE-Gaussian force is 1.06 as fitted to data enumerated in Ref.~\cite{Navarro_Corase}.
We have been equipped by the authors of Ref.~\cite{Navarro_OPE} with 50 sets of parameters $\{ V_{k,n}, a\}$ 
obtained by a correlated sampling from the multivariate normal distribution with a known 
covariance matrix (see~\cite{Navarro_sampling} for details). 
The OPE-Gaussian model, as having a similar structure to the AV18 force, but being fitted to the newer data
can be regarded as a refreshed version of the standard AV18 model.
In the $NN$ sector these two potentials lead to a slightly different description of
phase shifts, especially at energies above 150~MeV in the $^3F_2$ and $^3D_3$ partial waves~\cite{Navarro_OPE}. 
Thus it seems to be interesting to compare predictions for $Nd$ scattering given by both potentials.

Beside the OPE-Gaussian and the AV18 models we show in Sec.~\ref{results_various} predictions 
based on two chiral forces at N$^4$LO, derived by R.Machleidt and collaborators~\cite{Machleidt2017} 
and by E.Epelbaum and collaborators~\cite{imp1,imp2}. 
In the case of the first of these forces 
the non-local regularization, applied directly in momentum space, has been used. 
The regulator function is taken as $f(p',p)=\exp{(\,\rm{-}\,(\frac{p'}{\Lambda})^{2n}\,\rm{-}\,(\frac{p}{\Lambda})^{2n})}$,
where $n$ depends on regarded operators (e.g. $n=4$ for the one-pion exchange potential). 
Three values of the cutoff parameter $\Lambda$ (450, 500 and 550 MeV) were suggested  
for this potential and are also used in this paper.
In the case of the N$^4$LO potential and $\Lambda=500$~MeV the $\chi^2/{\rm data}=1.15$ 
for the combined neutron-proton and proton-proton data in the energy range 0-290~MeV~\cite{Machleidt2017}.
In this paper we show for the first time the predictions 
of this new chiral potential at N$^4$LO for the $Nd$ elastic scattering observables.
As mentioned above, in the approach of Refs.~\cite{imp1,imp2}
the semi-local regularization of nuclear forces is performed in coordinate space with the regulator function
$f(r)  =  [1-\exp{(\,\rm{-}\,(\frac{r}{R})^2)}]^6$, where $r$ is the distance between nucleons
and $R$ is the regulator parameter.
The authors of Ref.~\cite{imp1} suggested five values of the regulator $R=$0.8, 0.9, 1.0, 1.1, and 1.2~fm.
The best description of the $NN$ observables is achieved with $R$=0.9~fm and $R$=1.0~fm,
and leads to the 
$\chi^2/{\rm data} \approx 1.14$ at $R$=0.9~fm for the N$^4$LO force~\cite{Reinert} when 
using the ``3$\sigma$-self-consistent database" from Ref.~\cite{Navarro_Corase}. This value 
is comparable with the ones obtained for the semi-phenomenological potentials.

\section{The OPE-Gaussian predictions for Nd scattering and their statistical errors}
\label{results}

\subsection{Determination of statistical uncertainty in 3N system}

To determine the theoretical uncertainty arising from the 2$N$ potential parameters
we took the following steps:
\begin{enumerate}
\item
We prepared various sets of the potential parameters.

Actually, this step had been already taken by the Granada group as a part of their 
study of the statistical uncertainty of the $^3$H binding energy.
They provided us with fifty sets ($S_i$ with $i=1,\dots,50$) 
of 42 potential parameters 
(drawn from the multivariate normal distribution with known expectation values and 
covariance matrix) and one set of expectation values of potential parameters ($S_0$).
Such a relatively big sample of fifty-one sets allows us to obtain statistically meaningful conclusions.

\item
For each set $S_i$ ($i=0,1,\dots,50$) we calculated the deuteron wave function and the $t$ matrix, solved,
at each considered energy, 
the Faddeev equation~(\ref{eq_Fadd_NN_3N}),
calculated the scattering amplitude~(Eq.~(\ref{eq.U})) and finally computed observables.
As a result the angular dependence of
various scattering observables is known for each set of parameters $S_i$.

\end{enumerate}

The predictions obtained in such a way allow us to study: 
\begin{enumerate}[a)]
\item 
for a given energy $E$, an observable $O$, and a scattering angle $\theta$, 
the empirical probability density function of the observable $O(E,\theta)$ resulting when 
various sets $S_i, (i=1,\dots,50)$ are used;

\item for a given observable $O$, both the angular and energy dependencies 
of results based on various sets $S_i$.

\end{enumerate}

Based on these studies we can conclude on the measure of statistical uncertainties and quality of
elastic $Nd$ scattering data description.
This is a content of the next two subsections.

\subsection{Measure of statistical uncertainty}

\begin{figure}[hbpt]
\includegraphics[width=.6\textwidth,clip=true]{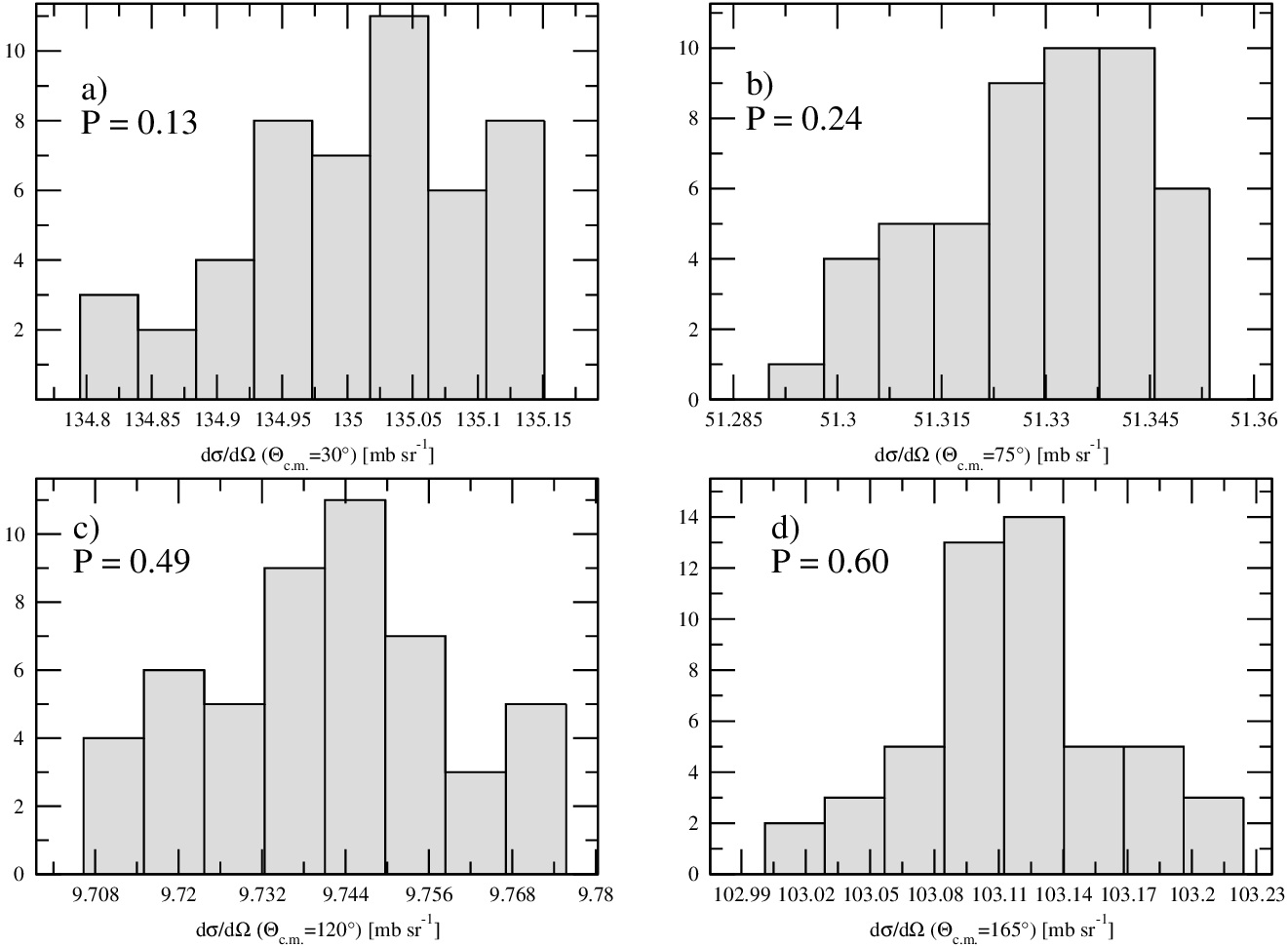}
\caption{The histograms and the P-values for the Shapiro-Wilk test for the elastic $Nd$ scattering differential 
cross section d$\sigma$/d$\Omega$~[mb~sr$^{-1}$] at the incoming nucleon laboratory energy $E$=13~MeV 
and the scattering angle: a) $\theta_{\rm{c.m.}} =30^\circ$, b) $\theta_{\rm{c.m.}} =75^\circ$, 
c) $\theta_{\rm{c.m.}} =120^\circ$, and d) $\theta_{\rm{c.m.}} =165^\circ$, 
obtained with 50 sets of the OPE-Gaussian potential parameters.}
\label{fig1}
\end{figure}

\begin{figure}[hbpt]
\includegraphics[width=.6\textwidth,clip=true]{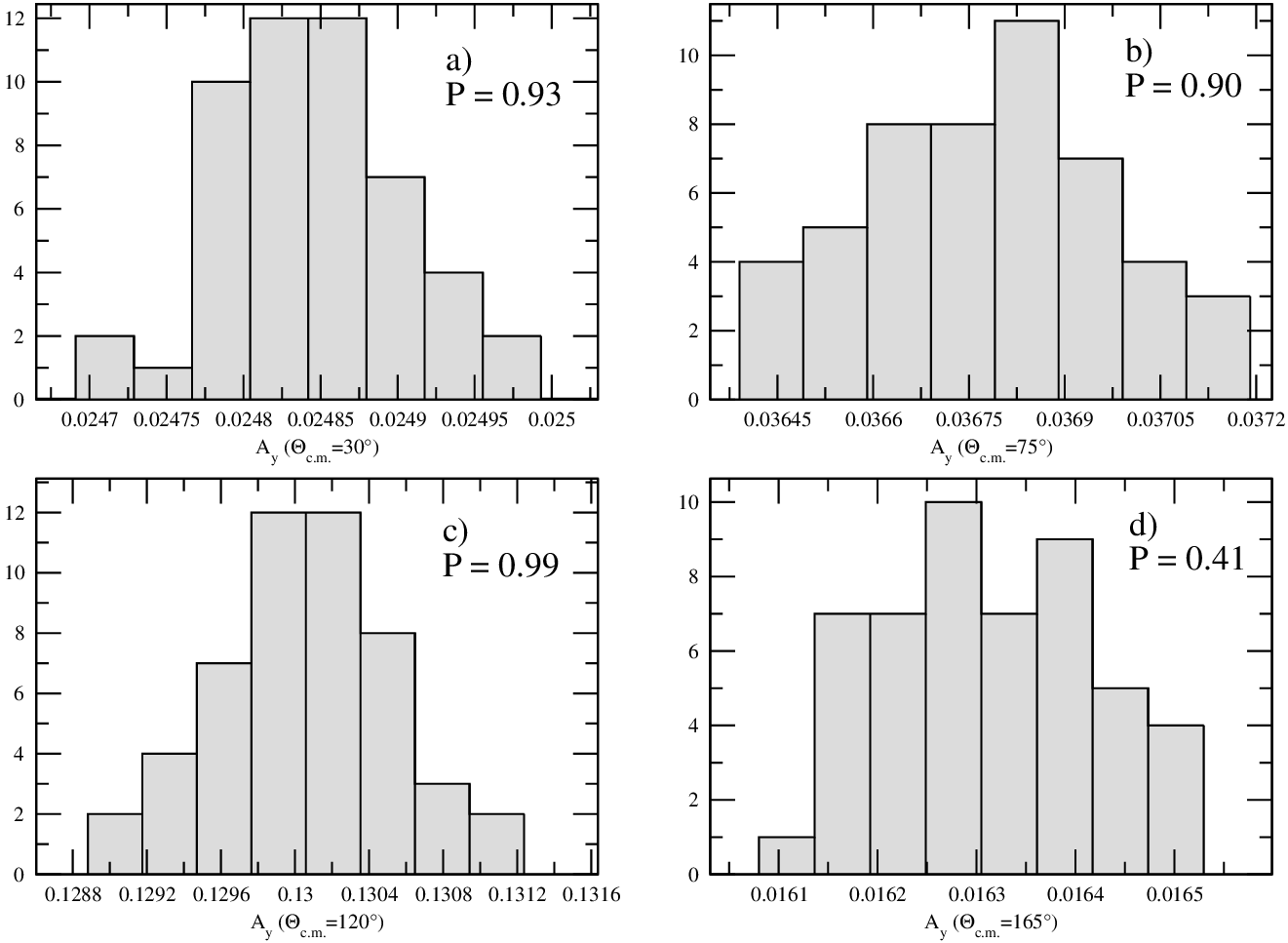}
\caption{The histograms and the P-values for the Shapiro-Wilk test for the nucleon analyzing power A$_{\rm{y}}$
in $Nd$ elastic scattering at the incoming nucleon laboratory energy $E$=13~MeV
and the scattering angle: a) $\theta_{\rm{c.m.}} =30^\circ$, b) $\theta_{\rm{c.m.}} =75^\circ$, 
c) $\theta_{\rm{c.m.}} =120^\circ$, and d) $\theta_{\rm{c.m.}} =165^\circ$, 
obtained with 50 sets of the OPE-Gaussian potential parameters.}
\label{fig2}
\end{figure}

\begin{figure}[hbpt]
\includegraphics[width=.6\textwidth,clip=true]{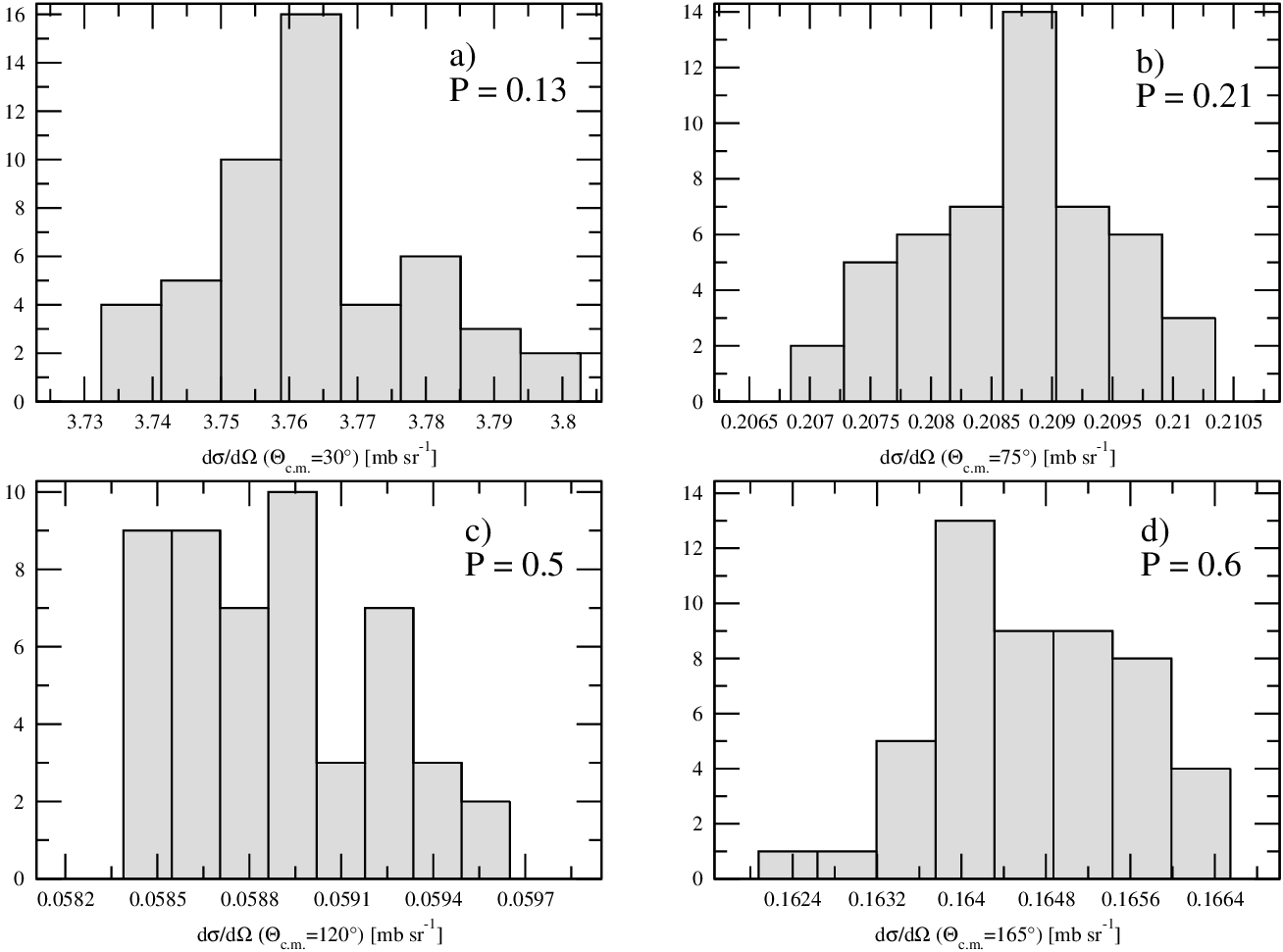}
\caption{The same as in Fig.~\ref{fig1} but at $E$=200 MeV.}
\label{fig3}
\end{figure}

\begin{figure}[hbpt]
\includegraphics[width=.6\textwidth,clip=true]{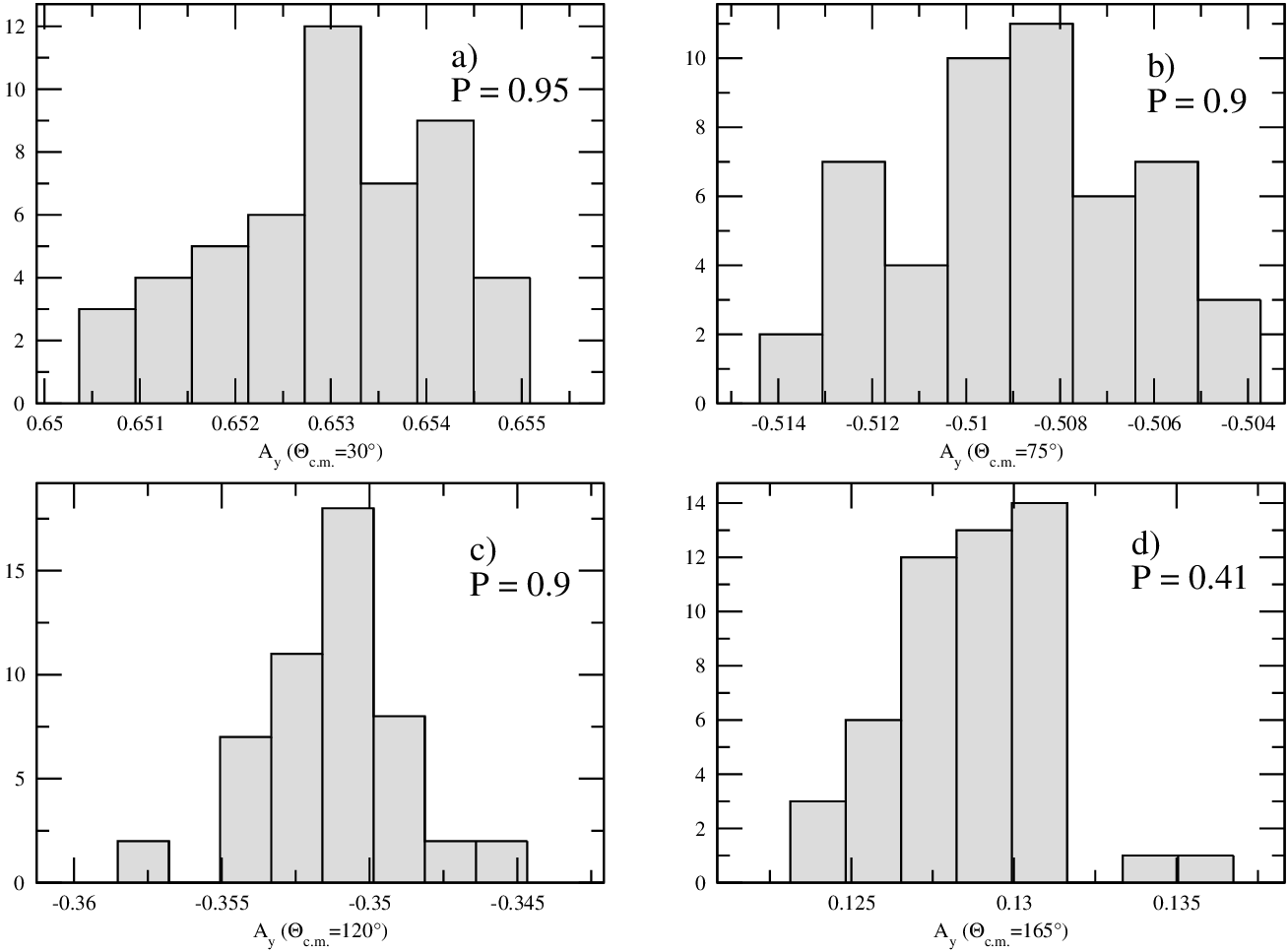}
\caption{The same as in Fig.~\ref{fig2} but at $E$=200 MeV.}
\label{fig4}
\end{figure}

Our first task is to choose an estimator of the theoretical uncertainties in question.
Due to a big complexity of calculations required to obtain the 3$N$ scattering observables we are not able 
a priori to determine analytically the probability distribution function of the resulting 3$N$ predictions
and consequently to choose the best estimator to describe the dispersion of results.
In Figs.~\ref{fig1} and~\ref{fig2} we show the empirical distributions (histograms) of the cross section $d\sigma/d\Omega$ 
and the nucleon analyzing power A$_{\rm{y}}$ at the nucleon laboratory energy $E$=13~MeV and at four c.m. scattering angles:
$\theta_{\rm{c.m.}}$ = 30$^\circ$, 75 $^\circ$, 120$^\circ$ and 165$^\circ$. 
The same observables at the same $\theta_{\rm{c.m.}}$ angles but at $E$=200 MeV are shown in Figs.~\ref{fig3} and~\ref{fig4}, respectively.
It is clear that the distribution of the predictions cannot be regarded as the normal distribution.
To obtain quantitative information on the distribution we have performed the Shapiro-Wilk test 
\cite{Shapiro-Wilk-Test}, which belongs to 
the strongest statistical tests of normality. 
As is seen from the obtained P-values 
(the smaller P-value the more unlikely the predictions are normally distributed)
given in Figs.~\ref{fig1}-\ref{fig4}, in many cases the resulting distributions of the
cross section and the nucleon analyzing power cannot be regarded with high confidence as normal distributions.
This restricts
a choice of the dispersion estimators - neither the commonly used confidence interval 
nor the usual estimators for 
the standard deviation can be used directly as they are tailored to the normal distribution.
Thus we considered the following estimators for the statistical error of the 
observable $O(E,\theta)$ (at a given energy and a scattering angle):
\begin{enumerate}  
\item 
$\frac12 \Delta_{100\%} \equiv \frac12 ( max_i(O_i)-min_i(O_i)\,)$, where the minimum and maximum are taken over all predictions 
based on different sets of the $NN$ potential parameters $S_i$, $i=1,2,\dots,50$;
\item
$\frac12 \Delta_{68\%} \equiv \frac12 ( max_i(O_i)-min_i(O_i)\,)$, where the minimum and maximum are taken over 
34 (68\% of 50) predictions based on different sets of the $NN$ potential parameters; The set of 34 observables is constructed by disposing of 
the 8 smallest and the 8 biggest predictions for the observable $O(E,\theta)$;
\item
$\frac12$IQR -- the half of standard estimator of the interquartile range being the difference 
between the third and the first quartile $IQR=Q_3-Q_1$.
For the sample of size 50 this corresponds to taking the half of difference between the predictions on 37th and 13th 
position in a sample sorted in the ascending order. The flexibility in applying this measure to the non-normal distribution is
a great asset to the IQR;
\item
$\sigma(O)$ -- the sample standard deviation $\sigma(O) = \sqrt{ \frac{1}{n-1} \sum_{i=1}^{n} (x_i-\bar{x})^2}$, where $\bar{x}$
is the usual mean value. 
The disadvantage of this estimator is that on formal grounds it cannot be applied to samples from an arbitrary 
probability distribution.
\end{enumerate}

The $\frac12 \Delta_{100\%}$ and the $\sigma(O)$ are sensitive to the possible outliers in the sample 
and thus taking them as estimators of dispersion can lead to overestimation of the statistical error.
On the other hand the IQR is calculated using only half of the elements in the sample and thus can lead to 
underestimation of the theoretical uncertainty.
Thus we decided to adapt the $\frac12 \Delta_{68\%}$ as an optimal measure of predictions' dispersion and 
consequently as an estimator of the theoretical uncertainty 
in question. The same choice has been made
in a study of the statistical error of the $^3H$ binding energy in Ref.~\cite{Navarro_sampling}.
The similarity to the standard deviation is one more advantage of $\frac12 \Delta_{68\%}$ since 
the comparison of the theoretical errors with the experimental (statistical) uncertainties, delivered usually 
in the form of standard deviations, is finally unavoidable.

However, in Tab.~\ref{tab1} we compare values of the above mentioned estimators for the $Nd$ elastic scattering 
differential cross section at three energies of the incoming nucleon and at four c.m. scattering angles. By definition   
$\frac12{\rm IQR} \leq \frac12 \Delta_{68\%} \leq \frac12 \Delta_{100\%}$ and indeed this is observed in Tab.~\ref{tab1}.
The magnitudes of the $\frac12 \Delta_{68\%}$ is very close to the measure based on the sample standard deviation
$\sigma(d\sigma/d\Omega)$ and in practice it does not matter which of these estimators is used.
The relative uncertainty (exemplified in the Tab.~\ref{tab1} for the sample standard deviation) remains 
below 1\% for all scattering angles at $E=13$~MeV and $E=65$~MeV, and only slightly exceeds it 
at $E=200$~MeV.  
In Tab.~\ref{tab1} we also show values of the differential cross section obtained with the central values of the OPE-Gaussian
potential parameters and mean values of predictions calculated separately for the 50 ($M_{100\%}$) or 34 ($M_{68\%}$) 
sets of parameters $S_i$. Also here in most of the cases $d\sigma/d\Omega(S_0) \approx M_{100\%} \approx M_{68\%}$,
what shows, that the predictions based on sets $S_i$ for $i\neq0$ cluster around $d\sigma/d\Omega(S_0)$ evenly.
The other observables behave in a similar way.  

\begin{table}[hbpt]
\begin{tabular}{|c|r|r|r|r|r|r|r|r|}
\hline
E [MeV] &$\theta_{\rm{c.m.}}$ [deg] & d$\sigma$/d$\Omega$(S$_0$) & $\frac12 \Delta_{100\%}$  & $\frac12 \Delta_{68\%}$ &  $\frac12$IQR & $\sigma(d\sigma/d\Omega)$ \;\;\; & $M_{100\%}$ & $M_{68\%}$ \\
\hline
           & 30  & 134.9970  & 0.1780  & 0.1025  & 0.0635  & 0.0954 (0.132\%)  & 135.0040  & 135.0100  \\
13.0       & 75  &  51.3274  & 0.0315  & 0.0153  & 0.0110  & 0.0149 (0.061\%)  &  51.3283  &  51.3295  \\
           & 120 &   9.7437  & 0.0347  & 0.0181  & 0.0118  & 0.0179 (0.356\%)  &   9.7421  &   9.7420  \\
           & 165 & 103.1210  & 0.1085  & 0.0420  & 0.0230  & 0.0462 (0.105\%)  & 103.1190  & 103.1190  \\
\hline
           & 30  &  23.7000  & 0.1785  & 0.0812  & 0.0569  & 0.0824 (0.753\%)  &  23.7137  &  23.7092  \\
65.0       & 75  &   2.3630  & 0.0134  & 0.0060  & 0.0040  & 0.0057 (0.568\%)  &   2.3630  &   2.3630  \\
           & 120 &   0.7787  & 0.0035  & 0.0015  & 0.0011  & 0.0016 (0.451\%)  &   0.7786  &   0.7785  \\
           & 165 &   4.7537  & 0.0174  & 0.0076  & 0.0060  & 0.0075 (0.366\%)  &   4.7532  &   4.7535  \\
\hline
           & 30  &   3.7626  & 0.0351  & 0.0164  & 0.0097  & 0.0162 (0.325\%)  &   3.7634  &   3.7625  \\
200.0      & 75  &   0.2088  & 0.0018  & 0.0008  & 0.0005  & 0.0008 (0.839\%)  &   0.2087  &   0.2087  \\
           & 120 &   0.0585  & 0.0006  & 0.0004  & 0.0003  & 0.0003 (1.069\%)  &   0.0589  &   0.0589  \\
           & 165 &   0.1645  & 0.0022  & 0.0009  & 0.0007  & 0.0009 (1.356\%)  &   0.1647  &   0.1647  \\
\hline
\end{tabular}
\caption{The differential cross section d$\sigma$/d$\Omega$ obtained with 
the expectation values of the OPE-Gaussian potential parameters (set $S_0$), 
various estimators of its dispersion (see text) and mean values taken from 50 ($M_{100\%}$) 
or 34 ($M_{68\%}$) predictions. In case of the sample standard deviation $\sigma(d\sigma/d\Omega)$ 
also the relative magnitude $\sigma(d\sigma/d\Omega)\;/\;(d\sigma/d\Omega(S_0)) * 100\%$ is shown in brackets. 
All predictions are given in [mb~sr$^{-1}$].}
\label{tab1}
\end{table}

\subsection{Nucleon-deuteron elastic scattering observables from the OPE-Gaussian model}

In the following we present predictions obtained with the OPE-Gaussian $NN$ interaction 
for various observables in the elastic neutron-deuteron scattering process at
incoming nucleon laboratory energies $E=13$~MeV, 65~MeV, and 200~MeV. We will focus on the 
elastic scattering cross section d$\sigma$/d$\Omega$, 
the nucleon vector analyzing power A$_{\rm{y}}$,
the nucleon to nucleon spin transfer coefficients K$_{\rm{y}}^{\rm{y'}}$,
and the spin correlation coefficients C$_{\rm{y,y}}$. However, we will also give examples for 
other observables.

\begin{figure}[bt]
\includegraphics[width=1\textwidth,clip=true]{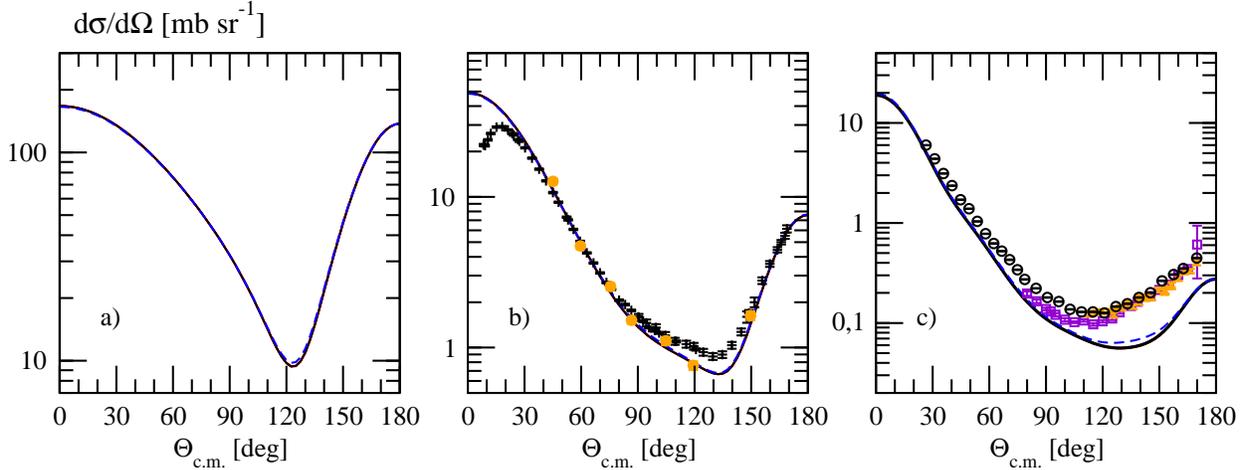}
\caption{
The $Nd$ elastic scattering cross section d$\sigma$/d$\Omega$~[mb~sr$^{-1}$] at the incoming nucleon 
laboratory energy a) $E$=13~MeV, b) $E$=65~MeV, and c) $E$=200~MeV as a function of the c.m.  
scattering angle $\theta_{\rm{c.m.}}$.
The black curve represents predictions obtained with the central values of the OPE-Gaussian parameters, the red band
reflects statistical uncertainty discussed in this subsection, and the blue dashed curve represents 
predictions based on the AV18 force. The data are in b) from Ref.~\cite{Shimizu} ($pd$ black pluses) 
and~\cite{Ruhl} ($nd$ orange circles) and in c) from Ref.~\cite{Adelberg} ($pd$, $E=198$~MeV, violet squares),
Ref.~\cite{Igo} ($pd$, $E=180$~MeV, orange x's), and Ref.~\cite{Ermisch} ($pd$, $E=198$~MeV, black circles).}
\label{fig5}
\end{figure}

The $Nd$ cross section is shown in Fig.~\ref{fig5}. Apart from the solid line which represents predictions based on 
the OPE-Gaussian force when the expectation values of its parameters (set $S_0$) are used, 
we also show the red band representing the range of predictions obtained with the same 34 sets $S_i$ 
as used to calculate $\frac{1}{2}\Delta_{68\%}$,
and the blue dashed curve showing results obtained with the AV18 interaction.
The nucleon-deuteron data (at the same or nearby energies) are also added for the sake of comparison.
The predictions based on the OPE-Gaussian force are in agreement with the predictions based on the AV18 potential.
Only small, ($\approx 3.9$\% at $E$=13~MeV and 
$\approx$ 3.5\% at $E$=200~MeV), differences are seen in the minimum of the cross section. 
Similarly to the AV18, the OPE-Gaussian model clearly underestimates the data at two higher energies 
reflecting the known fact of growing importance of a 3$N$ force~\cite{WitalaPRL, Kuros}. 
The statistical error arising from the uncertainty of the $NN$ force parameters is in all cases
very small and 
red bands are hardly visible in Fig.~\ref{fig5}.  

The OPE-Gaussian force delivers predictions which are very close to the AV18 results also for the most 
of the polarization observables at the energies studied here.
Likewise the dispersion of predictions remains small for most of the polarization observables. 
Below we discuss a few of them, choosing mainly ones with the largest statistical uncertainties.

\begin{figure}[t!]
\includegraphics[width=1\textwidth,clip=true]{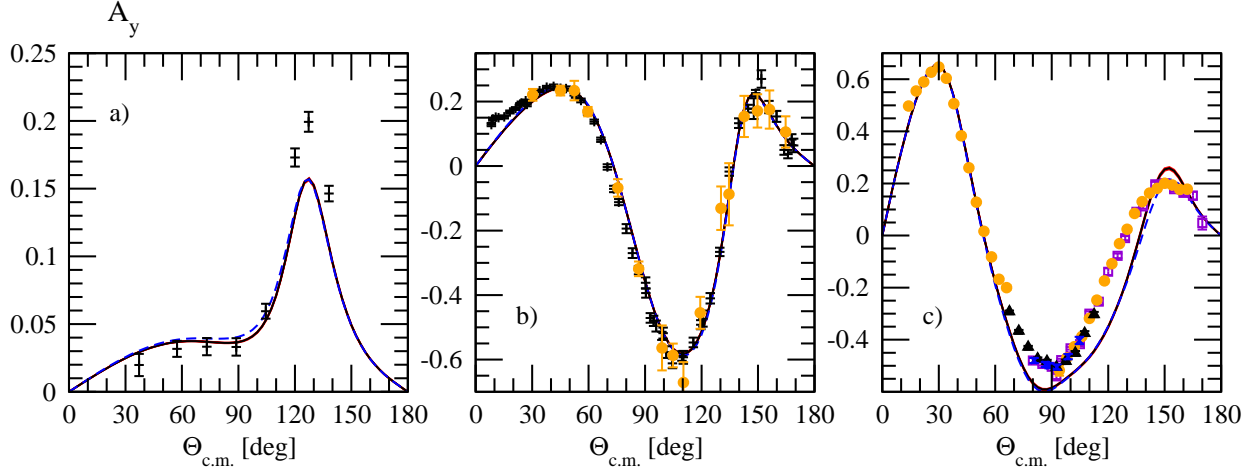}
\caption{The nucleon analyzing power A$_{\rm{y}}$ for $Nd$ elastic scattering at 
the same energies as used in Fig.~\ref{fig5} as a function of the c.m. 
scattering angle $\theta_{\rm{c.m.}}$. 
Curves and band as in Fig.~\ref{fig5}. The data in a) are from Ref.~\cite{Cub-ayn13} ($nd$ black pluses), 
in b) are from Ref.~\cite{Shimizu} ($pd$ black pluses) and Ref.~\cite{Ruhl} ($nd$ orange circles), and 
in c) are from Ref.~\cite{Adelberg} ($pd$ violet squares), Ref.~\cite{Przewoski} ($pd$ $E=200$~MeV orange circles), 
Ref.~\cite{Cadman} ($pd$ $E=197$~MeV black triangles up), and Ref.~\cite{Wells} ($pd$ blue x's).}
\label{fig6}
\end{figure}

Let us start, however, with the nucleon analyzing power A$_{\rm{y}}$, 
shown in Fig.~\ref{fig6}.
Here the uncertainties remain negligible at all energies and also the differences 
between predictions based on the OPE-Gaussian force and the ones obtained with the AV18 potential are tiny.
Thus we see that the OPE-Gaussian model does not deliver any hint on the nature of the A$_{\rm{y}}$ 
puzzle at $E=13$~MeV. 
 
\begin{figure}[h]
\includegraphics[width=1\textwidth,clip=true]{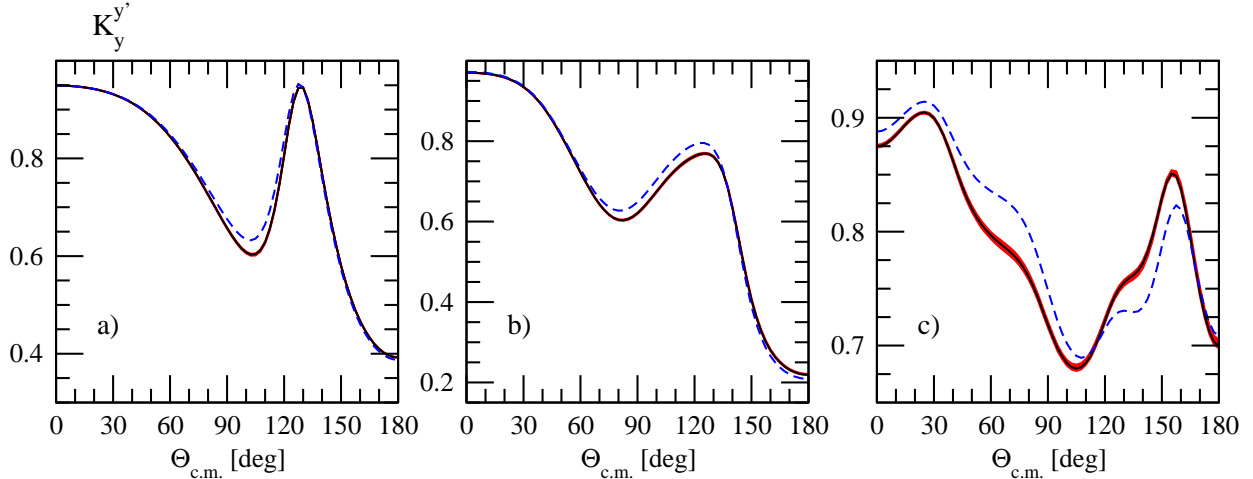}
\caption{The nucleon to nucleon spin transfer coefficient K$_{\rm{y}}^{\rm{y'}}$ at the
incoming nucleon laboratory energy a) $E$=13~MeV, b) $E$=65~MeV and c) $E$=~200 MeV
as a function of the c.m. scattering angle $\theta_{\rm{c.m.}}$.
See Fig.\ref{fig5} for a description of band and curves.}
\label{fig7}
\end{figure}

\begin{figure}[h]
\includegraphics[width=1\textwidth,clip=true]{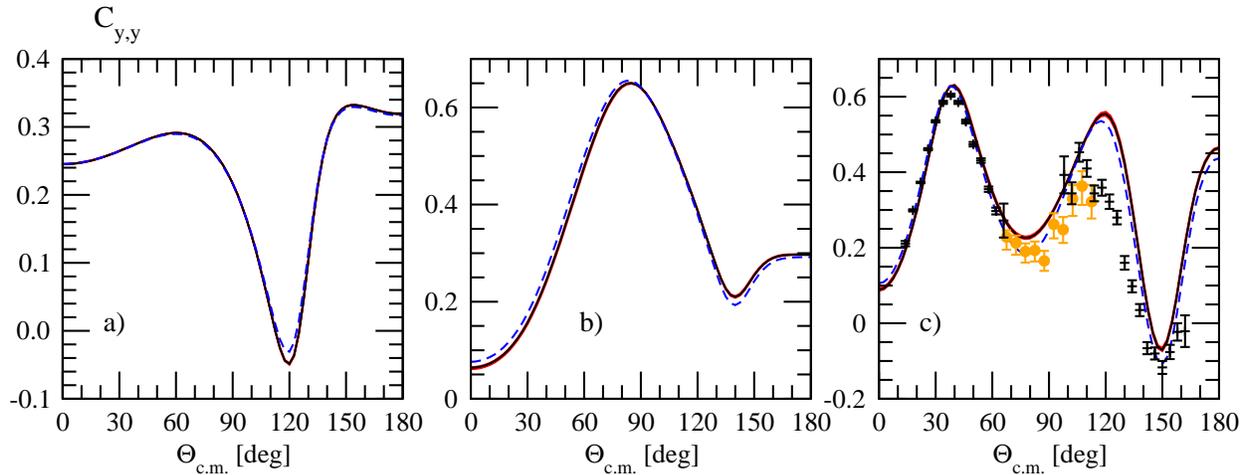}
\caption{The spin correlation coefficient C$_{\rm{y,y}}$ at the incoming nucleon laboratory
energy a) $E$=13~MeV, b) $E$=65~MeV and c) $E$=200~MeV as a function of the c.m.
scattering angle $\theta_{\rm{c.m.}}$. See Fig.\ref{fig5} for a description of band
and curves. In c) data are from Ref.~\cite{Cadman} ($pd$ $E=197$~MeV, orange circles) 
and Ref.~\cite{Przewoski} ($pd$ $E=200$~MeV, black pluses).}
\label{fig8}
\end{figure}

We have chosen the nucleon to nucleon spin transfer coefficient K$_{\rm{y}}^{\rm{y'}}$ and the spin correlation coefficient C$_{\rm{y,y}}$
to demonstrate, in Figs.~\ref{fig7} and~\ref{fig8}, respectively, changes of the statistical  
errors when increasing the reaction energy. 
For both observables dispersion of the results grows with energy, and while at lowest energy $E$=13~MeV
it is negligible, at $E$=200 MeV its size is bigger, although it remains small ( $\frac12 \Delta_{68\%} < 0.5\%$). 
In the case of C$_{\rm{y,y}}$
comparison with the data reveals that the spread of the OPE-Gaussian results is still smaller than
uncertainties of experimental results.  

\begin{figure}[h]
\includegraphics[width=.38\textwidth,clip=true]{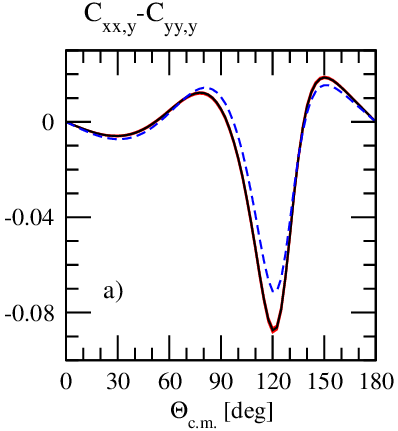}
\includegraphics[width=.38\textwidth,clip=true]{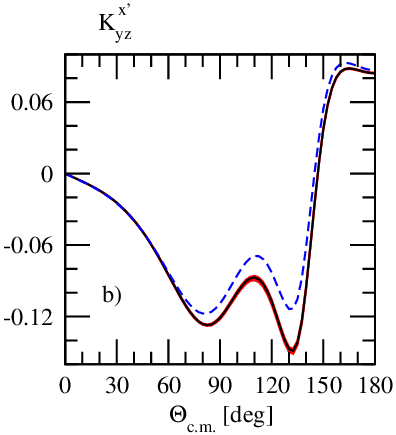}
\caption{The spin correlation coefficient C$_{\rm{xx,y}}$-C$_{\rm{yy,y}}$ at the incoming nucleon laboratory energy $E$=13~MeV (a)
and the deuteron to nucleon spin transfer coefficient K$_{\rm{yz}}^{\rm{x'}}$ at the
incoming nucleon laboratory energy
$E$=65~MeV (b)
as a function of the c.m. scattering angle $\theta_{\rm{c.m.}}$.
Curves and band are as in Fig.\ref{fig5}.}
\label{fig9}
\end{figure}

In Fig.~\ref{fig9} we show two observables for which  
the difference between the AV18 predictions and the OPE-Gaussian results is especially 
big already at the two lower energies. 
They are the spin correlation coefficient C$_{\rm{xx,y}}$-C$_{\rm{yy,y}}$ at $E$=13~MeV and
the deuteron to nucleon spin transfer coefficient K$_{\rm{yz}}^{\rm{x'}}$  at $E$=65~MeV.
The difference between both predictions amounts to $\approx 19$\% at the minimum of C$_{\rm{xx,y}}$-C$_{\rm{yy,y}}$,
while the statistical error of the OPE-Gaussian results is only $\approx 2$\%. 
For K$_{\rm{yz}}^{\rm{x'}}$ these differences 
amount to $\approx 23$\% and $\approx 3$\%, respectively.
We see that even in these two cases the statistical uncertainty remains much smaller than
uncertainty related to using various models of the $NN$ interaction. 

\begin{figure}[h]
\includegraphics[width=.66\textwidth,clip=true]{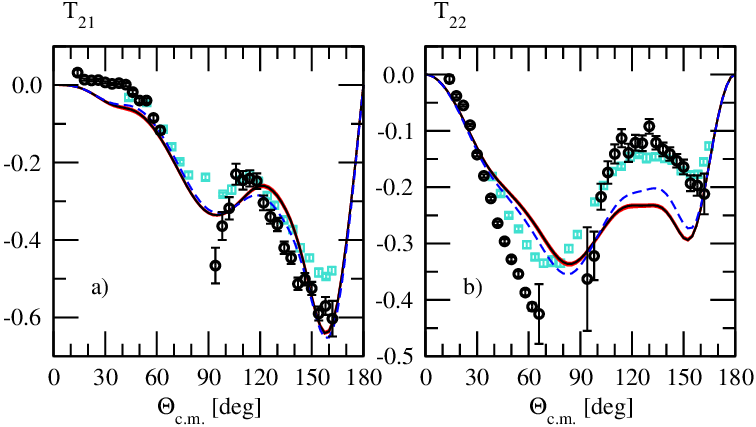}
\includegraphics[width=.33\textwidth,clip=true]{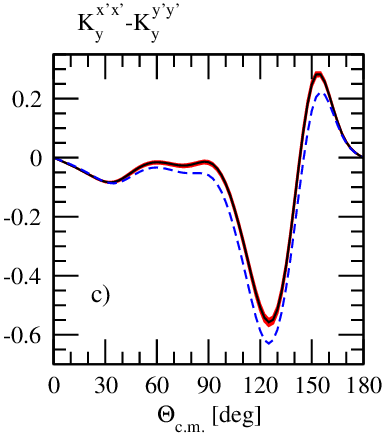}
\caption{The deuteron tensor analyzing
powers T$_{21}$ (a) and T$_{22}$ (b) and the nucleon to deuteron spin transfer 
coefficient K$_{\rm{y}}^{\rm{x'x'}}$-K$_{\rm{y}}^{\rm{y'y'}}$ (c)
for $E=200$~MeV as a function of the center of mass scattering angle $\theta_{\rm{c.m.}}$. 
See Fig.\ref{fig5} for description of bands and curves. The T$_{21}$ and T$_{22}$ data are from 
Ref.~\cite{Sekiguchi} ($pd$ $E=186.6$~MeV turquoise squares) and Ref.~\cite{Przewoski} ($pd$ $E=200$~MeV black circles)}
\label{fig10}
\end{figure}

The statistical errors grow with the reaction energy. Thus in Fig.~\ref{fig10} we show 
for $E=200$~MeV a few observables with the largest uncertainties. Beside the 
spin transfer coefficient K$_{\rm{y}}^{\rm{y'}}$ already shown in Fig.~\ref{fig7} 
they are the deuteron tensor analyzing powers T$_{21}$ and T$_{22}$ and the 
nucleon to deuteron spin transfer coefficient K$_{\rm{y}}^{\rm{x'x'}}$-K$_{\rm{y}}^{\rm{y'y'}}$.
While the bands representing the theoretical uncertainties are clearly visible,
they still remain small compared to the experimental errors for both analyzing powers.
The differences between predictions based on the AV18 potential 
and the OPE-Gaussian force 
are small. This is true also for the other Nd elastic scattering observables both at $E=200$~MeV
and at the lower energies,
so we conclude that the OPE-Gaussian force
yields a similar description of this process compared with the AV18 potential.
 
\section{Comparison of various theoretical uncertainties in Nd scattering}
\label{results_various}

It is interesting to compare the statistical error $\frac12 \Delta_{68\%}$ obtained in the previous section with the other 
uncertainties (like the uncertainty arising from using the various models of nuclear interaction,
the uncertainty introduced by the partial wave decomposition approach, the truncation errors of chiral predictions 
and the uncertainties originating in the cut-off dependence of chiral forces)
present in the elastic $Nd$ scattering studies and specifically in our approach.

The accuracy of predictions arising from the algorithms used in our numerical scheme, which comprises, 
among others, numerical integrations, interpolations and series summations,  
is well under control. This has been tested, e.g. by using various grids of mesh points,
or more generally by benchmark calculations involving different methods to treat 
$Nd$ scattering~\cite{COR90,HUB95,FRI90b,FRI95}. 
The main contribution to theoretical uncertainties 
comes in our numerical realization from using a truncated set of partial waves.
Typically we restrict ourselves to partial waves with the two-body total orbital momentum $j\leq 5$. 
Predictions for observables converge with increasing $j$, as was documented e.g. in~\cite{Glockle-raport}. 
In the following we compare the OPE-Gaussian predictions, shown in the previous section, 
based on all two-body channels up to $j=5$ with the predictions based on all channels up to $j=4$ only to remind 
the reader some facts about the convergence of our approach.
However, since the differences between (not shown here) predictions based 
on all channels up to $j=6$ and those with $j_{max}=5$ are, based on results with other NN potentials, 
smaller than this for 
$j_{max}=5$ and $j_{max}=4$ predictions, the latter difference very likely overestimates the uncertainty arising from our 
computational scheme.
A recent work~\cite{Topolnicki} compares predictions for the elastic $Nd$ scattering, based however only on 
the driving term of Eq.~(\ref{eq_Fadd_NN_3N}), obtained within the 
partial wave formalism with the ones from the "three-dimensional" approach, 
i.e. the approach which totally avoids the partial wave 
decomposition and uses momentum vectors. 
A very good agreement between the partial waves based results and the 
"three-dimensional" ones confirms that neglecting the higher partial waves 
in the calculations presented here practically does not affect our predictions.

Next, we would like to focus on the truncation errors and the cut-off dependence present in the 
chiral calculations and last but not least on the differences between predictions based on various 
models of the nuclear two-body interaction.

To estimate two types of theoretical uncertainties present when chiral potentials are used 
we calculated the elastic $Nd$ scattering 
observables using two $NN$ interactions at the N$^4$LO: one delivered by E.Epelbaum et al.~\cite{imp2} (E-K-M force) and
the other derived by D.R.Entem et al.~\cite{Machleidt2017} (E-M-N force). 
In the case of the E-K-M model the semi-local regularization with 
the cut-off parameter $R$ in the range between 0.8~fm and 1.2~fm is used and the  
breakdown scale of the $\chi$EFT is 0.4-0.6 GeV~\cite{imp2}. 
The E-M-N model uses the chiral breaking scale of 1~GeV and the cut-off parameter $\Lambda$ for non-local regularization 
lies between 450~MeV and 550~MeV~\cite{Machleidt2017}. 

The truncation errors $\delta(O)^{(i)}$ of an observable $O$ at
$i$-th order of the chiral expansion, with $i=0,2,3,\dots$, when only two-body interaction is used,
 can be estimated as~\cite{Binder}:
\begin{eqnarray}
\delta(O)^{(0)} &\geq& max \left( Q^2 \vert O^{(0)} \vert\,, \vert O^{(i \geq 0)} - O^{(j \geq 0)} \vert \right) \,,\nonumber \\
\delta(O)^{(2)} &=& max \left( Q^{3} \vert O^{(0)} \vert \,, Q \vert \Delta O^{(2)} \vert \,, \vert O^{(i \geq 2)} - O^{(j \geq 2)} \vert \right) \,, \nonumber \\
\delta(O)^{(i)} &=& max \left( Q^{i+1} \vert O^{(0)} \vert \,, Q^{i-1} \vert \Delta O^{(2)} \vert \,, Q^{i-2} \vert \Delta O^{(3)} \vert \right) \, {\rm for} \, i \geq 3\;,
\label{therrors}
\end{eqnarray}
where $Q$ denotes the chiral expansion parameter,
$\Delta O^{(2)} \equiv O^{(2)} - O^{(0)}$ and $\Delta O^{(i)} 
\equiv O^{(i)} - O^{(i-1)}$ for $i \geq 3$. In addition conditions:
$\delta(O)^{(2)} \geq Q \delta(O)^{(0)}$ and
$\delta(O)^{(i)} \geq Q \delta(O)^{(i-1)}$ for $i \geq 3$ are imposed on the truncation errors $\delta(O)^{(i)}$
in case when at higher orders 3N force is not included into calculations~\cite{Binder}.

The uncertainty arising from the cut-off dependence can be easily quantified - we just take the difference
between the minimal and the maximal prediction, separately for the E-K-M force and for the E-M-N model.
However, one has to be aware that in the case of the E-K-M force the cut-off values between 
$R=0.9$~fm and $R=1.0$~fm are preferred in the 2$N$ system. Thus in the following we separately discuss the whole range of 
regulator values ($0.8$~fm~$ \leq R \leq 1.2$~fm) and the range restricted to $0.9$~fm~$ \leq R \leq 1.0$~fm only.

To estimate the uncertainty arising from using various models of nuclear forces we do not introduce any separate 
measure but just show the differences between predictions obtained with various interactions. 
Admittedly, the authors of Ref.~\cite{Dobaczewski} suggest in such a case to calculate the
estimator of standard deviations, but it is valid only under assumptions of the same quality
of all interaction models, what is not clear in the case of the presented here calculations. 

\begin{figure}[h]
\includegraphics[width=.264\textwidth,clip=true]{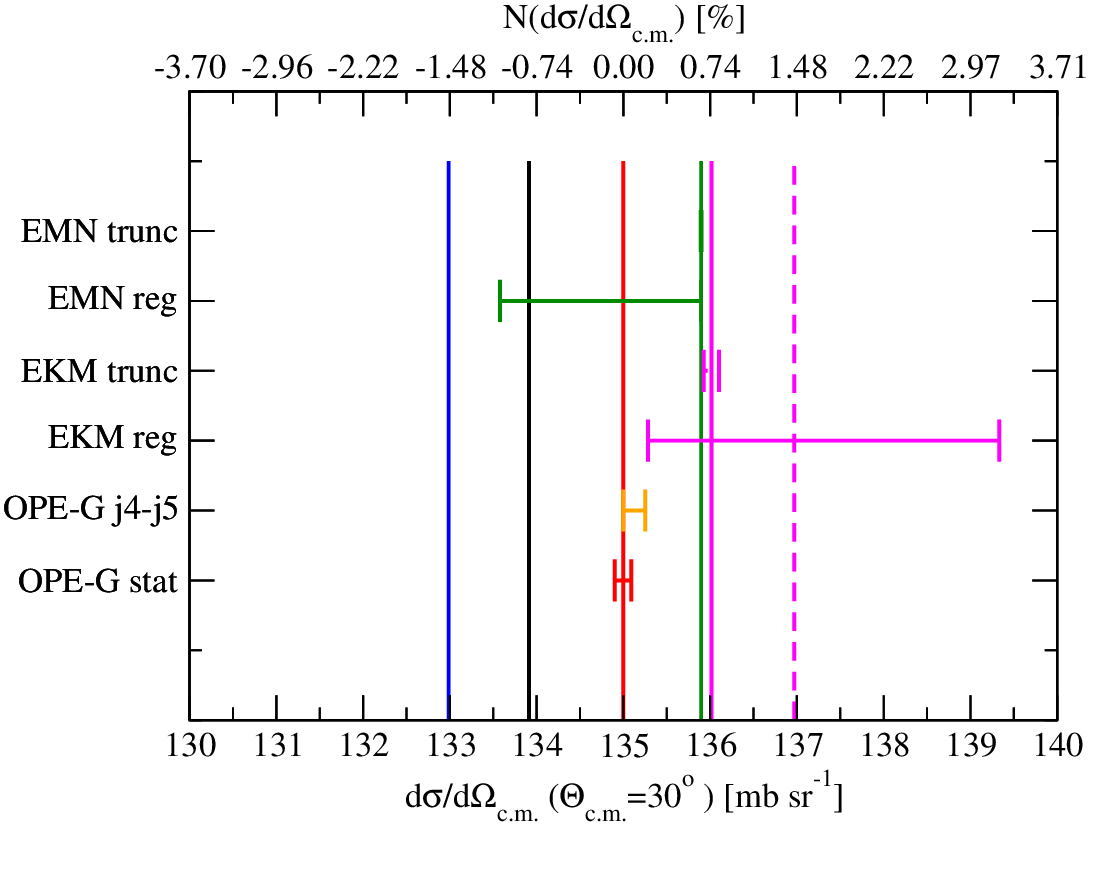}
\includegraphics[width=.23\textwidth,clip=true]{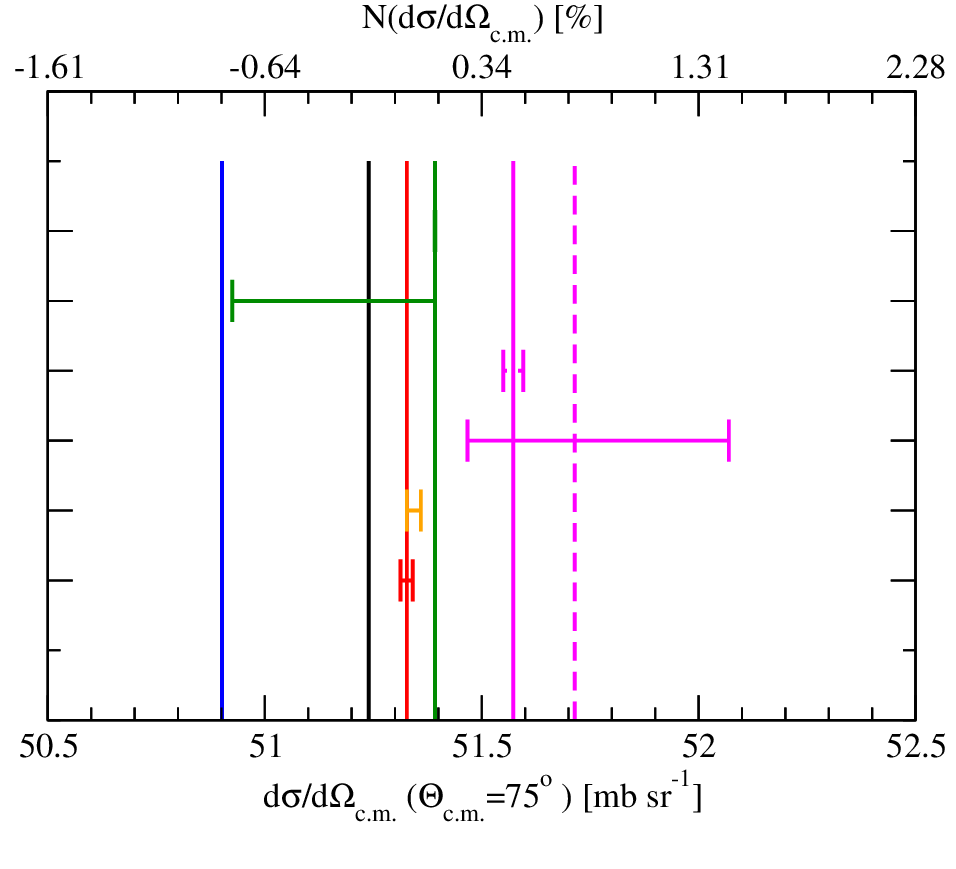}
\includegraphics[width=.23\textwidth,clip=true]{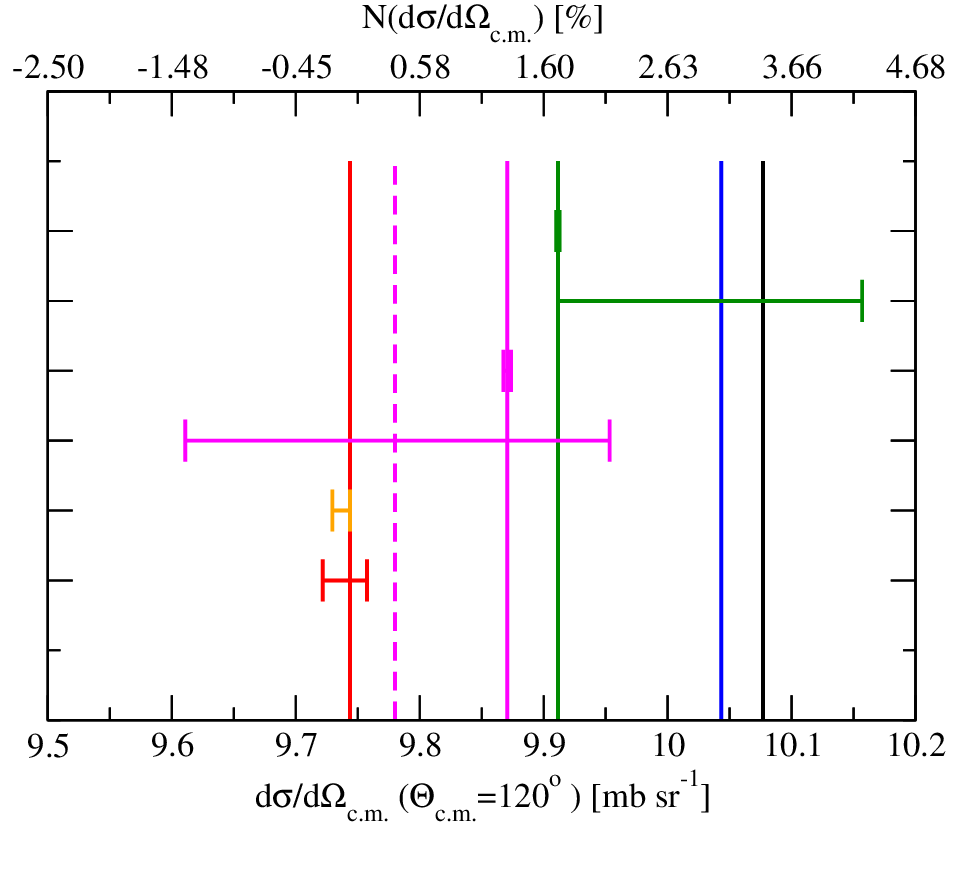}
\includegraphics[width=.23\textwidth,clip=true]{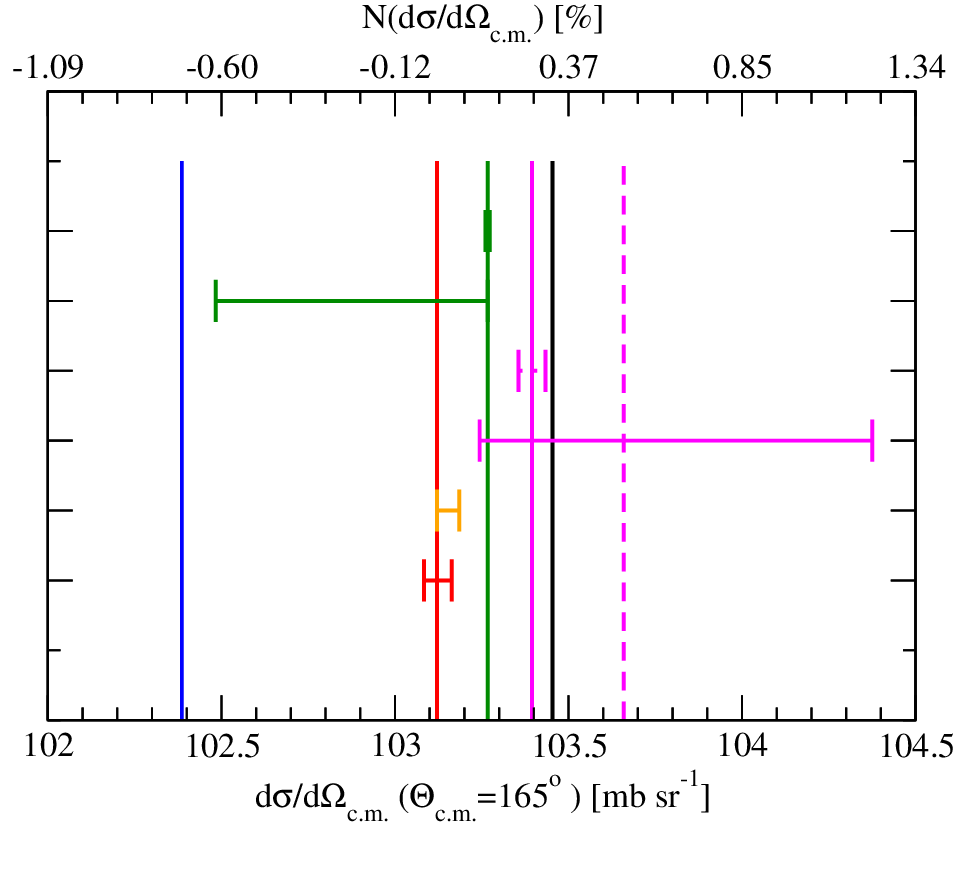}

\includegraphics[width=.264\textwidth,clip=true]{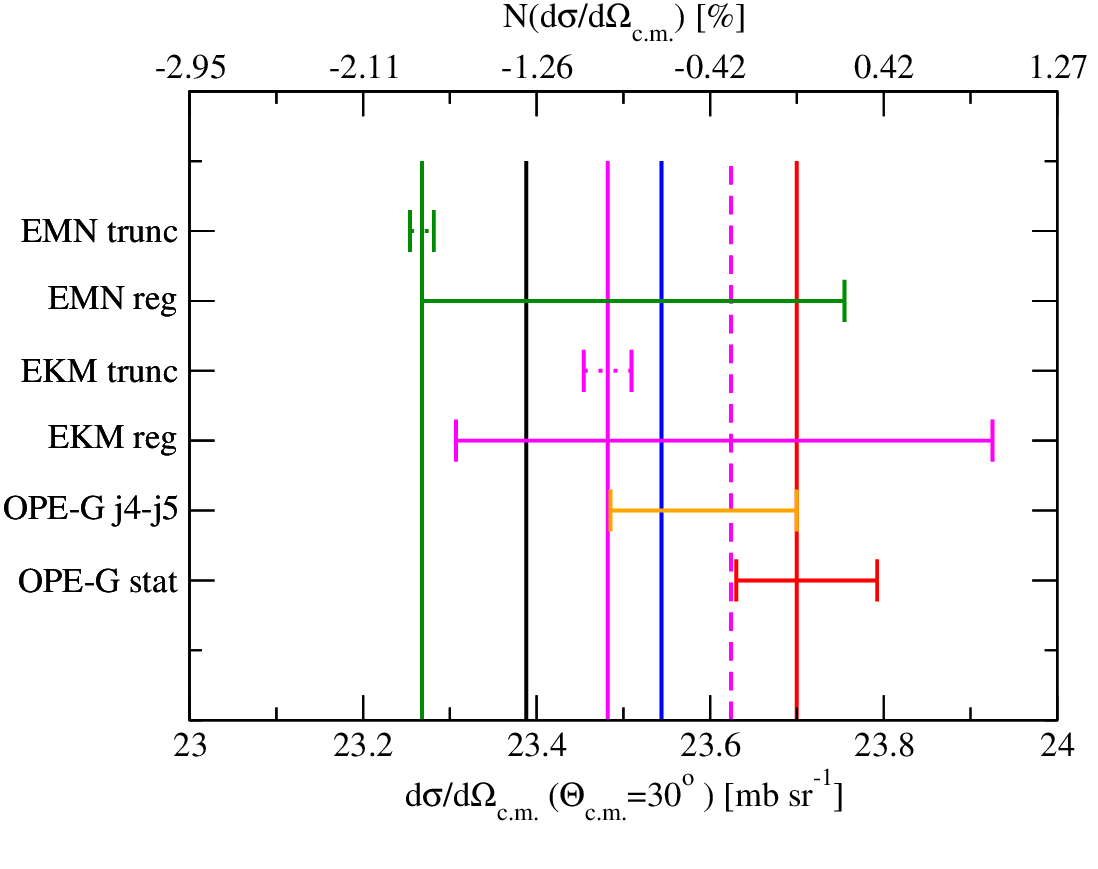}
\includegraphics[width=.23\textwidth,clip=true]{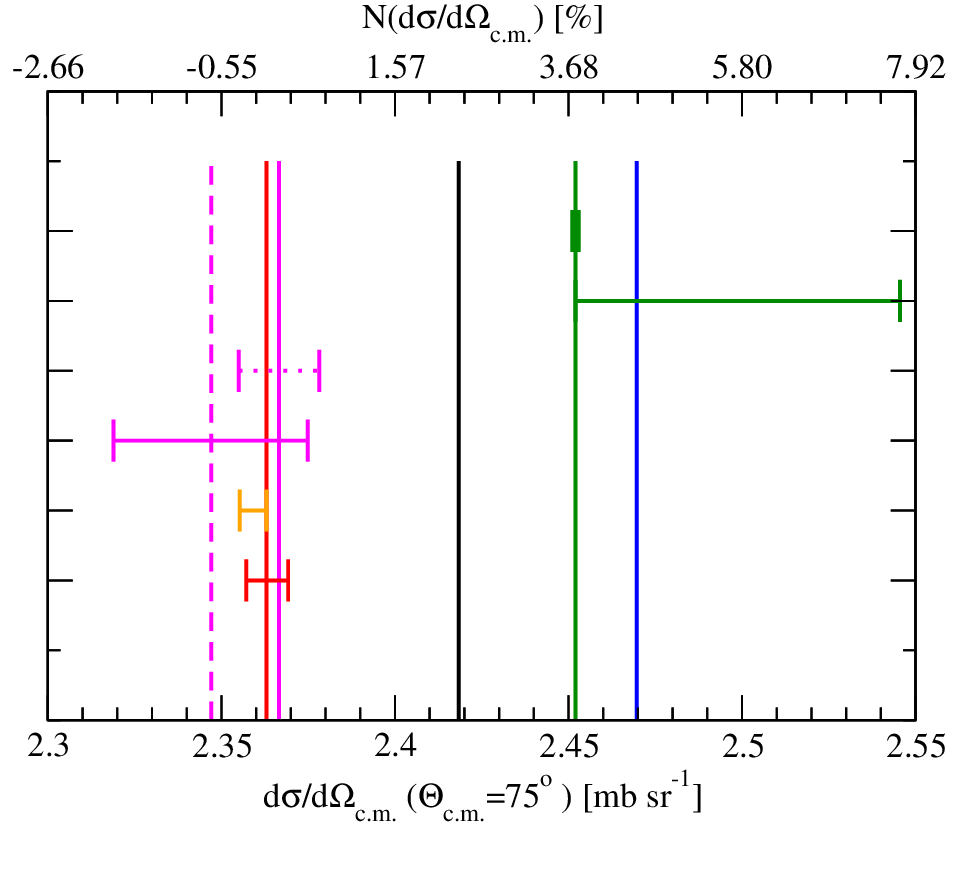}
\includegraphics[width=.23\textwidth,clip=true]{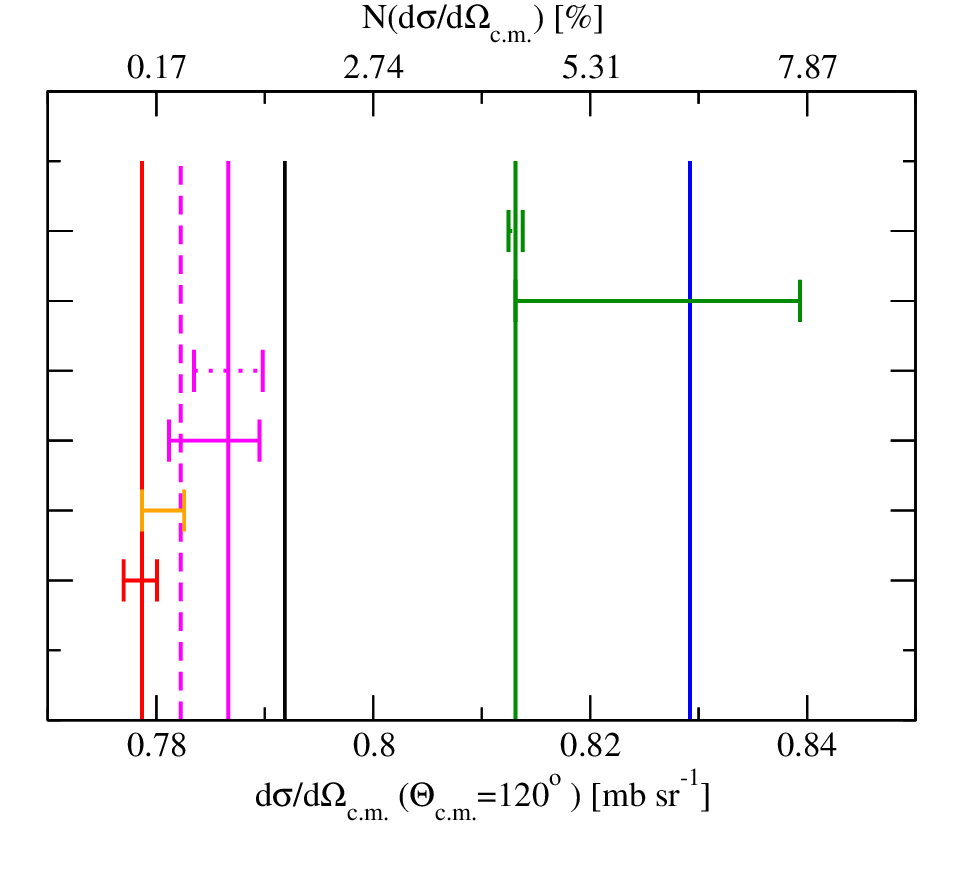}
\includegraphics[width=.23\textwidth,clip=true]{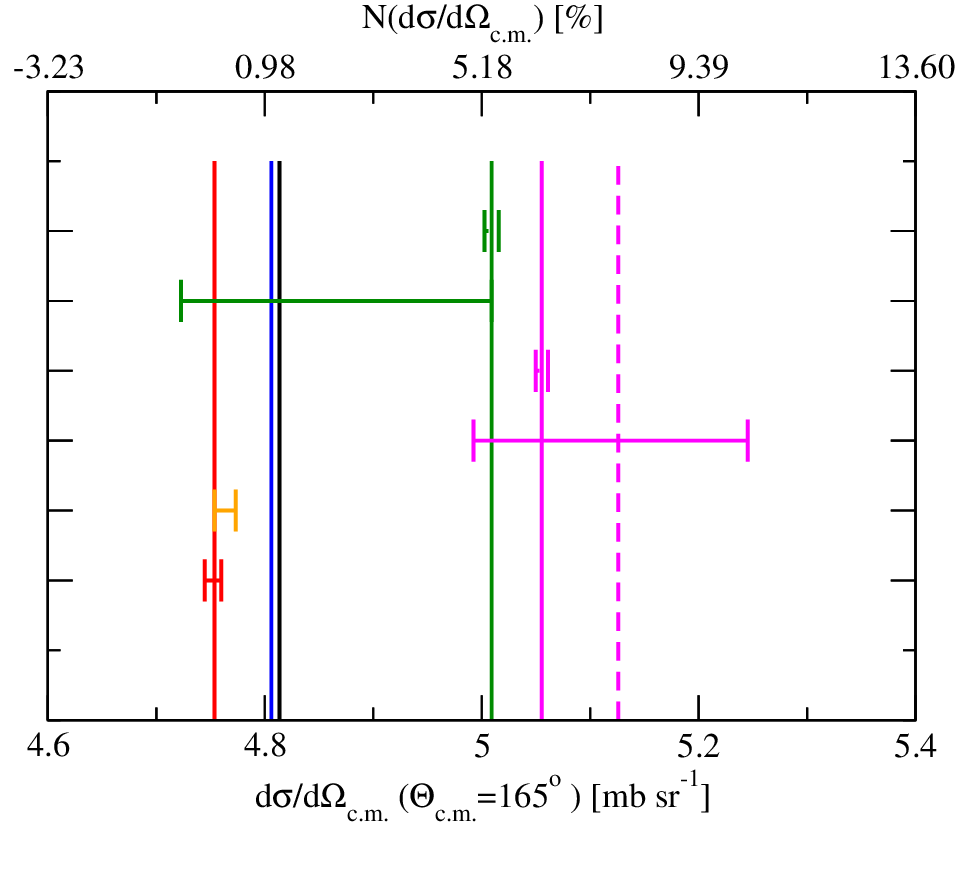}

\includegraphics[width=.264\textwidth,clip=true]{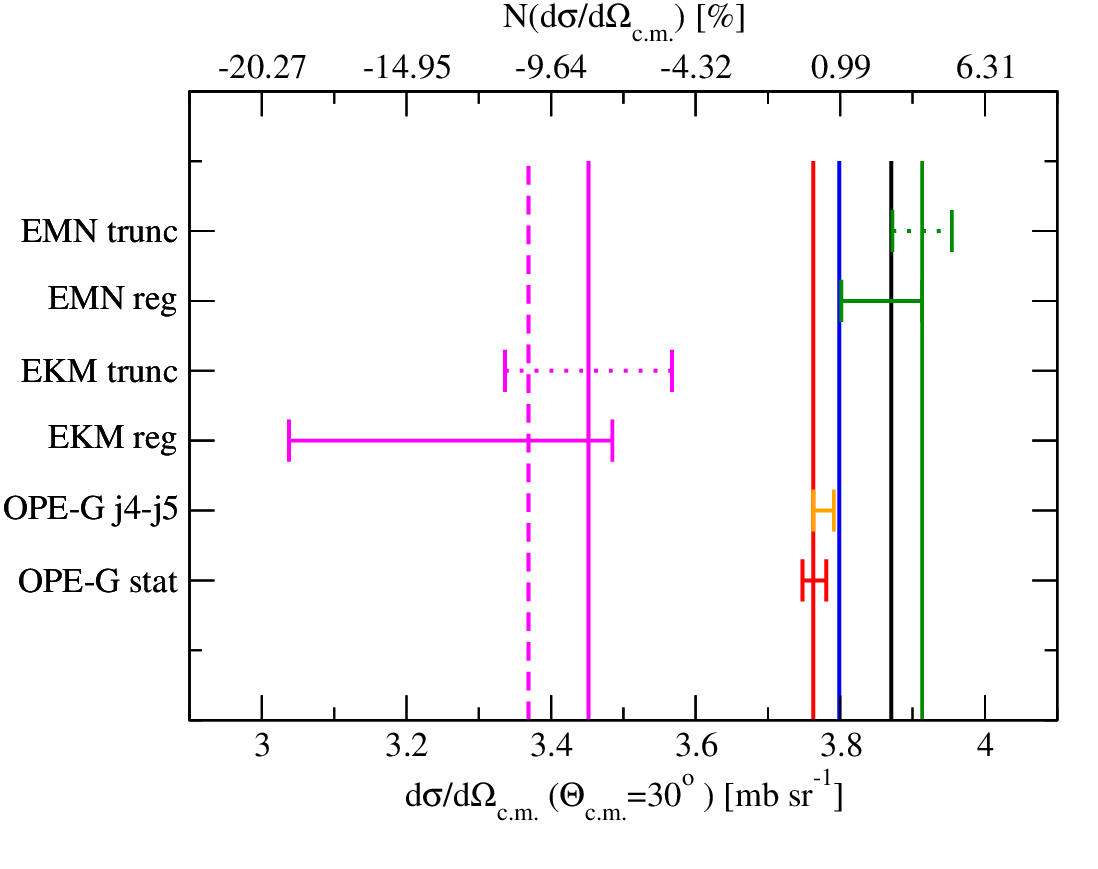}
\includegraphics[width=.23\textwidth,clip=true]{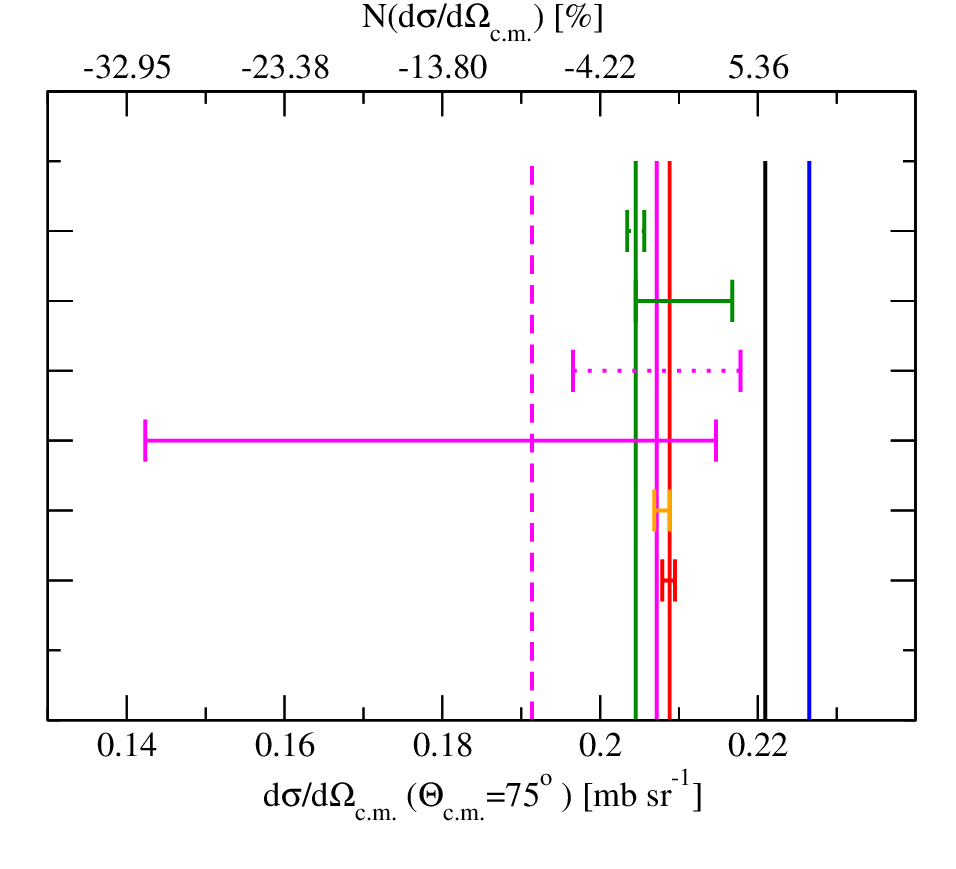}
\includegraphics[width=.23\textwidth,clip=true]{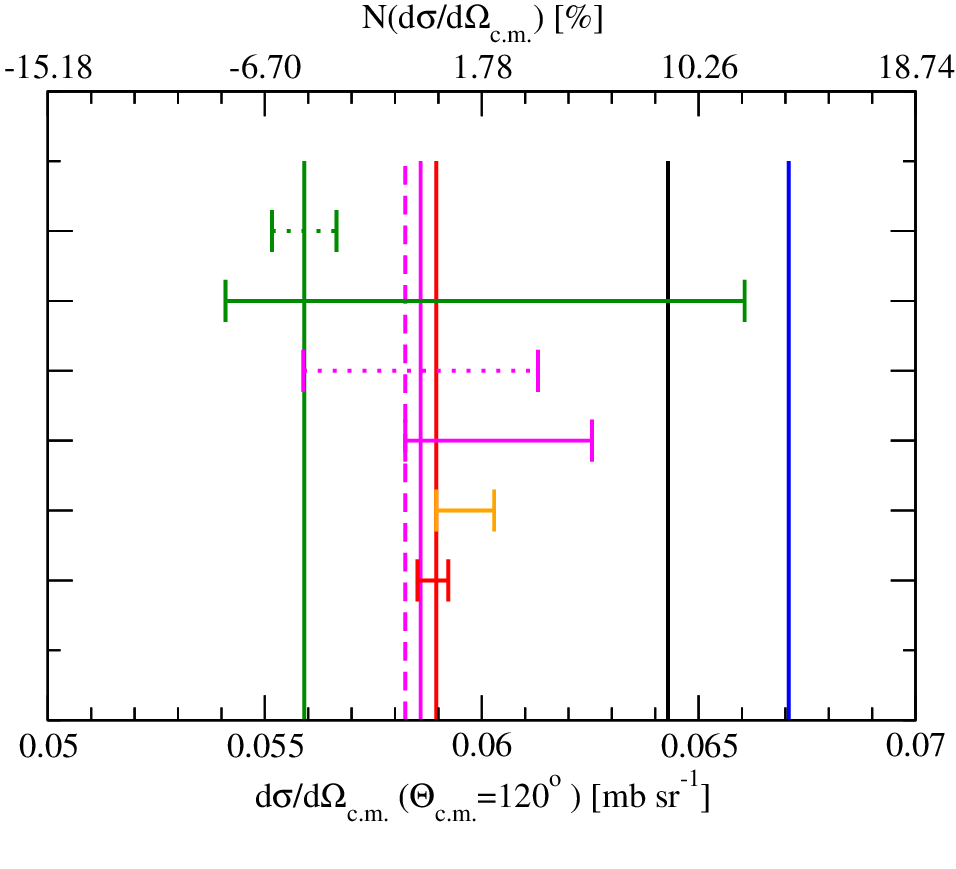}
\includegraphics[width=.23\textwidth,clip=true]{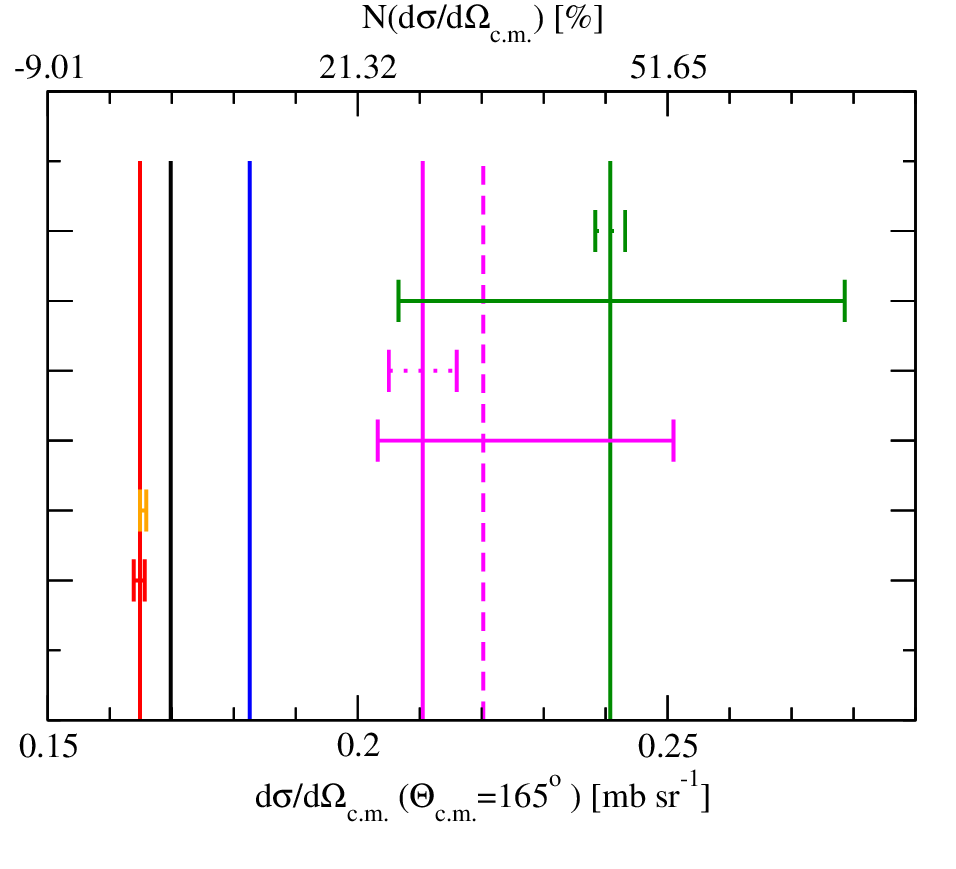}
\caption{The $Nd$ elastic scattering differential cross section and various theoretical uncertainties 
at four scattering angles: $\theta_{\rm{c.m.}}=30^{\circ}$ (1st column), 75$^{\circ}$ (2nd column),
120$^{\circ}$ (3rd column), and 165$^{\circ}$ (4th column) and at three scattering energies:
$E=13$~MeV (the upper row), $E=65$~MeV (the middle row), and $E=200$~MeV (the bottom row).
The x-axis at the bottom shows the values of the cross section, the x-axis at the top shows 
the relative difference of predictions with respect to the OPE-Gaussian results.
The vertical lines show the position of the cross section obtained with the AV18 force (black line), the OPE-Gaussian force (red line),
the E-M-N N$^4$LO $\Lambda=500$~MeV force (green line), the CD-Bonn (blue line), the E-K-M N$^4$LO $R$=0.9~fm force (magenta solid line), and
the E-K-M N$^4$LO $R$=1.0~fm force (magenta dashed line). 
The horizontal lines represent (from the bottom) the statistical error (red line), the difference between  
the OPE-Gaussian predictions with $j_{max}=5$ and $j_{max}=4$ (orange line) , the regulator dependence for the E-K-M 
force (magenta solid line), the truncation error for the E-K-M force (magenta dashed line), 
the regulator dependence for the E-M-N force (green solid line), and, at the top, the truncation 
error for the E-M-N potential (green dashed line), see text for details.
 }
\label{fig11}
\end{figure}

\begin{figure}[hbpt]
\includegraphics[width=.264\textwidth,clip=true]{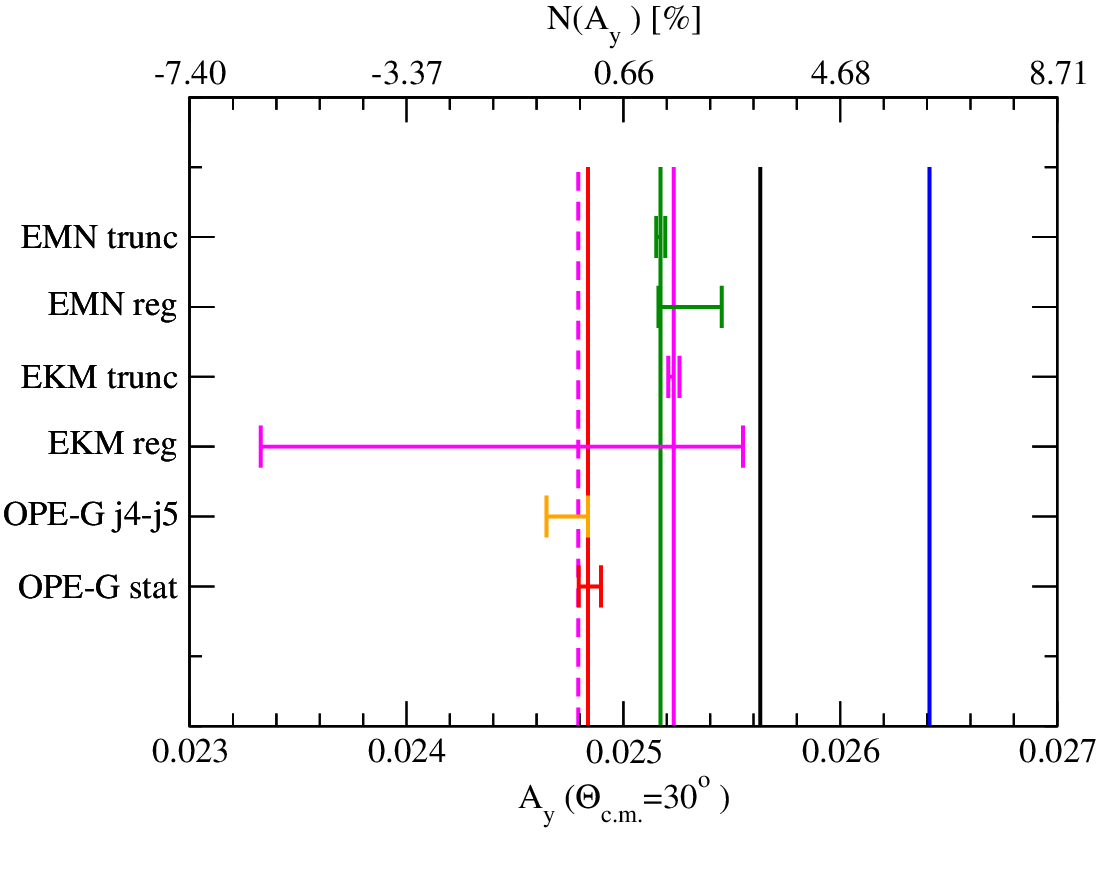}
\includegraphics[width=.23\textwidth,clip=true]{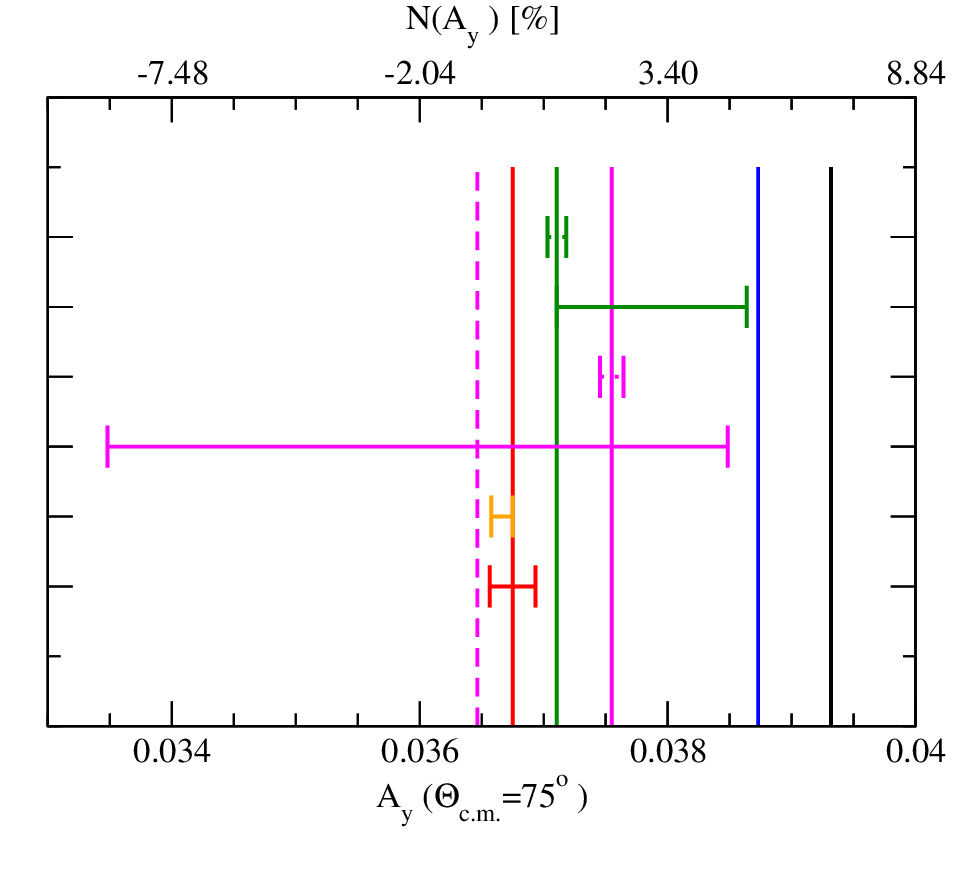}
\includegraphics[width=.23\textwidth,clip=true]{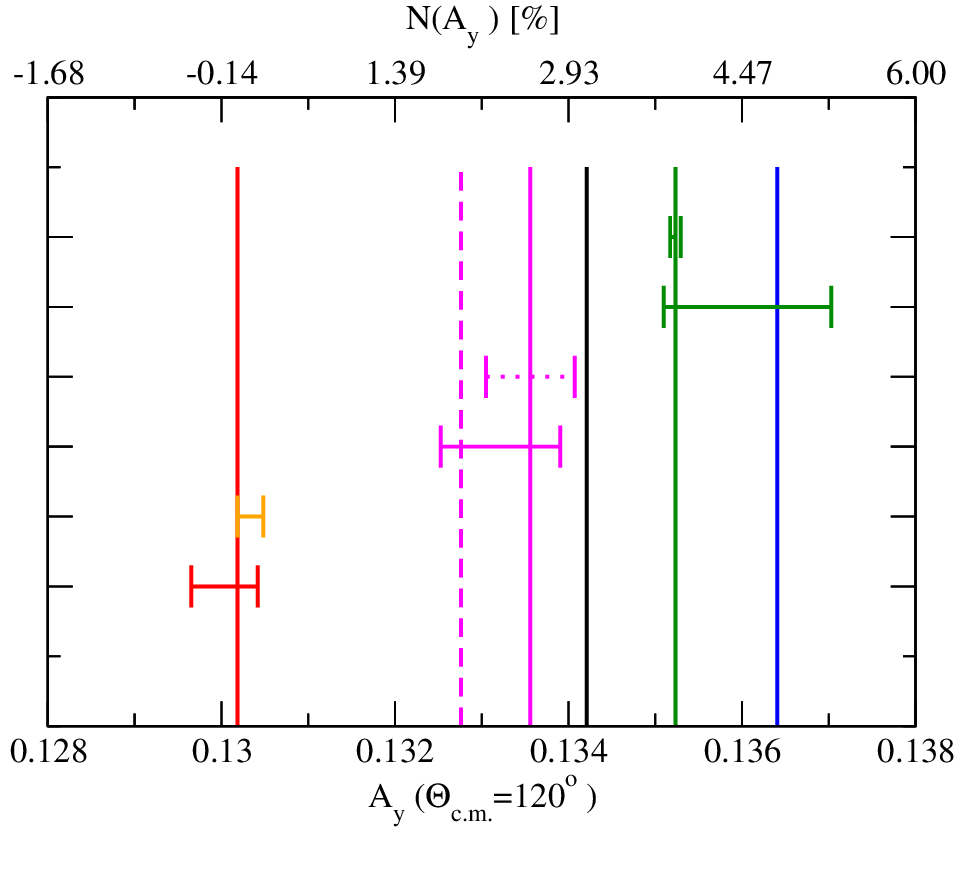}
\includegraphics[width=.23\textwidth,clip=true]{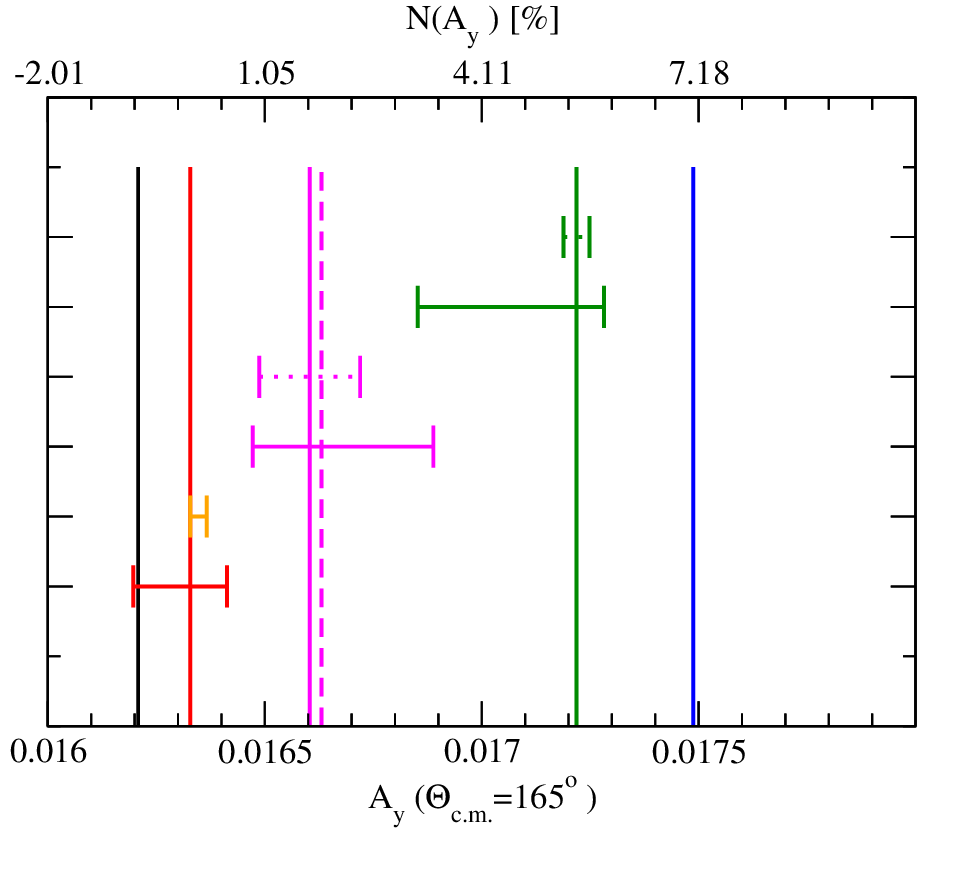}

\includegraphics[width=.264\textwidth,clip=true]{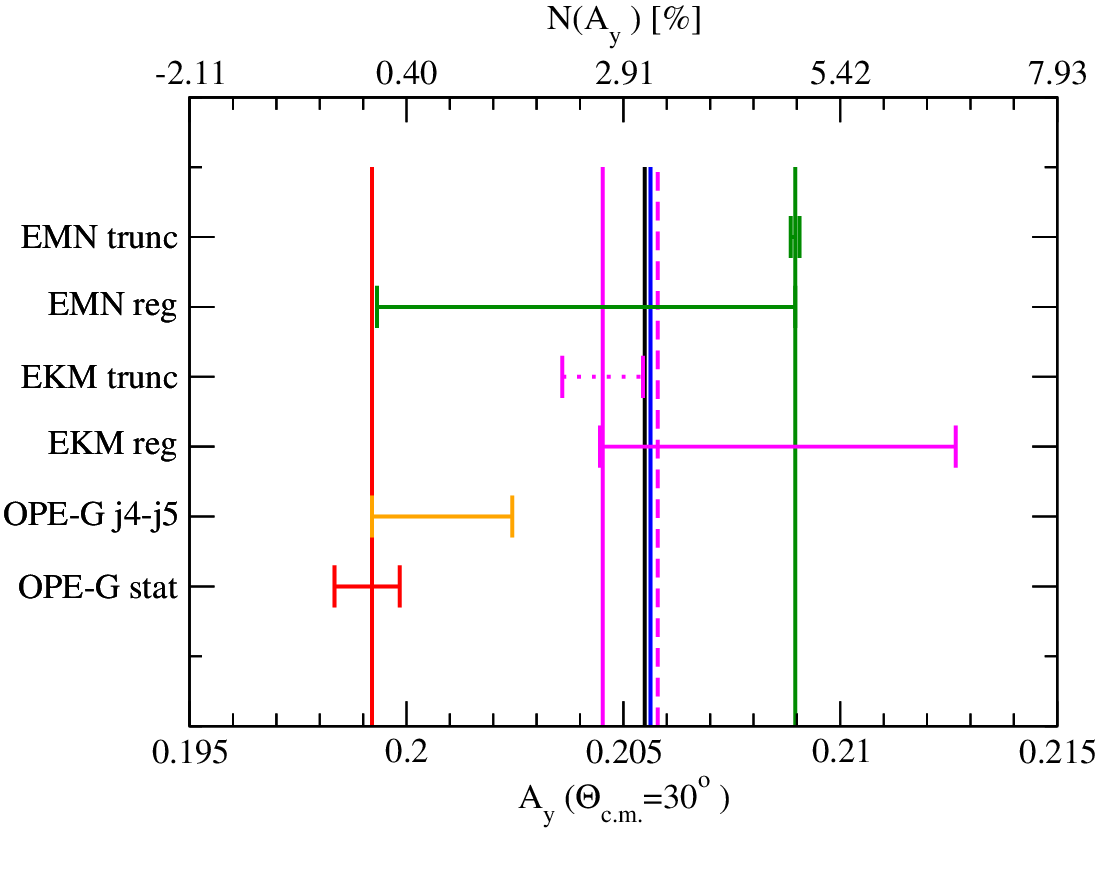}
\includegraphics[width=.23\textwidth,clip=true]{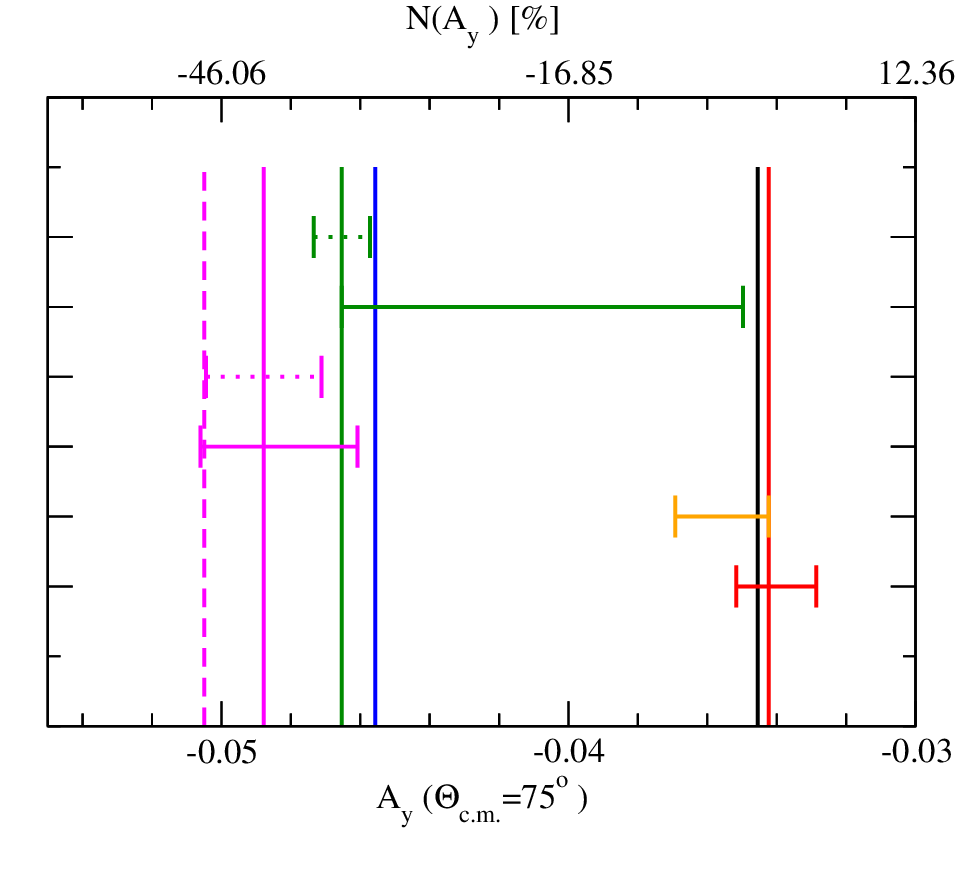}
\includegraphics[width=.23\textwidth,clip=true]{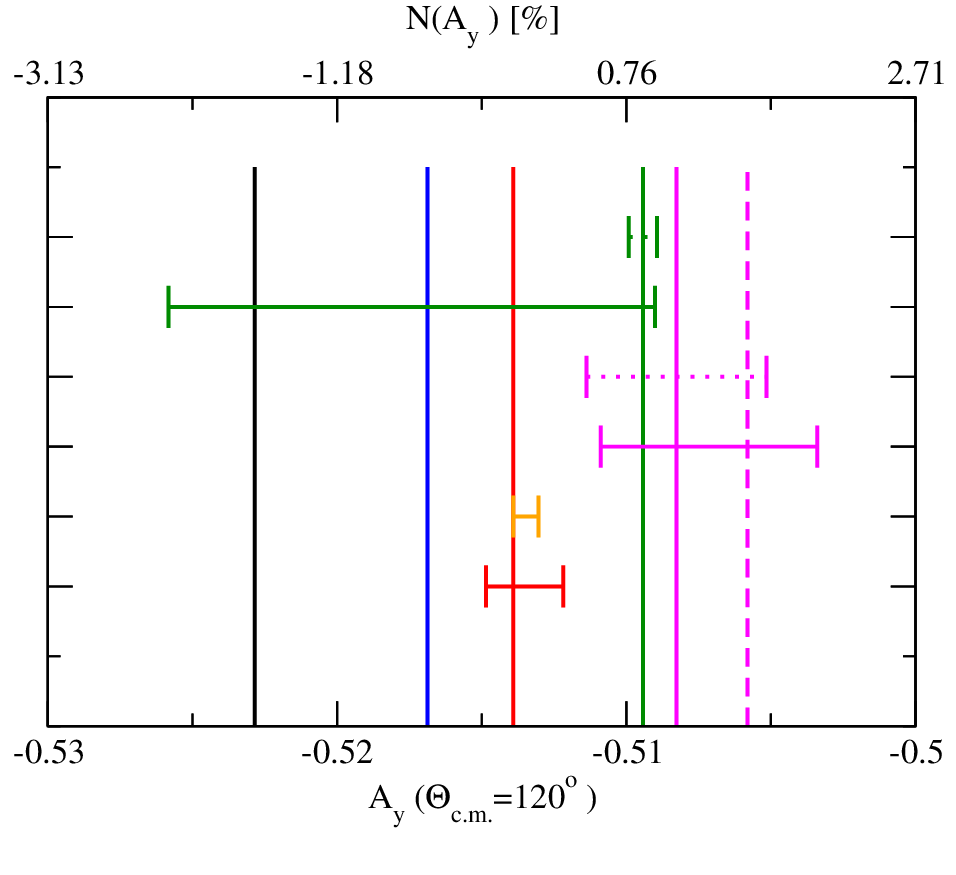}
\includegraphics[width=.23\textwidth,clip=true]{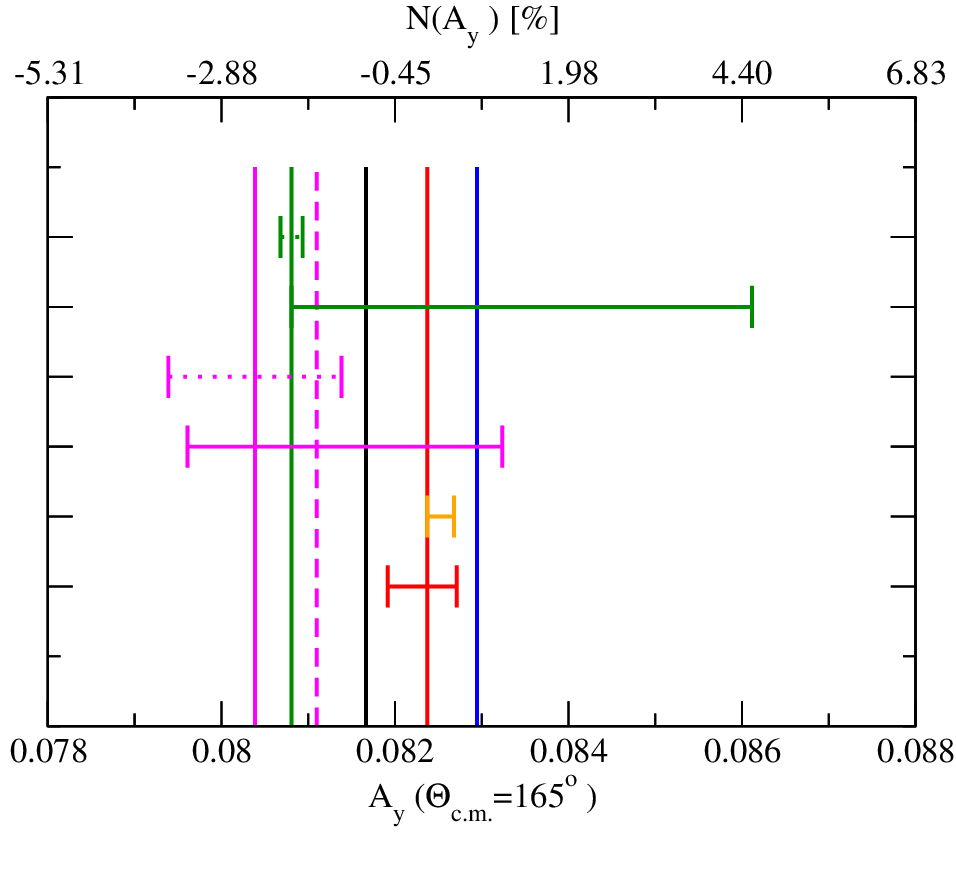}

\includegraphics[width=.264\textwidth,clip=true]{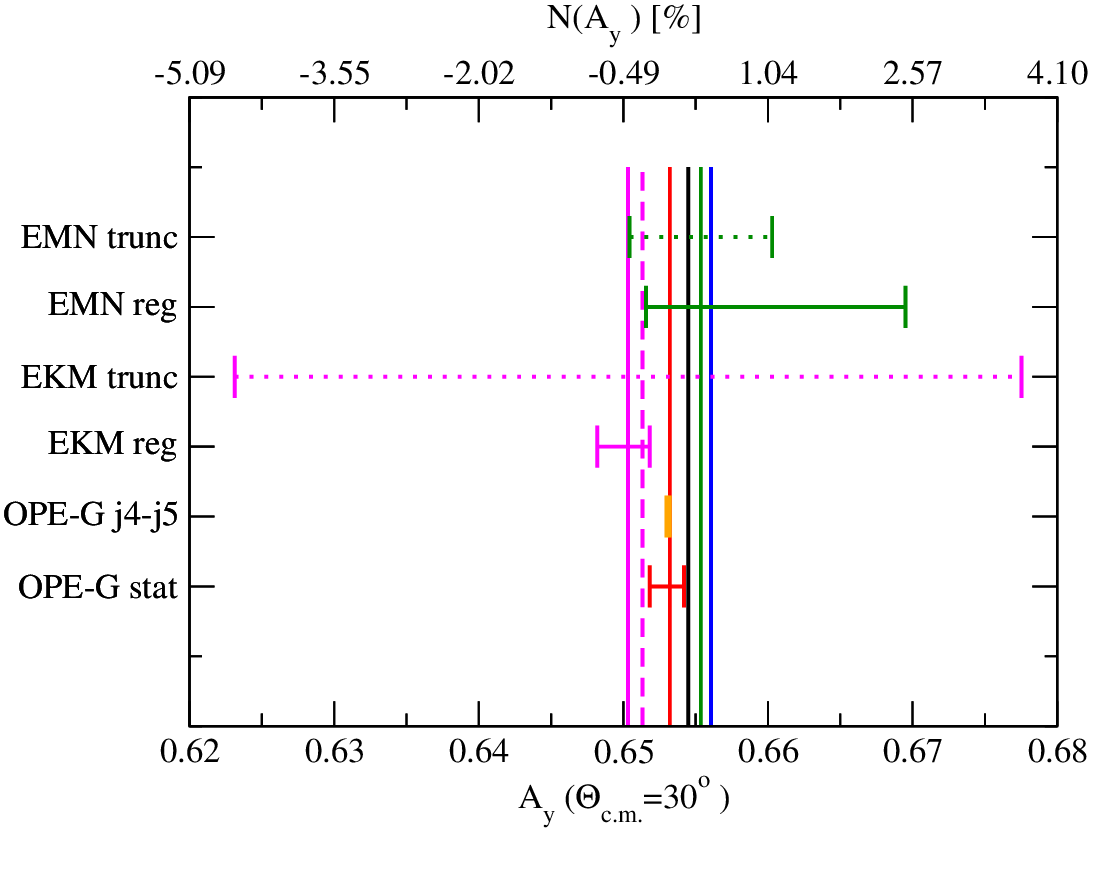}
\includegraphics[width=.23\textwidth,clip=true]{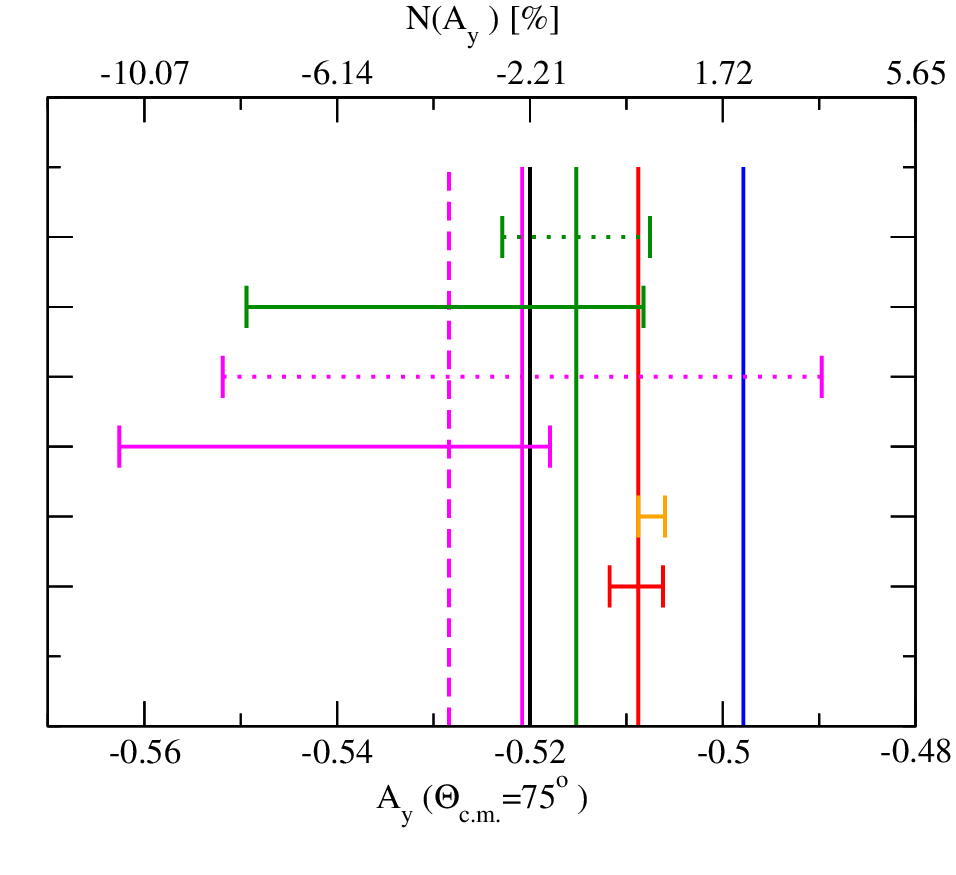}
\includegraphics[width=.23\textwidth,clip=true]{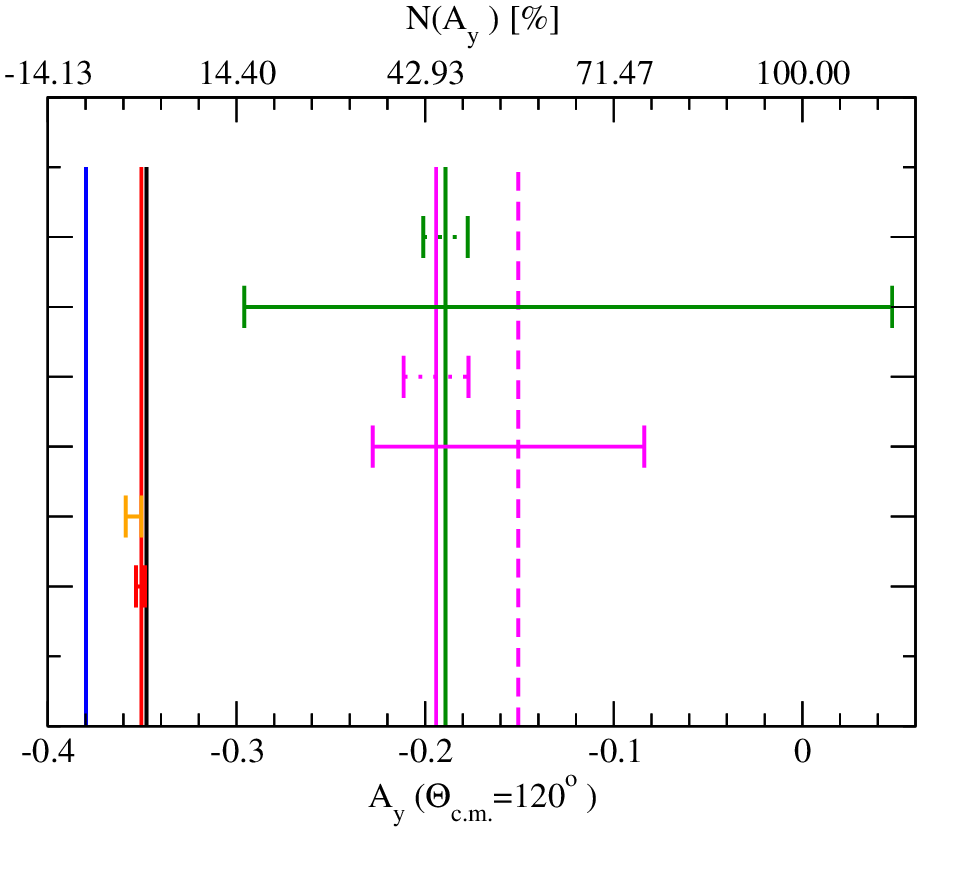}
\includegraphics[width=.23\textwidth,clip=true]{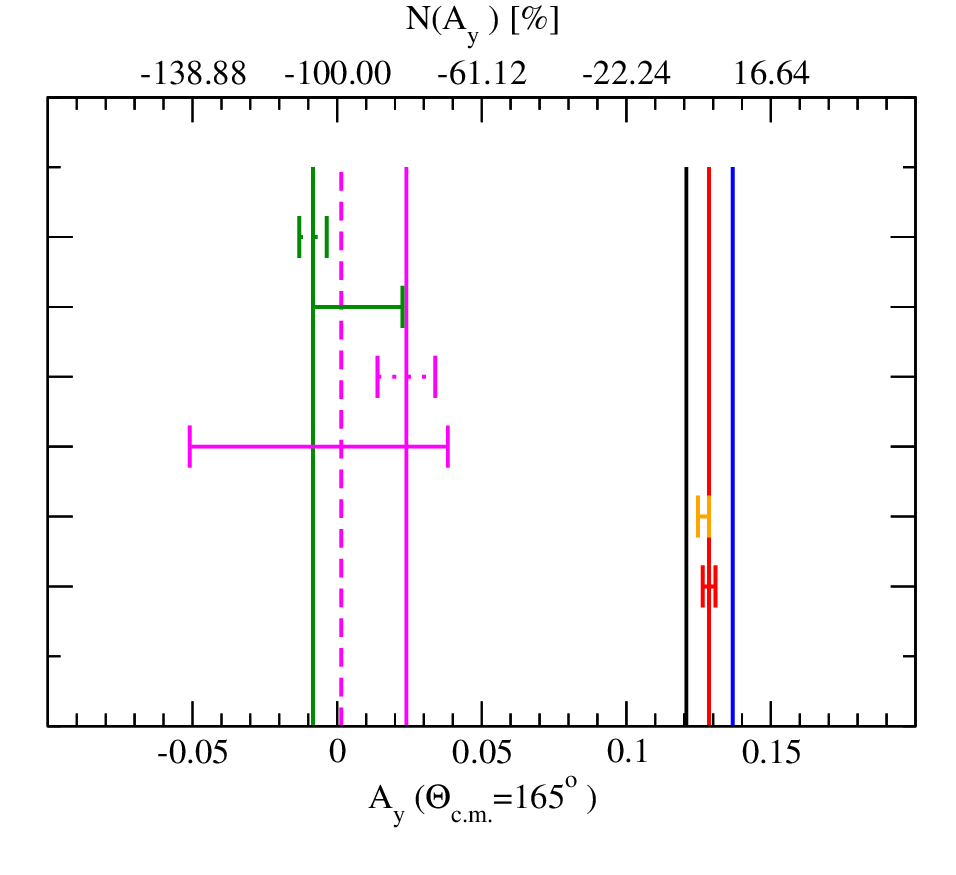}
\caption{The same as in Fig.~\ref{fig11} but for the nucleon analyzing power A$_{\rm{y}}$.}
\label{fig12}
\end{figure}

\begin{figure}[hbpt]
\includegraphics[width=.264\textwidth,clip=true]{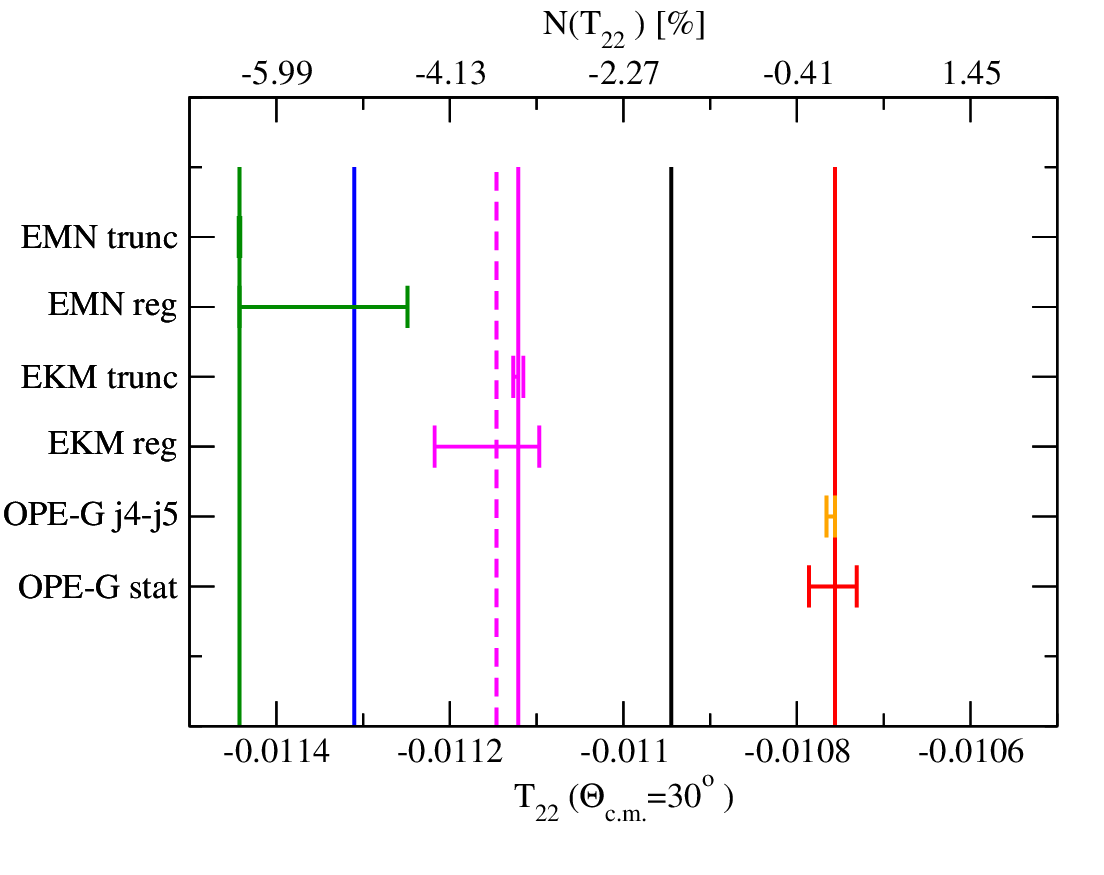}
\includegraphics[width=.23\textwidth,clip=true]{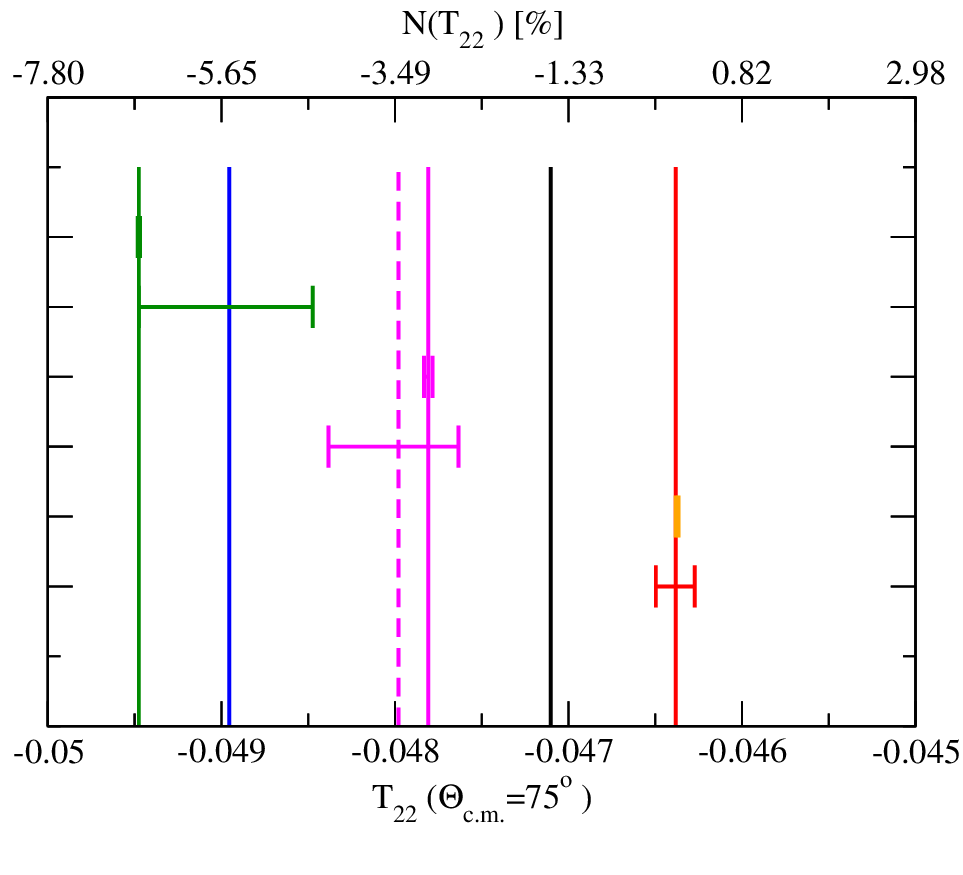}
\includegraphics[width=.23\textwidth,clip=true]{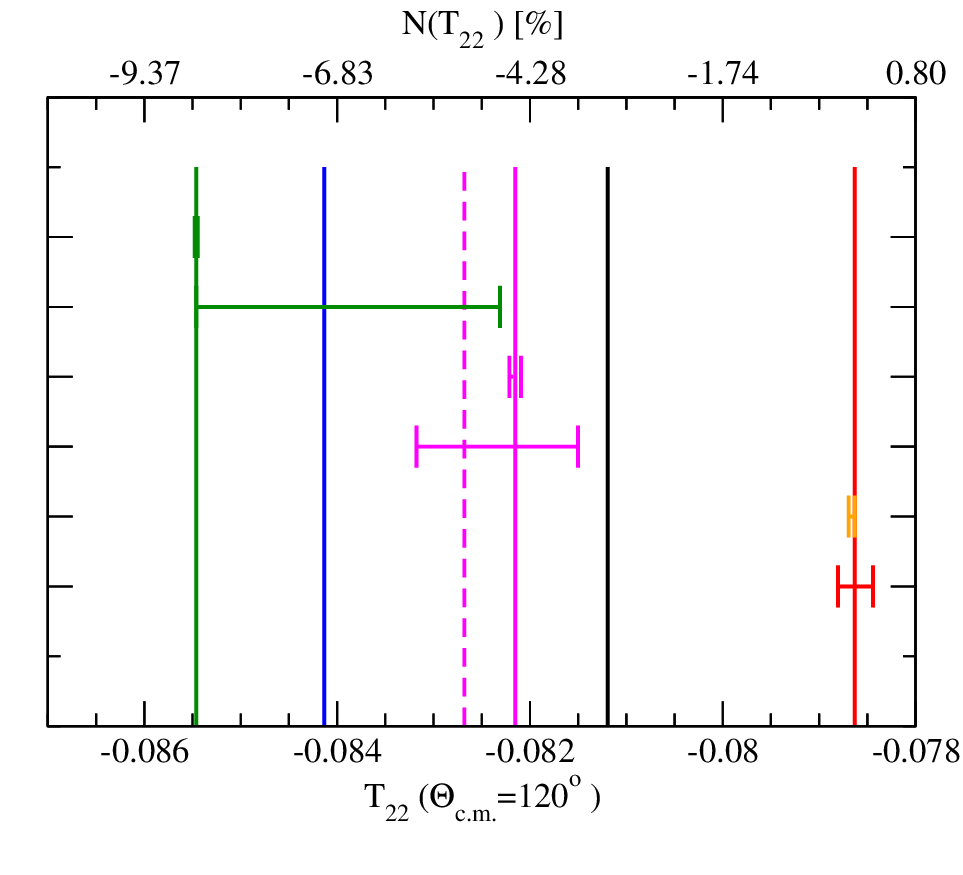}
\includegraphics[width=.23\textwidth,clip=true]{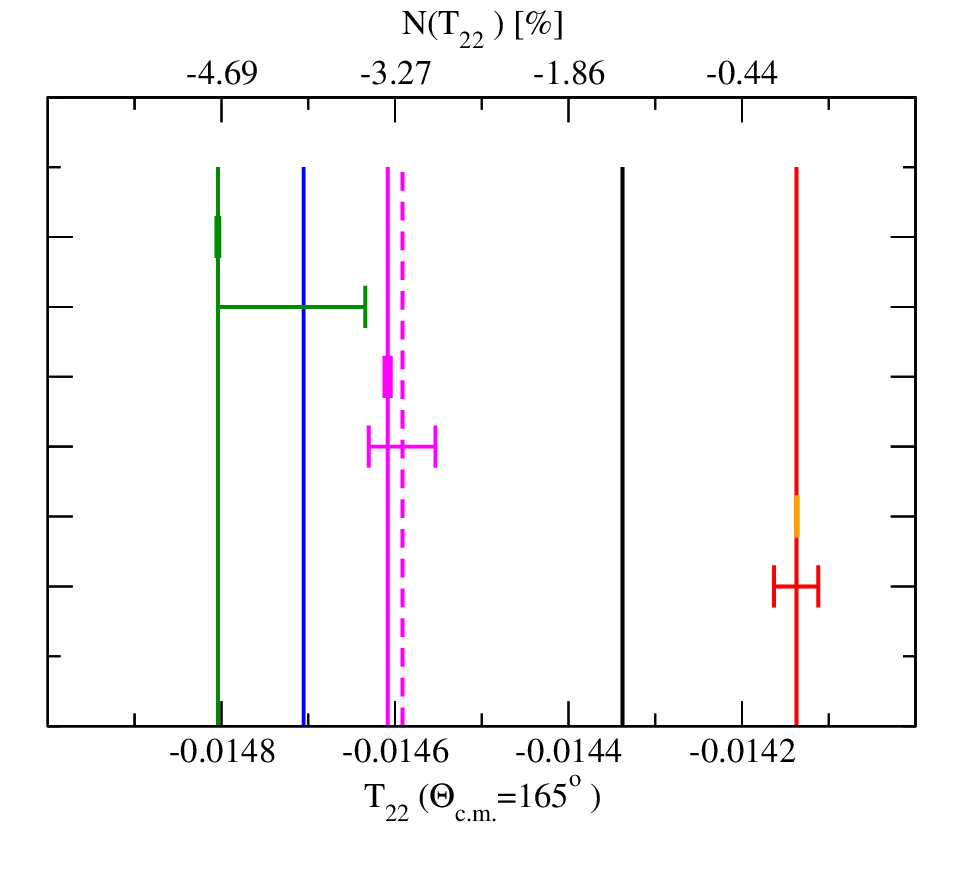}

\includegraphics[width=.264\textwidth,clip=true]{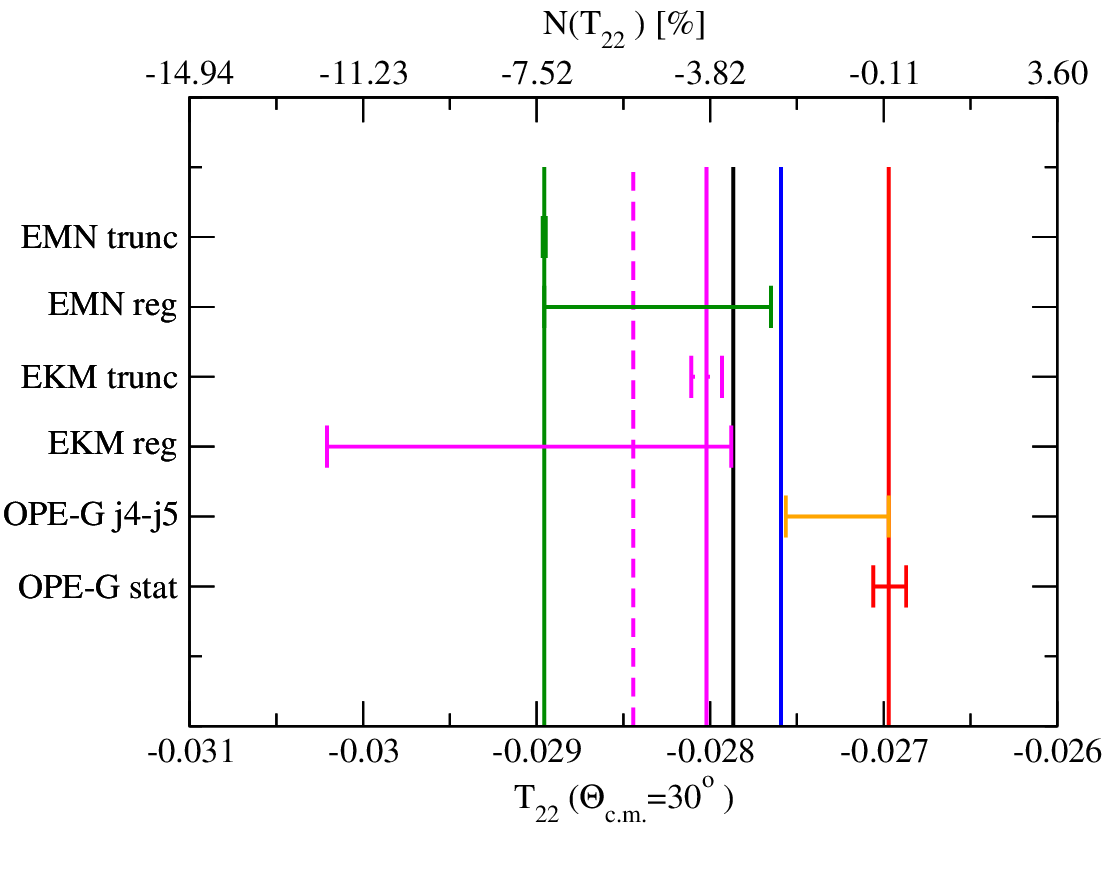}
\includegraphics[width=.23\textwidth,clip=true]{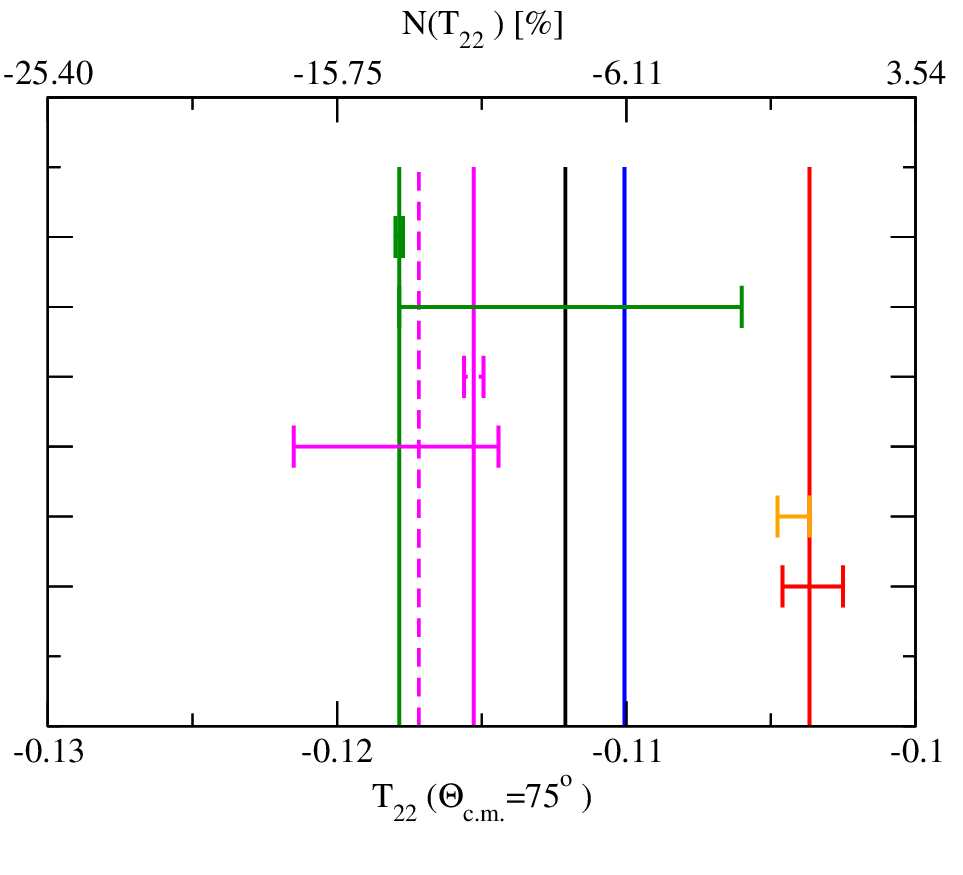}
\includegraphics[width=.23\textwidth,clip=true]{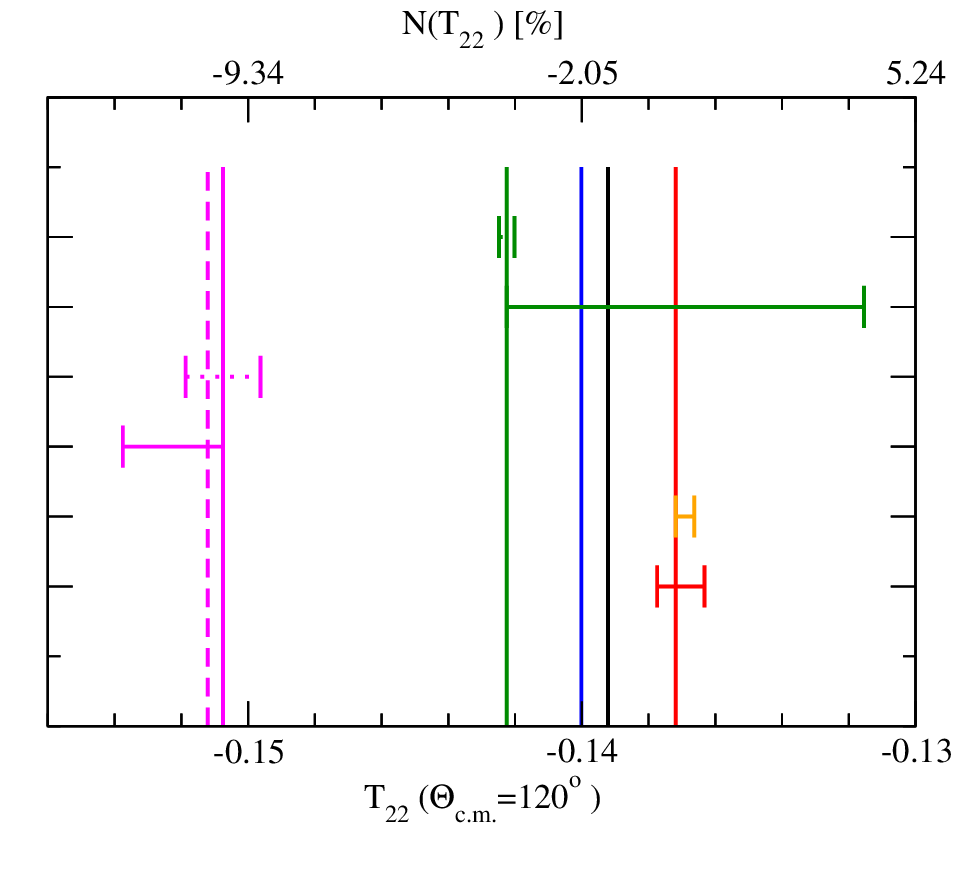}
\includegraphics[width=.23\textwidth,clip=true]{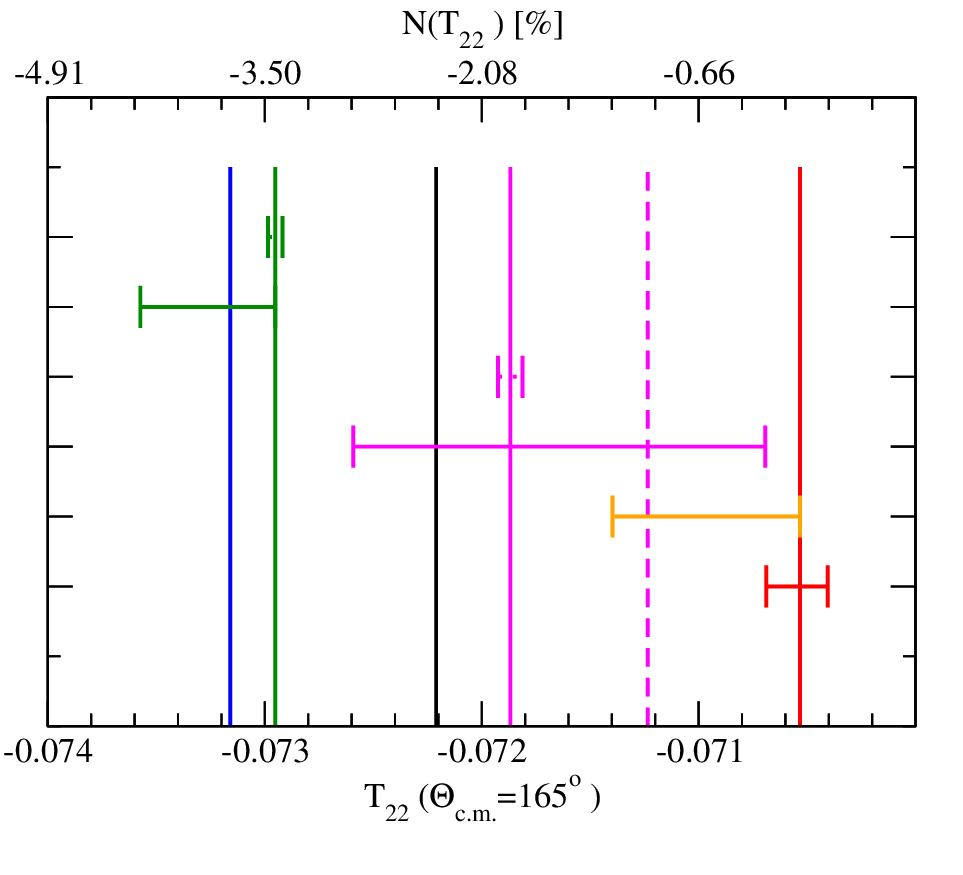}

\includegraphics[width=.264\textwidth,clip=true]{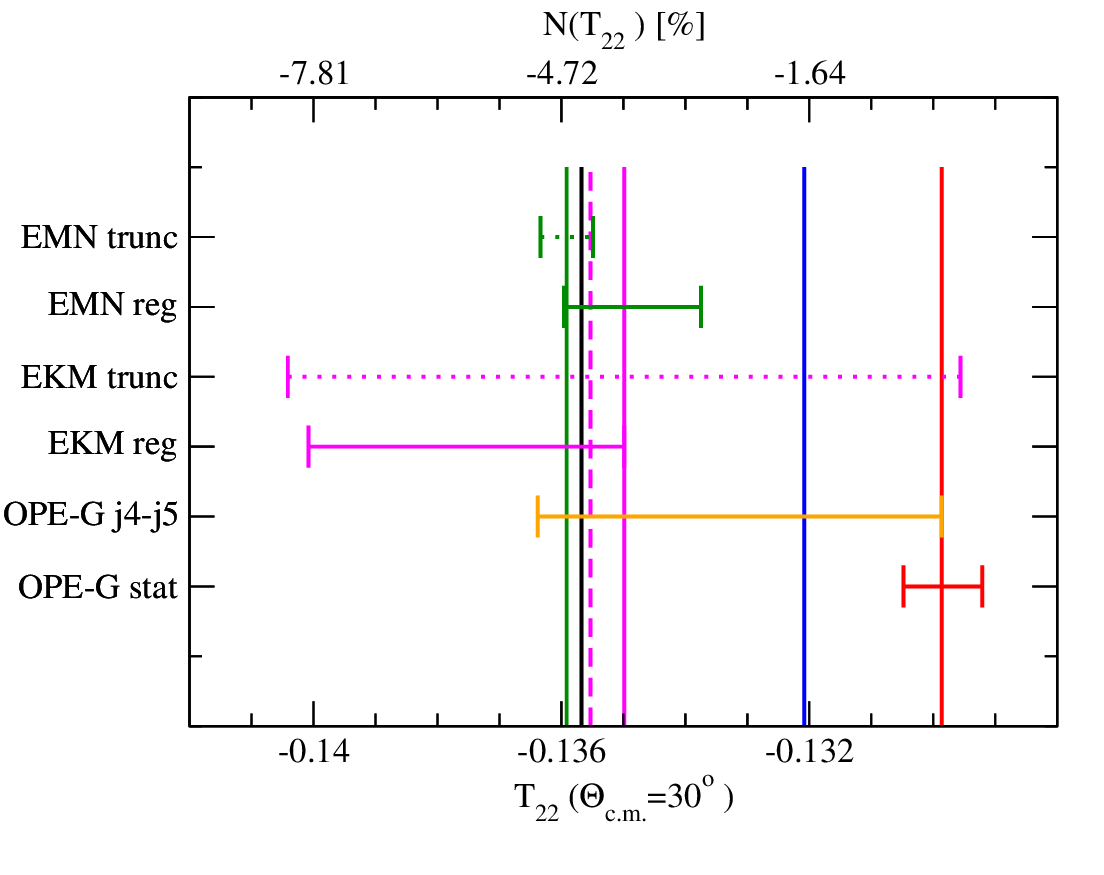}
\includegraphics[width=.23\textwidth,clip=true]{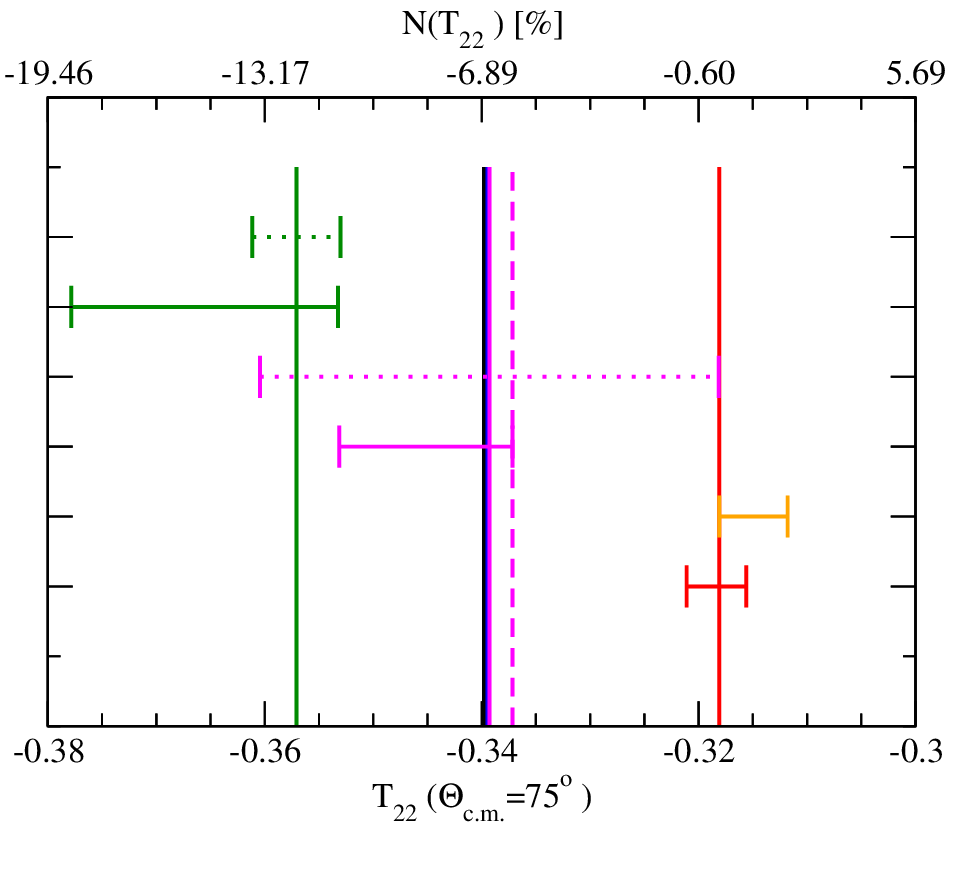}
\includegraphics[width=.23\textwidth,clip=true]{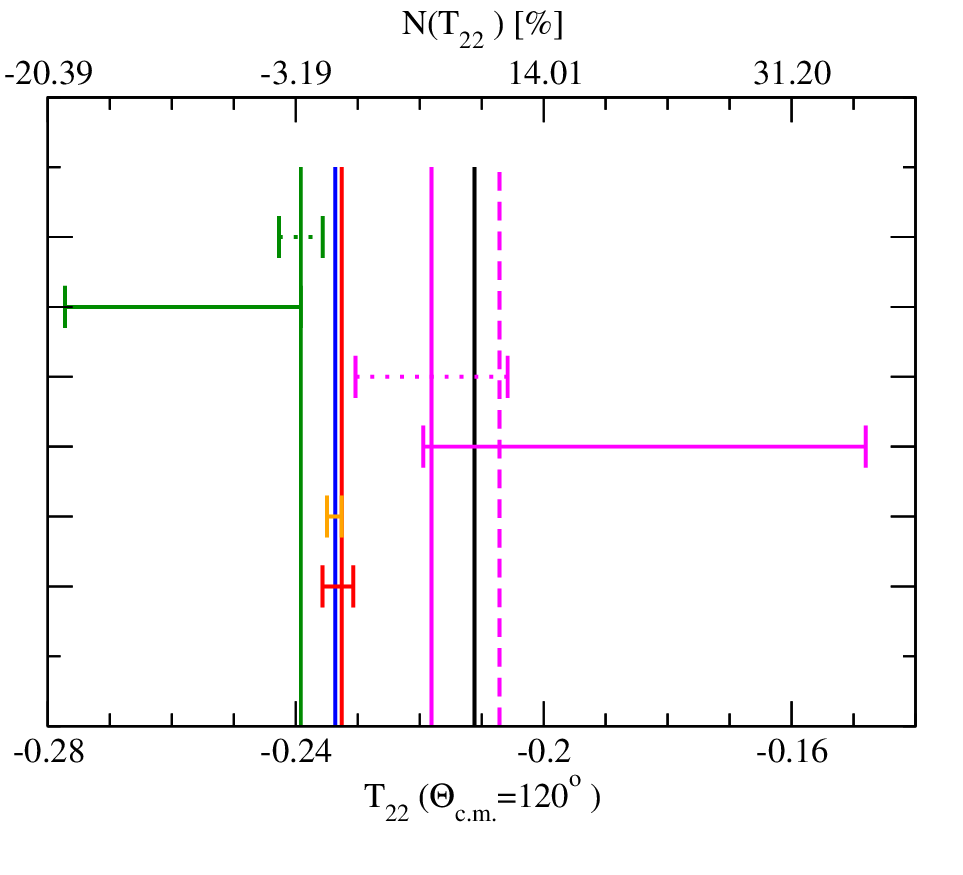}
\includegraphics[width=.23\textwidth,clip=true]{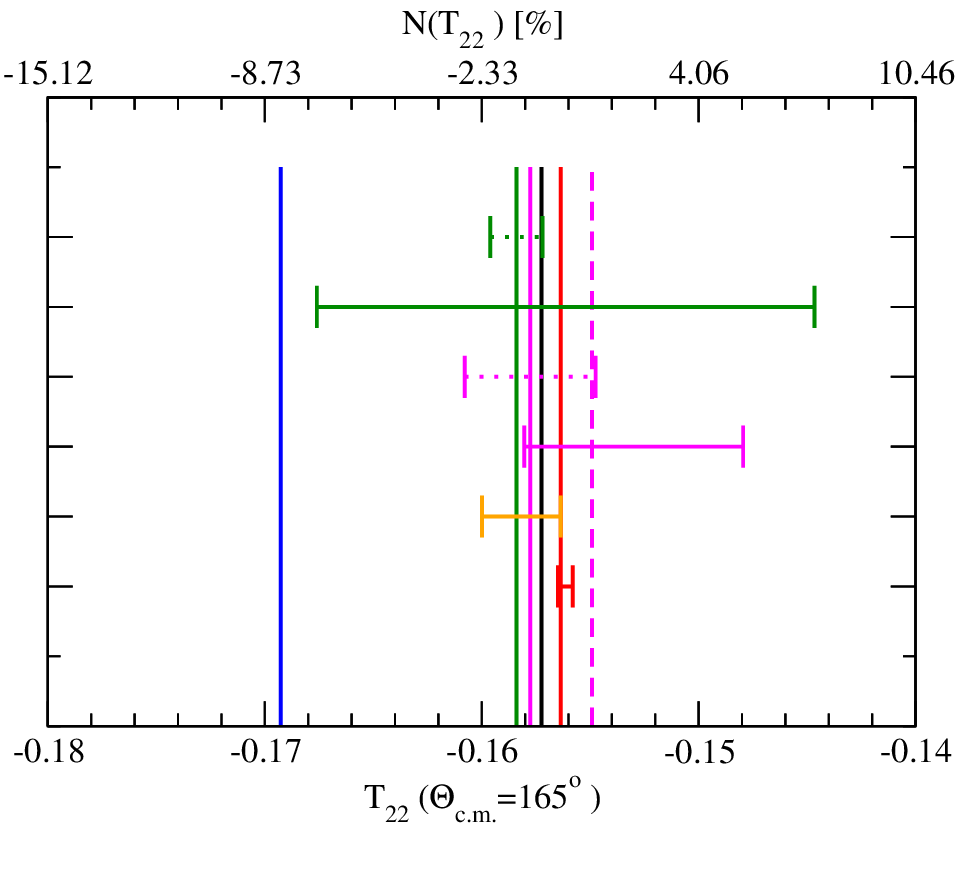}
\caption{The same as in Fig.~\ref{fig11} but for the deuteron tensor analyzing power T$_{22}$.}
\label{fig13}
\end{figure}

\begin{figure}[hbpt]
\includegraphics[width=.264\textwidth,clip=true]{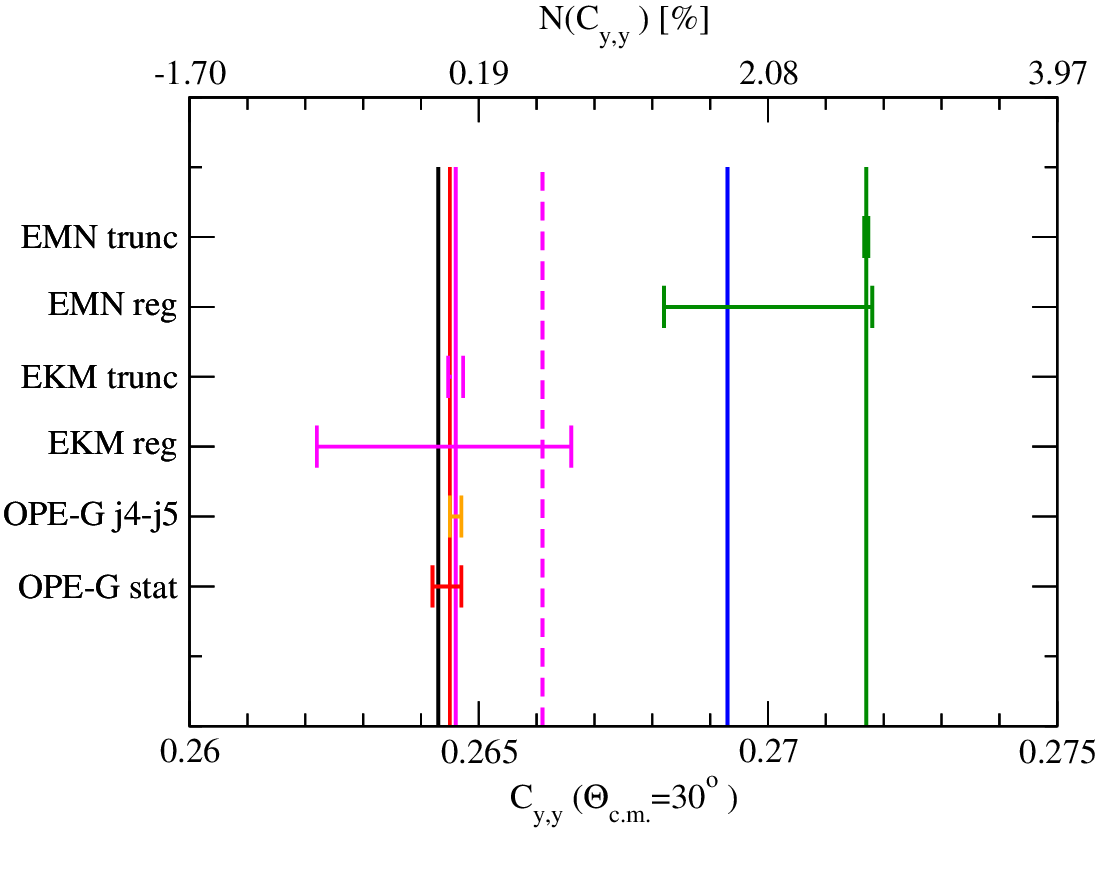}
\includegraphics[width=.23\textwidth,clip=true]{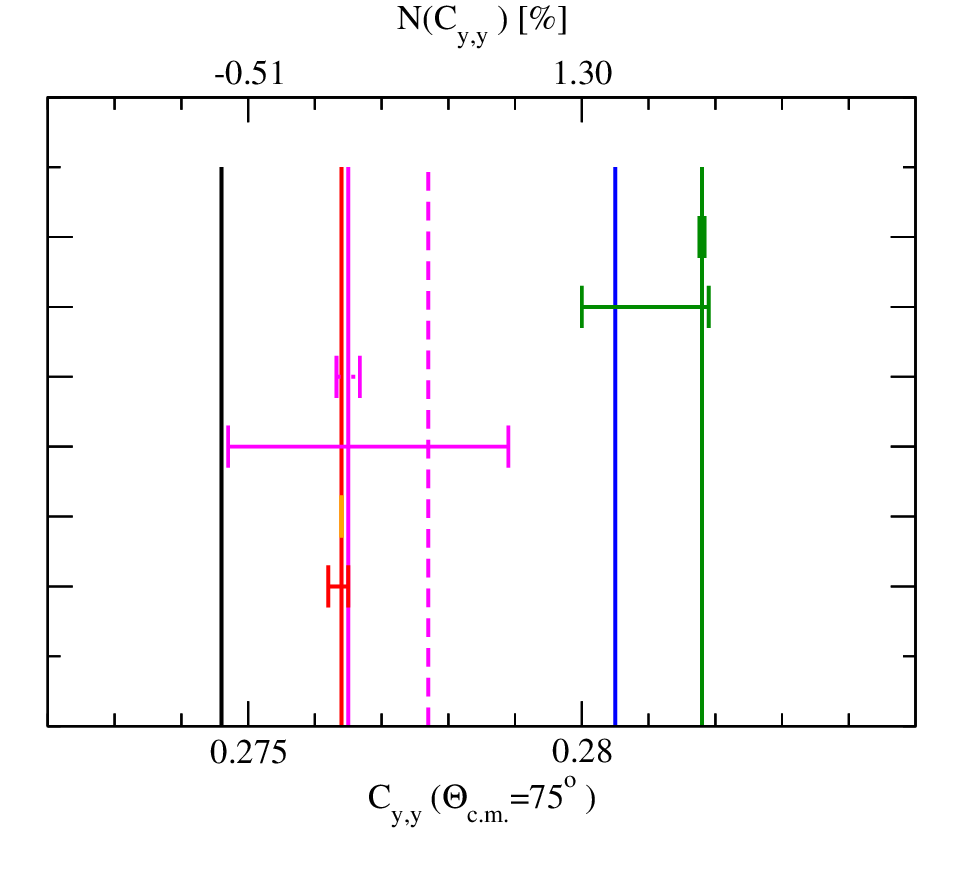}
\includegraphics[width=.23\textwidth,clip=true]{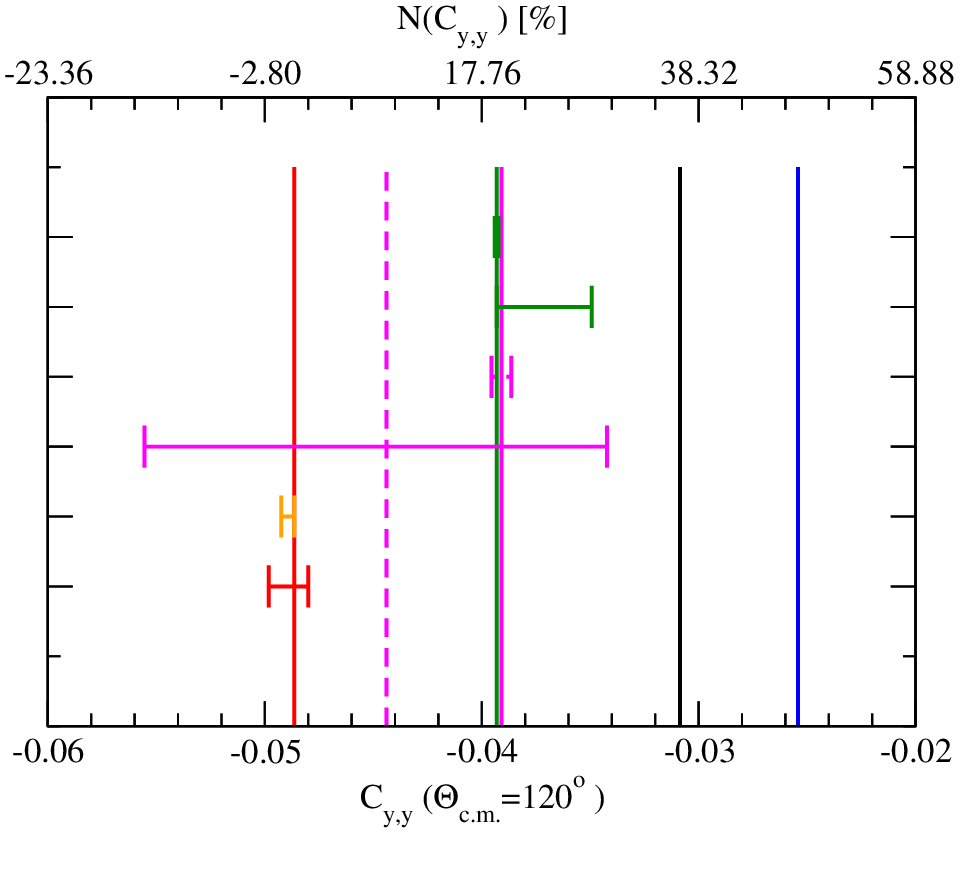}
\includegraphics[width=.23\textwidth,clip=true]{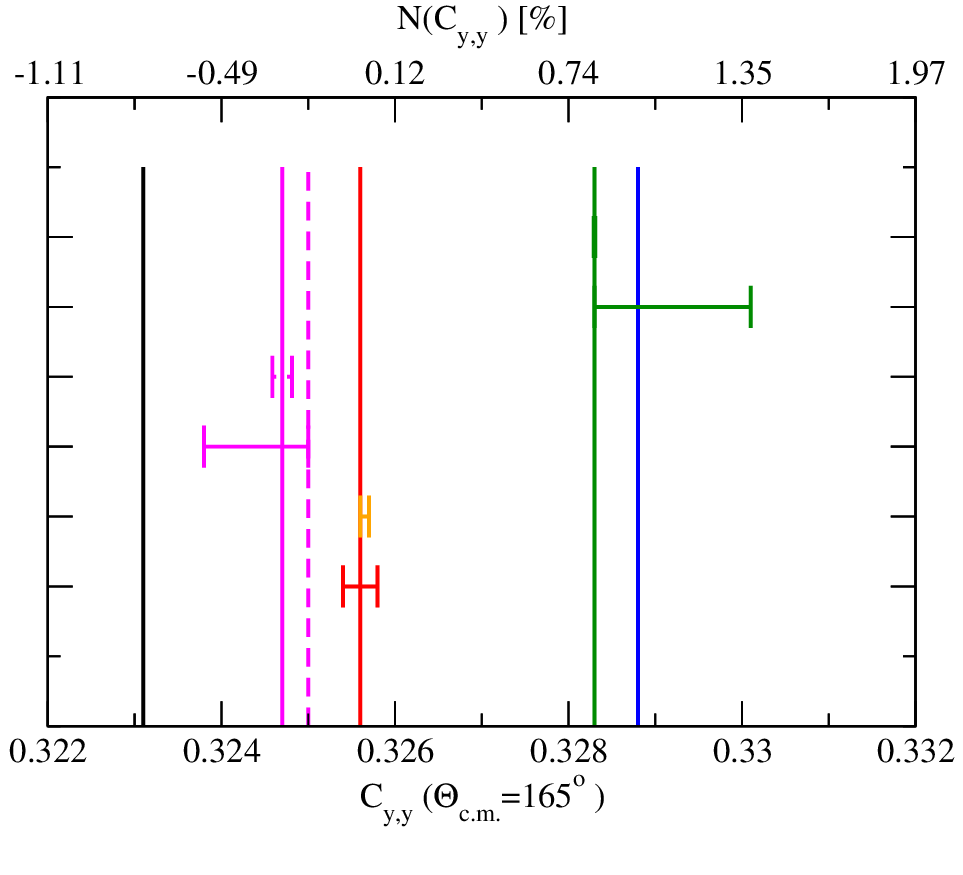}

\includegraphics[width=.264\textwidth,clip=true]{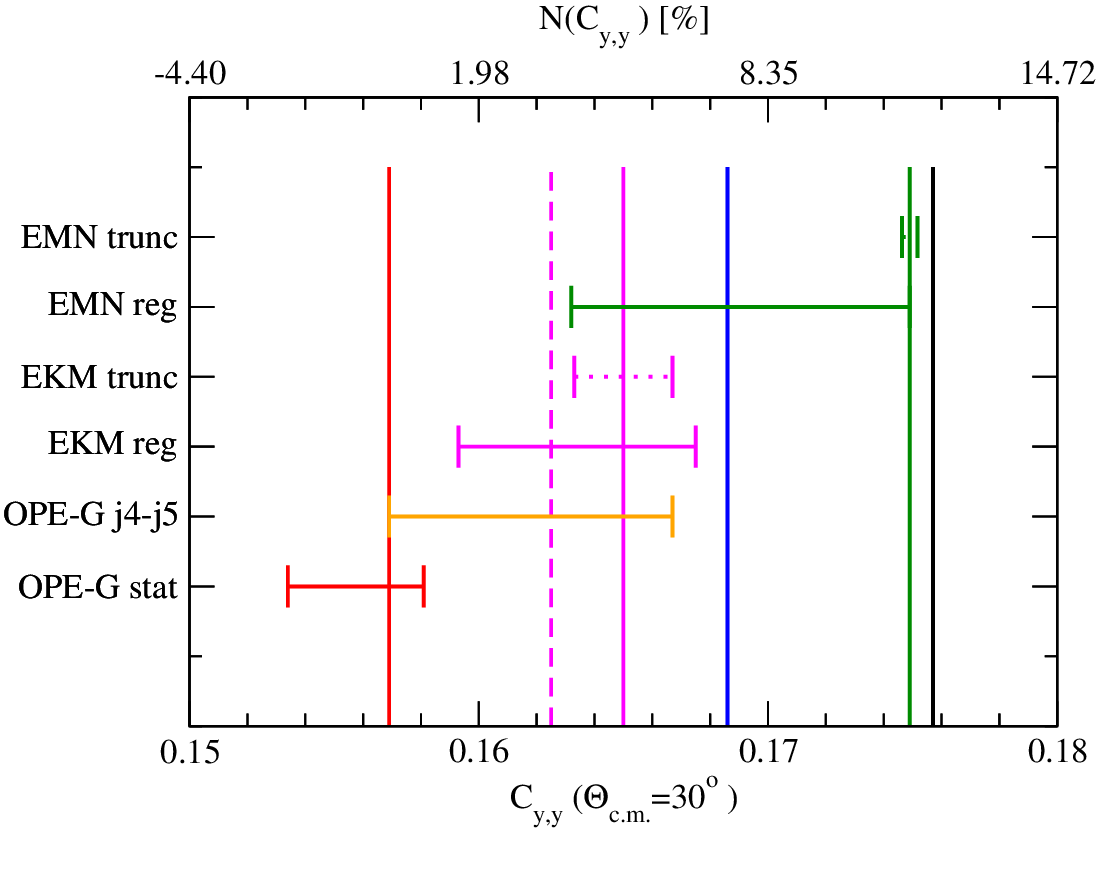}
\includegraphics[width=.23\textwidth,clip=true]{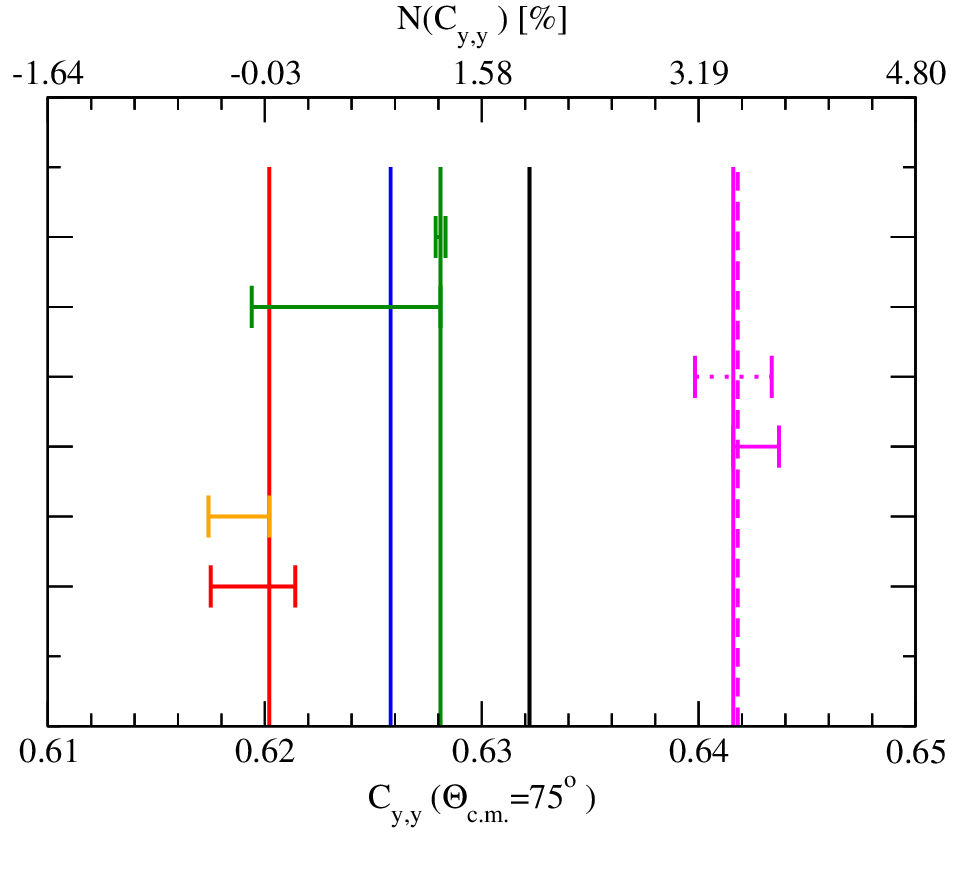}
\includegraphics[width=.23\textwidth,clip=true]{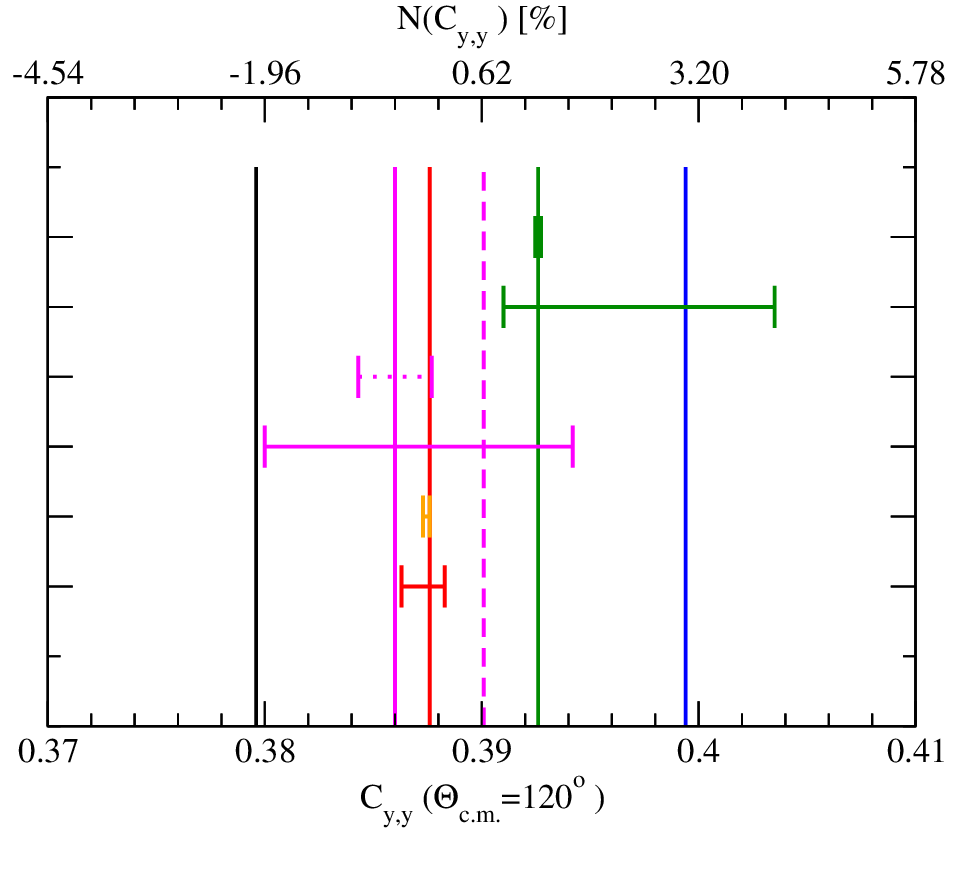}
\includegraphics[width=.23\textwidth,clip=true]{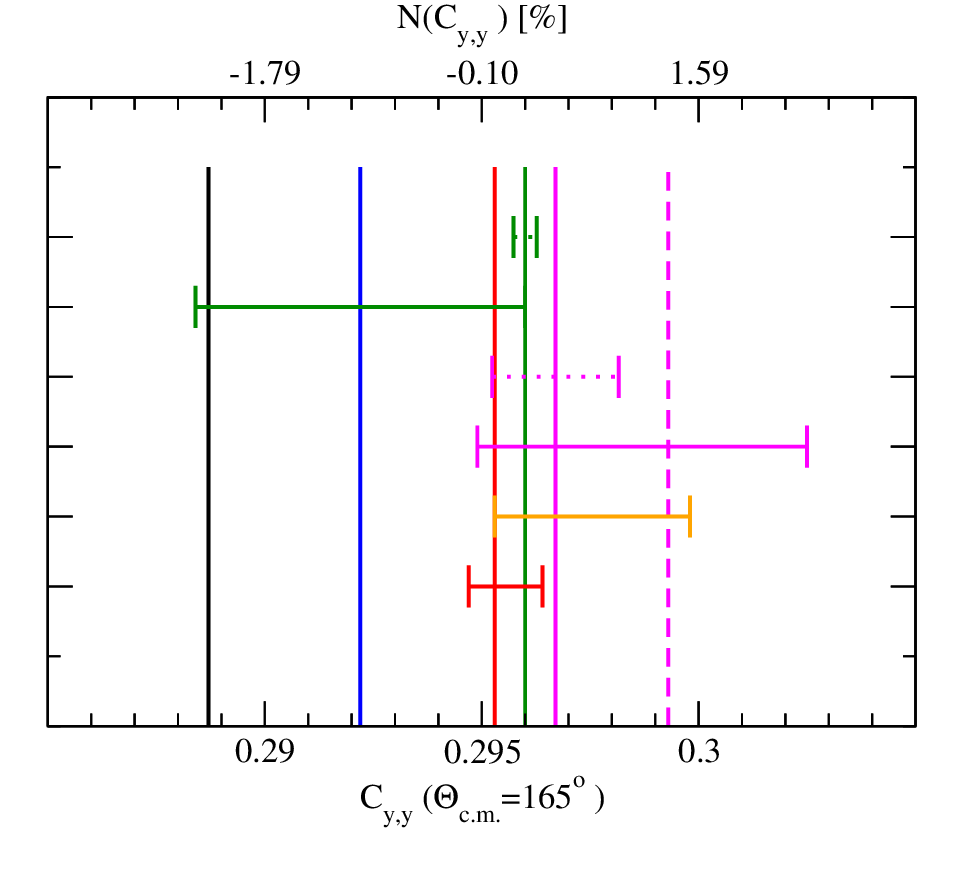}

\includegraphics[width=.264\textwidth,clip=true]{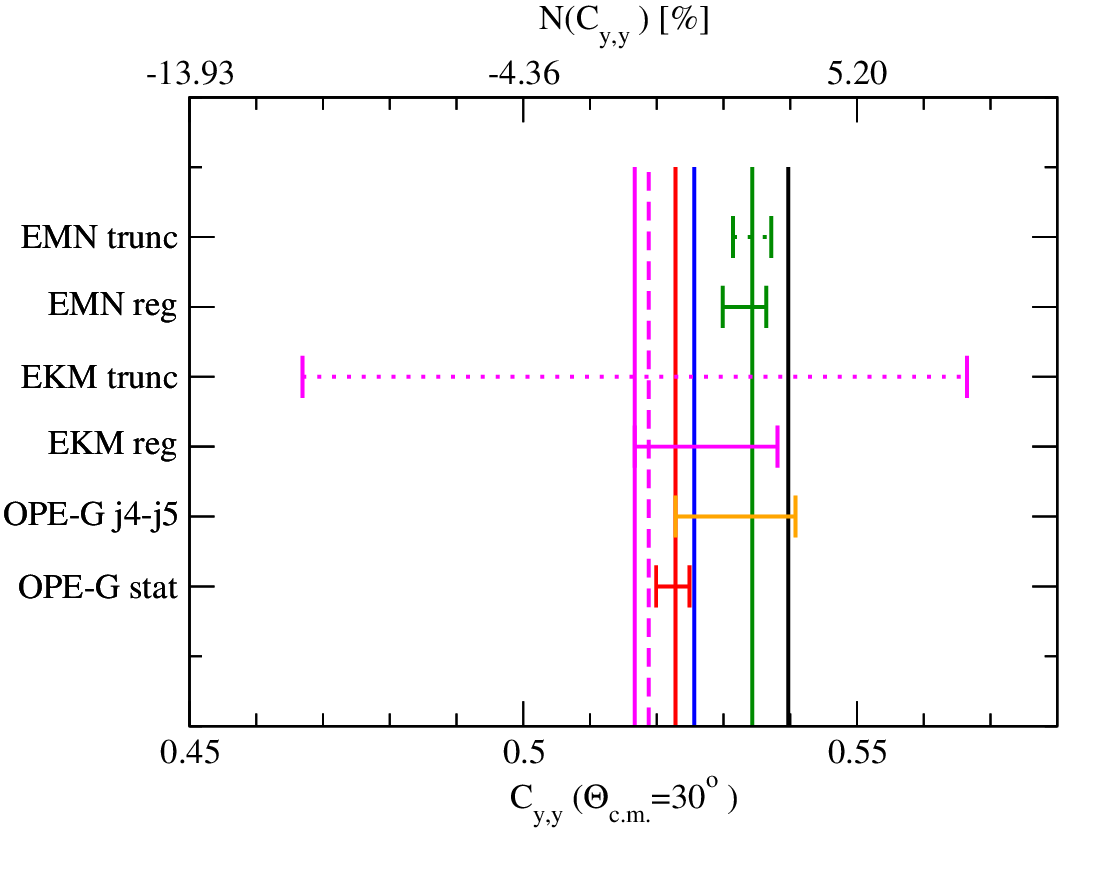}
\includegraphics[width=.23\textwidth,clip=true]{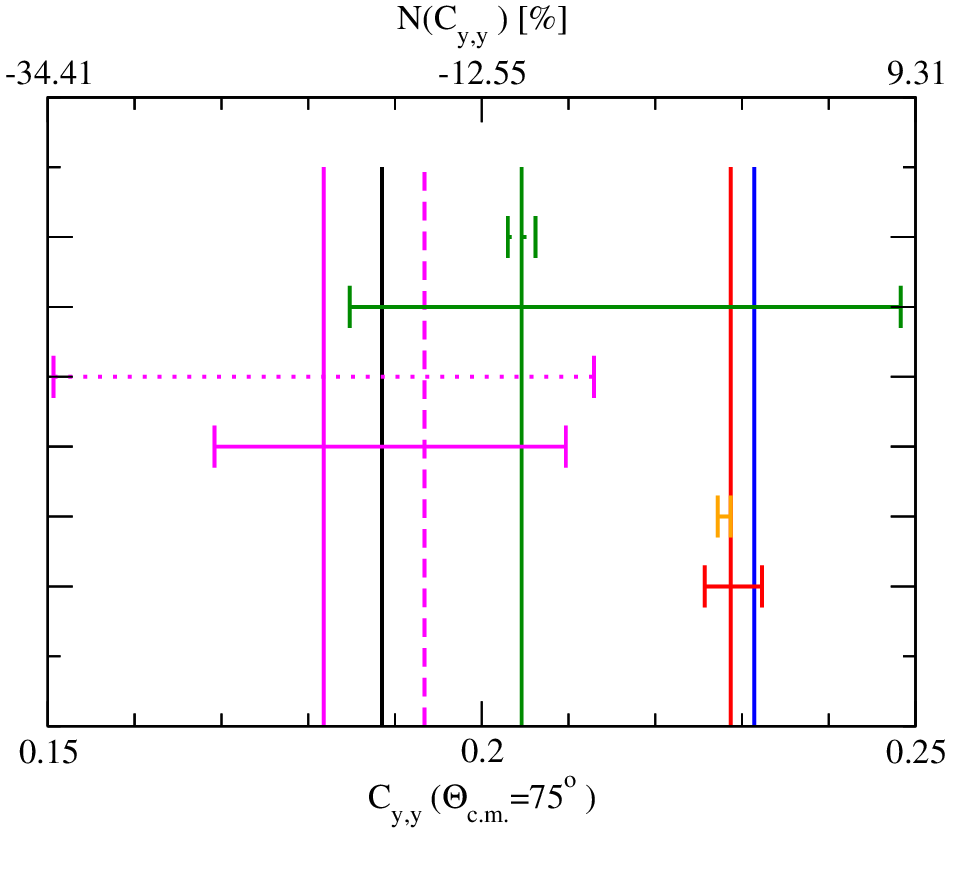}
\includegraphics[width=.23\textwidth,clip=true]{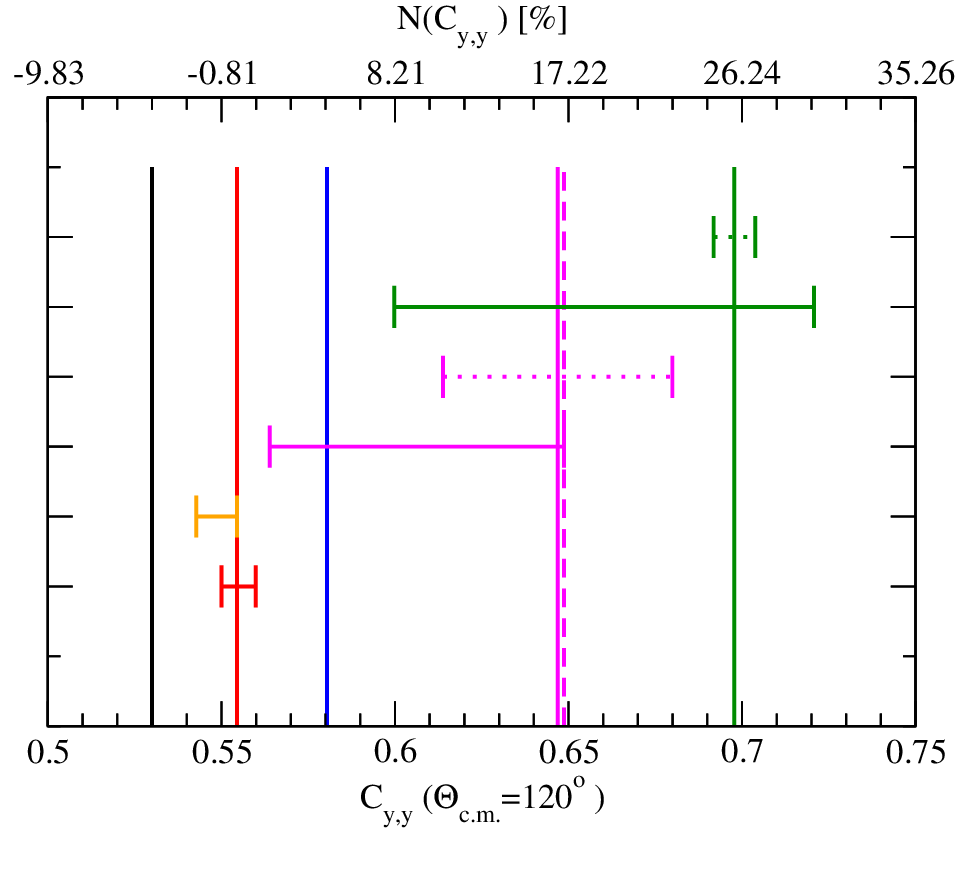}
\includegraphics[width=.23\textwidth,clip=true]{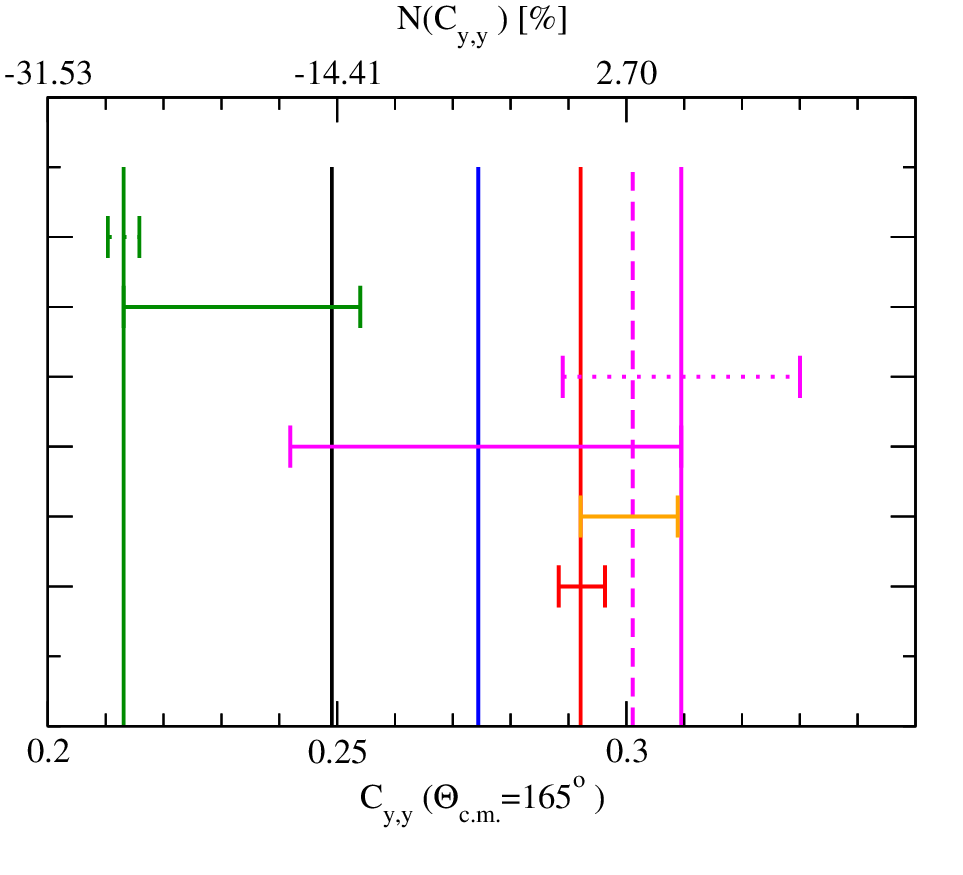}
\caption{The same as in Fig.~\ref{fig11} but for the spin correlation coefficient C$_{\rm{y,y}}$.}
\label{fig14}
\end{figure}

The systematical review of various uncertainties for the differential cross section d$\sigma$/d$\Omega$, 
the nucleon analyzing power A$_{\rm{y}}$, the deuteron tensor analyzing power T$_{22}$, 
and the spin correlation coefficient C$_{\rm{y,y}}$ is given in Figs.~\ref{fig11}-\ref{fig14}, respectively.
In these figures, in each subplot, the predicted value of the observable is given at the bottom horizontal axis and 
the vertical lines are used to mark predictions based on different $NN$ forces and length of these lines has no meaning.
The top horizontal axis shows the percentage relative difference $N(O)$
with respect to the OPE-Gaussian 
prediction and its ticks are calculated as
$\tilde x = (\frac{x-O_{\rm{OPE-Gaussian}}}{O_{\rm{OPE-Gaussian}}})*100*sign(O_{\rm{OPE-Gaussian}})$ where $x$ are the ticks values shown at the 
bottom axis. In addition, for the sake of figures' clarity, the  $\tilde x$'s are rounded to the two digits only.
Note, that the magnitude of such a relative difference depends on the magnitude of the OPE-Gaussian prediction 
and can increase to infinity as the OPE-Gaussian prediction approaches zero. 
The OPE-Gaussian results (at the central values of the parameters) are represented by vertical red lines, the AV18 ones by
the black line, the CD-Bonn predictions by the blue line, the  E-K-M N$^4$LO $R=0.9$~fm results by the magenta solid line,
the E-K-M N$^4$LO $R=1.0$~fm ones by the magenta dashed line, 
and the E-M-N N$^4$LO $\Lambda$=500~MeV ones by the green line.  
Horizontal lines represent magnitudes of various theoretical uncertainties and starting from the bottom they are:
statistical error for the OPE-Gaussian model (the red line), difference between OPE-Gaussian predictions based on the 
$j_{max}=5$ and $j_{max}=4$ calculations (the orange line), regulator dependence for the E-K-M N$^4$LO force in range 
$R$=0.8-1.2~fm (the solid magenta line), the truncation error for the E-K-M N$^4$LO $R=0.9$~fm model (the dashed 
magenta line), regulator dependence for the E-M-N  N$^4$LO force in range $\Lambda$=450-550~MeV
(the solid green line), and truncation error for the E-M-N N$^4$LO $\Lambda$=500~MeV potential (the dashed green line).     
Further, subplots in various rows in Figs.~\ref{fig11}-\ref{fig14} show predictions at different incoming 
nucleon lab. energies, which are $E=13$~MeV (top), $E=65$~MeV (middle) and $E=200$~MeV (bottom). Finally, the 
various columns show predictions at different scattering angles: 30$^{\circ}$, 75$^{\circ}$, 120$^{\circ}$, 
and 165$^{\circ}$ moving from the left to the right. 

An analysis of Figs.~\ref{fig11}-\ref{fig14} leads to the following conclusions:

\begin{enumerate}
\item
In general, all models investigated here provide similar results, which differ only by a few percent at lower energies but
differences 
between predictions 
grow with the increasing energy. 
There is no single
model 
which gives systematically the smallest or the biggest value. There are also no two models, whose predictions 
for all the cases lie close to each other. Note, the above statements describe general trends 
but exceptions from this pattern for specific observables and angles are possible.

\item
At all energies the dominant theoretical uncertainty is the one 
arising from using various models of the nuclear interaction.

\item 
The statistical errors for the OPE-Gaussian predictions are small (and with no practical importance) 
for all the considered observables and energies.

\item
The difference between $j_{max}=5$ and $j_{max}=4$ predictions, as expected, grows with energy, however,
it remains small, when compared to other uncertainties, even at $E=200$~MeV (with the only exceptions of the T$_{22}$ at 200 MeV and C$_{\rm{y,y}}$ at 65 MeV).
Thus the uncertainty bound with partial wave decomposition and numerical performance is also negligible.

\item
The OPE-Gaussian predictions based on the central values are always inside the range given by the statistical errors.
The E-K-M results show monotonic behaviour of the predicted observables with the regulator value.
In the case of the E-M-N force the middle value of regulator ($\Lambda = 500$~MeV) delivers extreme (among the E-M-N ones) 
predictions in many cases.

\item 
The difference between predictions based on the two chiral N$^4$LO models used (E-K-M and E-M-N) is not smaller than 
the difference between any other pair of predictions based on different $NN$ potentials. 
This suggests that there are 
substantial 
differences in the construction each of these models. Thus it seems mandatory
to regard these models independently, as two different models of nuclear forces.

\item
In numerous cases the two chiral approaches deliver results separated from each other by more than the estimated 
uncertainty for their predictions. This again points to 
differences between both chiral potentials  
(and/or to an underestimation of the corresponding total theoretical uncertainties).

\item
In the case of both chiral models, the dominant uncertainty at lower energies arises from the cut-off dependence.
This uncertainty is much bigger than the remaining types of errors, except for differences between various models. 
At higher energies the truncation errors are 
also important in some specific cases, e.g. the differential cross section at $\theta_{\rm{c.m.}}=120^{\circ}$ at $E=200$~MeV.
In the case of A$_{\rm{y}}$ at $E$=200~MeV and smaller angles, the truncation errors exceed the regulator dependence for the E-K-M potential.

\item 
In the case of the N$^4$LO E-K-M potential, the difference between predictions for $R=0.9$~fm and $R=1.0$~fm 
(so at the two preferred 
values of the regulator in the $NN$ system) is of the same size as the typical difference between any other pair of predictions, 
what shows strong sensitivity of the observables to the regulator parameter.

\item 
Comparing the cutoff dependence of both chiral models we can conclude, that the dispersion of their predictions 
behaves for both models in a correlated way, i.e. a big cutoff dependence for the E-M-N force usually appears 
together with a big cutoff dependence for the E-K-M potential. 

\item 
The truncation errors for the E-M-N force are smaller than these for the E-K-M interaction. The reason for this is 
the bigger value of the chiral breaking scale in the E-M-N approach, which results in different values of $Q$
parameter in Eq.~(\ref{therrors}). 

\end{enumerate}  

\begin{figure}[h]
\includegraphics[width=.43\textwidth,clip=true]{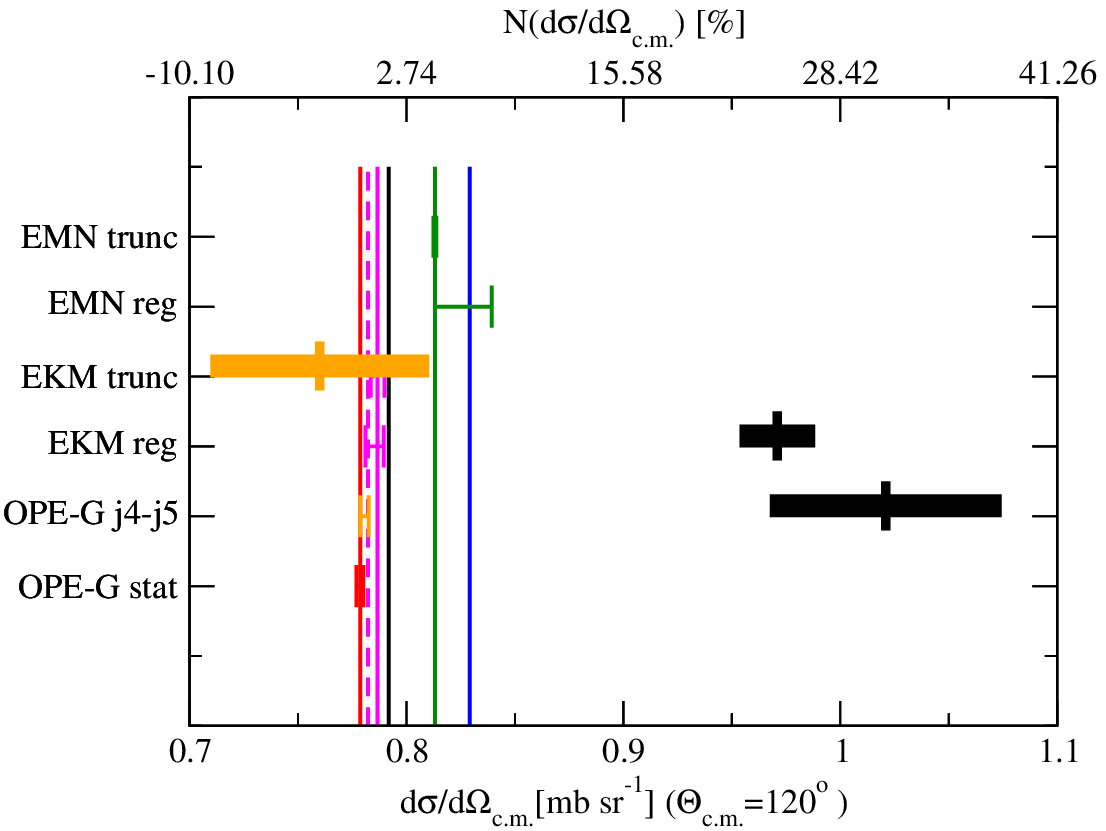}
\includegraphics[width=.43\textwidth,clip=true]{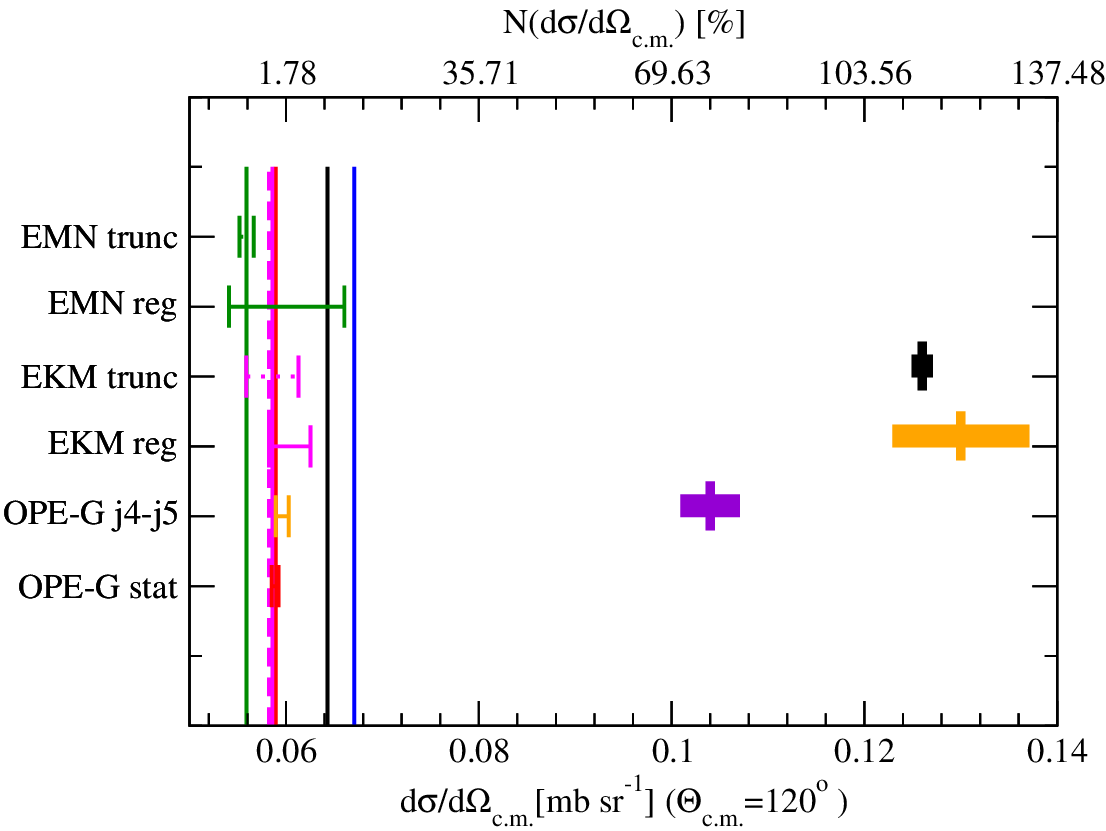}
\includegraphics[width=.43\textwidth,clip=true]{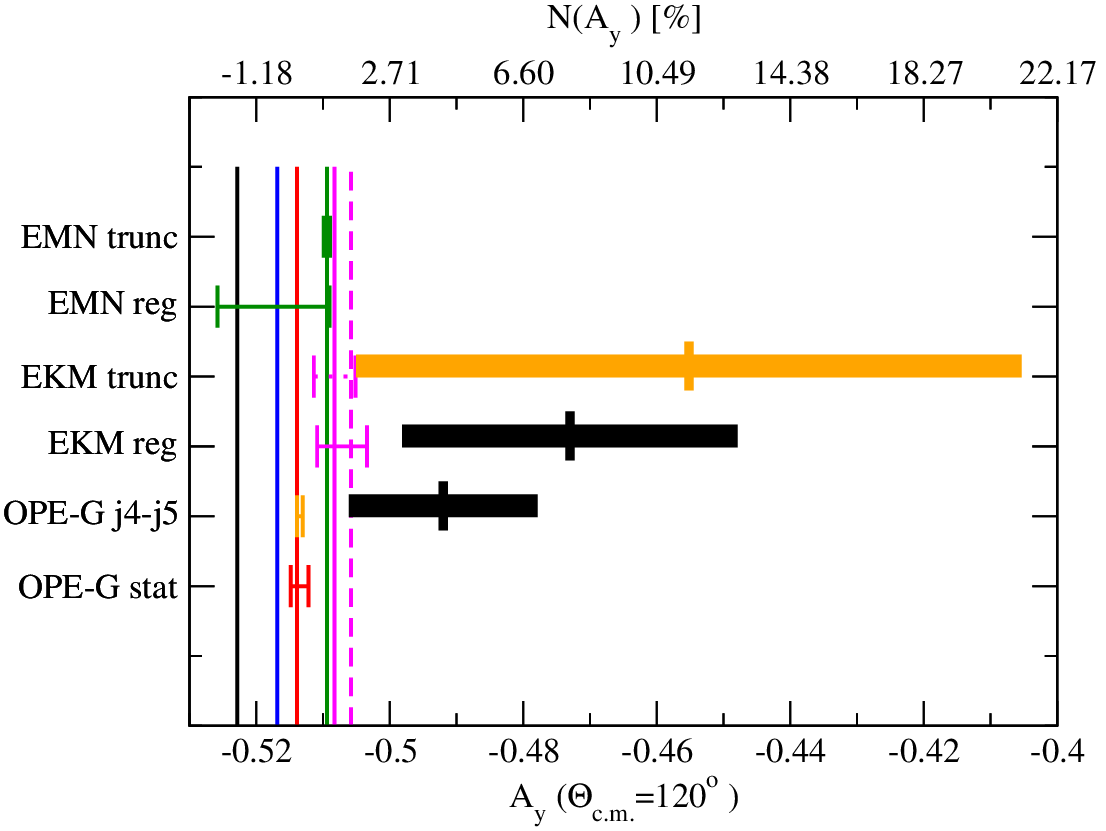}
\includegraphics[width=.43\textwidth,clip=true]{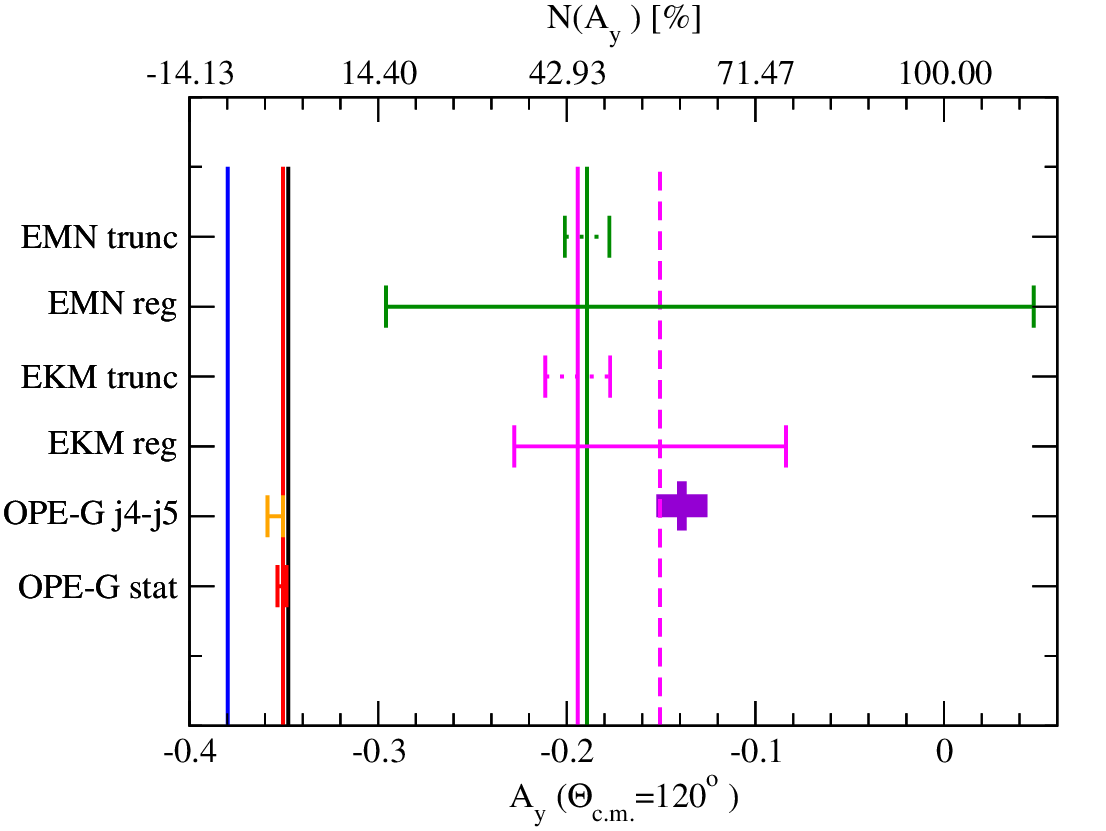}
\caption{The same as in Fig.~\ref{fig11} ($d\sigma/d\Omega$, the upper row) and Fig.~\ref{fig12} (A$_{\rm{y}}$, 
the bottom row) at $\theta_{\rm{c.m.}}=120^{\circ}$
for $E$=65~MeV (left) and $E$=200~MeV (right) but supplemented by the experimental points at angles 
near $\theta_{\rm{c.m.}}=120^{\circ}$. Vertical and thin horizontal lines are as in Fig.~\ref{fig11},
and filled rectangles represent experimental data and their statistical errors, as in Figs.~\ref{fig5} and~\ref{fig6}.
}
\label{fig15}
\end{figure}

Next, it is interesting to compare the size of the theoretical errors presented in Figs.~\ref{fig11}-\ref{fig14}
to experimental errors of available data.
In order not to leave the reader with the impression that the modern theoretical models of nuclear interactions
yield a chaotic description of the $Nd$ scattering observables, in Fig.~\ref{fig15} we compare, in a few examples,
previously presented predictions with the experimental results. This establishes an absolute scale in which
one has to peer at the problem of discrepancies between various theoretical models.

Examples given in Fig.~\ref{fig15} show various possible locations 
of theoretical predictions and data.
The differential cross section at $E=65$~MeV and $E=200$~MeV at the scattering angle $\theta_{\rm{c.m.}}=120^{\circ}$
is shown in the upper row and the analyzing power A$_{\rm{y}}$ for the same angle and energies is displayed below.
In the case of the cross section we see that at $E=65$~MeV there are discrepancies between various theoretical 
predictions and the data of different measurements. While the theoretical predictions are close one to each other, 
the data are scattered. One experimental point 
overlaps within its statistical error with some of the predictions, another one would be in agreement 
with predictions within 3$\sigma$ distance and the remaining experimental point is further 
from the data by more than its 3$\sigma$ uncertainty.
At $E=200$~MeV a clear discrepancy between all predictions, which again are close together, and all 
data is observed. This discrepancy can be traced back to action of 3$N$ force at higher energies~\cite{WitalaPRL,Kuros}.
The picture is more complex for the analyzing power. Here, at $E=65$~MeV the experimental data and 
predictions differ by more than experimental error but they already agree within the 2$\sigma$ range. At $E=200$
the experimental statistical error is much smaller than the distances between various theoretical predictions and 
the uncertainties related to the chiral forces. Such a mixed pattern clearly calls for further work on reducing both the 
theoretical and experimental uncertainties to 
avoid misleading conclusions
about 
the properties of the nuclear interactions. 
The presented here examples at one scattering angle only show that it is much more reliable to draw 
conclusions based on a comparison of predictions
with data in a wider range of scattering angles and at different energies.
Especially, these examples do not contradict strong  
effects of the 3$N$ force  in the minimum of the differential cross section at higher energies~\cite{WitalaPRL,Kuros}.
Such conclusions are based on a systematic comparison of predictions with the data at numerous scattering angles and energies.

\section{Summary}
\label{Summary}

We have employed the OPE-Gaussian potential of the Granada group to describe the elastic $Nd$
scattering at energies up to 200 MeV.
The OPE-Gaussian potential is one of the first
models of nuclear forces for which the covariance matrix of its free
parameters is known.
This gives an excellent opportunity to study the propagation
of uncertainties from the 2$N$ potential parameters to 3$N$ observables.
Therefore, for the same process, we also studied the  
statistical errors of our predictions.

The description of data delivered by the OPE-Gaussian force is in quantitative
agreement with picture obtained using other $NN$ potentials, especially the AV18 model,
which resembles by construction the OPE-Gaussian potential.
We found only small discrepancies between predictions of these forces, especially at the highest 
energy investigated here, $E=200$~MeV, which can very probably originate from a slightly different behaviour of the phase shifts
for the AV18 and the OPE-Gaussian potentials at energies above $\approx 150$~MeV. 
It should be noted that the procedure of fixing
free parameters 
for the OPE-Gaussian 
force has been performed with big 
care for statistical correctness and covers new 2$N$
data not included when fixing the AV18's parameters.

In order to obtain the theoretical uncertainty of our predictions arising from 
the uncertainty of the $NN$ potential parameters, we employed the statistical approach:
we computed the $Nd$ scattering observables using fifty sets of the 
OPE-Gaussian potential parameters obtained from a
suitable multivariate probability distribution. Next, we 
investigated a distribution of our results and adopted one of estimators 
of their dispersion, the $\frac12 \Delta_{68\%}$, as a measure of the theoretical statistical uncertainty.
We also compared such statistical uncertainties for different observables with various types 
of theoretical errors, including the truncation errors and a dispersion due to 
using various models of the nuclear interaction.
A comparison of uncertainties for the $Nd$ elastic scattering cross section 
and a few polarization observables for the
OPE-Gaussian model with other types of theoretical uncertainties leads to important
conclusions about currently used models of 2$N$ forces.
First, all models of the $NN$ interaction considered here deliver qualitatively
and quantitatively similar predictions for the $Nd$ elastic scattering observables.
None of the interactions yields predictions 
systematically different 
from others 
and also no systematic grouping of predictions is observed.
Secondly, we have found that in the case of the chiral forces,
at small and medium energies, which are their natural domain of applicability,
the dependence of predictions on the
values of regulators dominates over another types of theoretical errors.
At the highest investigated energy $E=200$~MeV which is at the 
limit of applicability of
chiral forces, the truncation errors
become important. 
It follows that during a derivation 
of the chiral models a
constant attention should be paid to the regularization methods applied.
Current attempts to solve this problem
result in a range of regulator parameters too broad to make the
chiral forces such a precise tool in studies of nuclear reactions as desired and expected. 
It would be very interesting to check if this conclusion remains valid
after taking into account also consistent 3$N$ interaction at the investigated here order (N$^4$LO)
of chiral expansion.

Altogether the presented results clearly show that the modern nuclear experiments and 
theoretical approaches for the $Nd$ scattering achieved similar precision. 
Having in mind that many investigations are currently focused 
on studying subtle details of underlying phenomena, there is a need to further improve precision 
both in theoretical as well as in experimental studies.
From the theoretical side a continuous progress in deriving consistent $NN$ and 3$N$ forces from the $\chi$EFT
gives hope that this goal will be achieved.  

\acknowledgments

We thank Dr.~E.~Ruiz Arriola and Dr.~R.~Navarro P\'erez for sending us sets of parameters
for the OPE-Gaussian model and for the valuable discussions.
This work is a part of the LENPIC project and was supported by the Polish National Science
Center under Grants No. 2016/22/M/ST2/00173
and 2016/21/D/ST2/01120.
The numerical calculations were partially performed on the supercomputer cluster of the JSC,
J\"ulich, Germany.

\end{document}